\documentclass[longauth]{aa}   
\usepackage{graphicx}           
\usepackage{url}
\usepackage{epstopdf}
\usepackage{color}
\usepackage{textcomp}
\usepackage{gensymb}
\usepackage{multirow}
\usepackage{ragged2e}
\usepackage{url}
\usepackage{amsmath}
\usepackage{booktabs}
\usepackage{comment}
\usepackage{afterpage}
\usepackage{mathtools}
\usepackage{tabularx}
\usepackage{bold-extra}
\usepackage{xspace}
\usepackage{relsize}
\usepackage{newunicodechar}
\usepackage{longtable}
\usepackage{enumerate}
\usepackage{amssymb}
\usepackage{natbib}
\usepackage{soul}
\usepackage[encapsulated]{CJK}
\usepackage{ucs}
\usepackage[utf8]{inputenc}
\usepackage{txfonts}
\usepackage[breaklinks=true]{hyperref}

\usepackage{orcidlink}
\let\orcid\orcidlink
\makeatletter
\renewcommand*\maketitle{%
  \thispagestyle{firstpage}
\begingroup
    \if@wideboxfn
    \setlength\bibindent{1.4\parindent}
    \else
    \setlength\bibindent{\parindent}
    \fi
    \renewcommand*\thefootnote{\@fnsymbol\c@footnote}%
    \renewcommand\@makefntext[1]{%
    \ifaa@longfn\hsize\textwidth\fi
    \noindent
    \hb@xt@\bibindent{\hss\@makefnmark\enspace}##1}
  \ifaa@twocolumn
  \begin{aa@strip}
    \aa@maketitle
    \@thanks
  \end{aa@strip}
  \else
    \begingroup
      \let\thanks\footnote
      \aa@maketitle
    \endgroup
  \fi
\endgroup
  \setcounter{footnote}{0}%
}

\renewcommand*\aa@pageof{, page \thepage{} of \pageref*{LastPage}}
\makeatother

\usepackage[encapsulated]{CJK}
\usepackage{ucs}

\usepackage[utf8]{inputenc} 
\newcommand{\cntext}[1]{\begin{CJK}{UTF8}{gbsn}#1\end{CJK}}

\newcommand{\lsim}{~\rlap{$<$}{\lower 1.0ex\hbox{$\sim$}}}
\newcommand{\gsim}{~\rlap{$>$}{\lower 1.0ex\hbox{$\sim$}}}

\newcommand{\pas}{$\rlap{.}^{\prime\prime}$}

\newcommand{\dg}{$^{\circ}$}

\newcommand{\radmsq}{rad~m$^{-2}$}

\newcommand{\casa}{{\sc CASA}}

\definecolor{darkgreen}{rgb}{0,.3,0}

\begin{document}

\title{First polarization study of the M87 jet and active galactic nuclei at submillimeter wavelengths with ALMA}

\titlerunning{First polarization study of the M87 jet and AGNs at submillimeter wavelengths with ALMA}

\author{
Ciriaco Goddi\orcid{0000-0002-2542-7743}\inst{\ref{inst1},\ref{inst2},\ref{inst3}}\and
Douglas F. Carlos\orcid{0000-0002-1340-7702}\inst{\ref{inst2}}\and
Geoffrey B. Crew\orcid{0000-0002-2079-3189}\inst{\ref{inst4}}\and
Lynn D. Matthews\orcid{0000-0002-3728-8082}\inst{\ref{inst4}}\and
Hugo Messias\orcid{0000-0002-2985-7994}\inst{}\and
Alejandro Mus\orcid{0000-0003-0329-6874}\inst{\ref{inst1},\ref{inst6}}\and
Iván Martí-Vidal\orcid{0000-0003-3708-9611}\inst{\ref{inst7},\ref{inst8}}\and
Ezequiel Albentosa-Ruíz\orcid{0000-0002-7816-6401}\inst{\ref{inst7}}\and
Mariafelicia De Laurentis\orcid{0000-0002-9945-682X}\inst{\ref{inst9},\ref{inst10}}\and
Elisabetta Liuzzo\orcid{0000-0003-0995-5201}\inst{\ref{inst11}}\and
Nicola Marchili\orcid{0000-0002-5523-7588}\inst{\ref{inst11},\ref{inst12}}\and
Kazi L. J. Rygl\orcid{0000-0003-4146-9043}\inst{\ref{inst11}}\and
Kazunori Akiyama\orcid{0000-0002-9475-4254}\inst{\ref{inst4},\ref{inst13},\ref{inst14}}\and
Antxon Alberdi\orcid{0000-0002-9371-1033}\inst{\ref{inst15}}\and
Walter Alef\inst{\ref{inst12}}\and
Juan Carlos Algaba\orcid{0000-0001-6993-1696}\inst{\ref{inst16}}\and
Richard Anantua\orcid{0000-0003-3457-7660}\inst{\ref{inst17},\ref{inst18},\ref{inst14},\ref{inst19}}\and
Keiichi Asada\orcid{0000-0001-6988-8763}\inst{\ref{inst20}}\and
Rebecca Azulay\orcid{0000-0002-2200-5393}\inst{\ref{inst7},\ref{inst8},\ref{inst12}}\and
Uwe Bach\orcid{0000-0002-7722-8412}\inst{\ref{inst12}}\and
Anne-Kathrin Baczko\orcid{0000-0003-3090-3975}\inst{\ref{inst21},\ref{inst12}}\and
David Ball\inst{\ref{inst22}}\and
Mislav Baloković\orcid{0000-0003-0476-6647}\inst{\ref{inst23}}\and
Bidisha Bandyopadhyay\orcid{0000-0002-2138-8564}\inst{\ref{inst24}}\and
John Barrett\orcid{0000-0002-9290-0764}\inst{\ref{inst4}}\and
Michi Bauböck\orcid{0000-0002-5518-2812}\inst{\ref{inst25}}\and
Bradford A. Benson\orcid{0000-0002-5108-6823}\inst{\ref{inst26},\ref{inst27}}\and
Dan Bintley\inst{\ref{inst28},\ref{inst29}}\and
Lindy Blackburn\orcid{0000-0002-9030-642X}\inst{\ref{inst14},\ref{inst19}}\and
Raymond Blundell\orcid{0000-0002-5929-5857}\inst{\ref{inst19}}\and
Katherine L. Bouman\orcid{0000-0003-0077-4367}\inst{\ref{inst30}}\and
Geoffrey C. Bower\orcid{0000-0003-4056-9982}\inst{\ref{inst31},\ref{inst32}}\and
Michael Bremer\inst{\ref{inst33}}\and
Roger Brissenden\orcid{0000-0002-2556-0894}\inst{\ref{inst14},\ref{inst19}}\and
Silke Britzen\orcid{0000-0001-9240-6734}\inst{\ref{inst12}}\and
Avery E. Broderick\orcid{0000-0002-3351-760X}\inst{\ref{inst34},\ref{inst35},\ref{inst36}}\and
Dominique Broguiere\orcid{0000-0001-9151-6683}\inst{\ref{inst33}}\and
Thomas Bronzwaer\orcid{0000-0003-1151-3971}\inst{\ref{inst37}}\and
Sandra Bustamante\orcid{0000-0001-6169-1894}\inst{\ref{inst38}}\and
John E. Carlstrom\orcid{0000-0002-2044-7665}\inst{\ref{inst39},\ref{inst27},\ref{inst40},\ref{inst41}}\and
Andrew Chael\orcid{0000-0003-2966-6220}\inst{\ref{inst42}}\and
Chi-kwan Chan\orcid{0000-0001-6337-6126}\inst{\ref{inst22},\ref{inst43},\ref{inst44}}\and
Dominic O. Chang\orcid{0000-0001-9939-5257}\inst{\ref{inst14},\ref{inst19}}\and
Koushik Chatterjee\orcid{0000-0002-2825-3590}\inst{\ref{inst14},\ref{inst19}}\and
Shami Chatterjee\orcid{0000-0002-2878-1502}\inst{\ref{inst45}}\and
Ming-Tang Chen\orcid{0000-0001-6573-3318}\inst{\ref{inst31}}\and
Yongjun Chen (\cntext{陈永军})\orcid{0000-0001-5650-6770}\inst{\ref{inst46},\ref{inst47}}\and
Xiaopeng Cheng\orcid{0000-0003-4407-9868}\inst{\ref{inst48}}\and
Ilje Cho\orcid{0000-0001-6083-7521}\inst{\ref{inst48},\ref{inst49},\ref{inst15}}\and
Pierre Christian\orcid{0000-0001-6820-9941}\inst{\ref{inst50}}\and
Nicholas S. Conroy\orcid{0000-0003-2886-2377}\inst{\ref{inst51},\ref{inst19}}\and
John E. Conway\orcid{0000-0003-2448-9181}\inst{\ref{inst21}}\and
Thomas M. Crawford\orcid{0000-0001-9000-5013}\inst{\ref{inst27},\ref{inst39}}\and
Alejandro Cruz-Osorio\orcid{0000-0002-3945-6342}\inst{\ref{inst52},\ref{inst53}}\and
Yuzhu Cui (\cntext{崔玉竹})\orcid{0000-0001-6311-4345}\inst{\ref{inst54},\ref{inst55}}\and
Brandon Curd\orcid{0000-0002-8650-0879}\inst{\ref{inst17},\ref{inst14},\ref{inst19}}\and
Rohan Dahale\orcid{0000-0001-6982-9034}\inst{\ref{inst15}}\and
Jordy Davelaar\orcid{0000-0002-2685-2434}\inst{\ref{inst56},\ref{inst57}}\and
Roger Deane\orcid{0000-0003-1027-5043}\inst{\ref{inst58},\ref{inst59},\ref{inst60}}\and
Jessica Dempsey\orcid{0000-0003-1269-9667}\inst{\ref{inst28},\ref{inst29},\ref{inst61}}\and
Gregory Desvignes\orcid{0000-0003-3922-4055}\inst{\ref{inst12},\ref{inst62}}\and
Jason Dexter\orcid{0000-0003-3903-0373}\inst{\ref{inst63}}\and
Vedant Dhruv\orcid{0000-0001-6765-877X}\inst{\ref{inst25}}\and
Indu K. Dihingia\orcid{0000-0002-4064-0446}\inst{\ref{inst55}}\and
Sheperd S. Doeleman\orcid{0000-0002-9031-0904}\inst{\ref{inst14},\ref{inst19}}\and
Sergio A. Dzib\orcid{0000-0001-6010-6200}\inst{\ref{inst12}}\and
Ralph P. Eatough\orcid{0000-0001-6196-4135}\inst{\ref{inst64},\ref{inst12}}\and
Razieh Emami\orcid{0000-0002-2791-5011}\inst{\ref{inst19}}\and
Heino Falcke\orcid{0000-0002-2526-6724}\inst{\ref{inst37}}\and
Joseph Farah\orcid{0000-0003-4914-5625}\inst{\ref{inst65},\ref{inst66}}\and
Vincent L. Fish\orcid{0000-0002-7128-9345}\inst{\ref{inst4}}\and
Edward Fomalont\orcid{0000-0002-9036-2747}\inst{\ref{inst67}}\and
H. Alyson Ford\orcid{0000-0002-9797-0972}\inst{\ref{inst22}}\and
Marianna Foschi\orcid{0000-0001-8147-4993}\inst{\ref{inst15}}\and
Raquel Fraga-Encinas\orcid{0000-0002-5222-1361}\inst{\ref{inst37}}\and
William T. Freeman\inst{\ref{inst68},\ref{inst69}}\and
Per Friberg\orcid{0000-0002-8010-8454}\inst{\ref{inst28},\ref{inst29}}\and
Christian M. Fromm\orcid{0000-0002-1827-1656}\inst{\ref{inst70},\ref{inst53},\ref{inst12}}\and
Antonio Fuentes\orcid{0000-0002-8773-4933}\inst{\ref{inst15}}\and
Peter Galison\orcid{0000-0002-6429-3872}\inst{\ref{inst14},\ref{inst71},\ref{inst72}}\and
Charles F. Gammie\orcid{0000-0001-7451-8935}\inst{\ref{inst25},\ref{inst51},\ref{inst73}}\and
Roberto García\orcid{0000-0002-6584-7443}\inst{\ref{inst33}}\and
Olivier Gentaz\orcid{0000-0002-0115-4605}\inst{\ref{inst33}}\and
Boris Georgiev\orcid{0000-0002-3586-6424}\inst{\ref{inst22}}\and
Roman Gold\orcid{0000-0003-2492-1966}\inst{\ref{inst74},\ref{inst75},\ref{inst76}}\and
Arturo I. Gómez-Ruiz\orcid{0000-0001-9395-1670}\inst{\ref{inst77},\ref{inst78}}\and
José L. Gómez\orcid{0000-0003-4190-7613}\inst{\ref{inst15}}\and
Minfeng Gu (\cntext{顾敏峰})\orcid{0000-0002-4455-6946}\inst{\ref{inst46},\ref{inst79}}\and
Mark Gurwell\orcid{0000-0003-0685-3621}\inst{\ref{inst19}}\and
Kazuhiro Hada\orcid{0000-0001-6906-772X}\inst{\ref{inst80},\ref{inst81}}\and
Daryl Haggard\orcid{0000-0001-6803-2138}\inst{\ref{inst82},\ref{inst83}}\and
Ronald Hesper\orcid{0000-0003-1918-6098}\inst{\ref{inst84}}\and
Dirk Heumann\orcid{0000-0002-7671-0047}\inst{\ref{inst22}}\and
Luis C. Ho (\cntext{何子山})\orcid{0000-0001-6947-5846}\inst{\ref{inst85},\ref{inst86}}\and
Paul Ho\orcid{0000-0002-3412-4306}\inst{\ref{inst20},\ref{inst29},\ref{inst28}}\and
Mareki Honma\orcid{0000-0003-4058-9000}\inst{\ref{inst81},\ref{inst87},\ref{inst88}}\and
Chih-Wei L. Huang\orcid{0000-0001-5641-3953}\inst{\ref{inst20}}\and
Lei Huang (\cntext{黄磊})\orcid{0000-0002-1923-227X}\inst{\ref{inst46},\ref{inst79}}\and
David H. Hughes\inst{\ref{inst77}}\and
Shiro Ikeda\orcid{0000-0002-2462-1448}\inst{\ref{inst13},\ref{inst89},\ref{inst90},\ref{inst91}}\and
C. M. Violette Impellizzeri\orcid{0000-0002-3443-2472}\inst{\ref{inst92},\ref{inst67}}\and
Makoto Inoue\orcid{0000-0001-5037-3989}\inst{\ref{inst20}}\and
Sara Issaoun\orcid{0000-0002-5297-921X}\inst{\ref{inst19},\ref{inst57}}\and
David J. James\orcid{0000-0001-5160-4486}\inst{\ref{inst93},\ref{inst94}}\and
Buell T. Jannuzi\orcid{0000-0002-1578-6582}\inst{\ref{inst22}}\and
Michael Janssen\orcid{0000-0001-8685-6544}\inst{\ref{inst37},\ref{inst12}}\and
Britton Jeter\orcid{0000-0003-2847-1712}\inst{\ref{inst20}}\and
Wu Jiang (\cntext{江悟})\orcid{0000-0001-7369-3539}\inst{\ref{inst46}}\and
Alejandra Jiménez-Rosales\orcid{0000-0002-2662-3754}\inst{\ref{inst37}}\and
Michael D. Johnson\orcid{0000-0002-4120-3029}\inst{\ref{inst14},\ref{inst19}}\and
Svetlana Jorstad\orcid{0000-0001-6158-1708}\inst{\ref{inst95}}\and
Adam C. Jones\inst{\ref{inst27}}\and
Abhishek V. Joshi\orcid{0000-0002-2514-5965}\inst{\ref{inst25}}\and
Taehyun Jung\orcid{0000-0001-7003-8643}\inst{\ref{inst48},\ref{inst96}}\and
Ramesh Karuppusamy\orcid{0000-0002-5307-2919}\inst{\ref{inst12}}\and
Tomohisa Kawashima\orcid{0000-0001-8527-0496}\inst{\ref{inst97}}\and
Garrett K. Keating\orcid{0000-0002-3490-146X}\inst{\ref{inst19}}\and
Mark Kettenis\orcid{0000-0002-6156-5617}\inst{\ref{inst98}}\and
Dong-Jin Kim\orcid{0000-0002-7038-2118}\inst{\ref{inst99}}\and
Jae-Young Kim\orcid{0000-0001-8229-7183}\inst{\ref{inst100},\ref{inst12}}\and
Jongsoo Kim\orcid{0000-0002-1229-0426}\inst{\ref{inst48}}\and
Junhan Kim\orcid{0000-0002-4274-9373}\inst{\ref{inst101}}\and
Motoki Kino\orcid{0000-0002-2709-7338}\inst{\ref{inst13},\ref{inst102}}\and
Jun Yi Koay\orcid{0000-0002-7029-6658}\inst{\ref{inst20}}\and
Prashant Kocherlakota\orcid{0000-0001-7386-7439}\inst{\ref{inst53}}\and
Yutaro Kofuji\inst{\ref{inst81},\ref{inst88}}\and
%Patrick M. Koch\orcid{0000-0003-2777-5861}\inst{\ref{inst20}}\and
Shoko Koyama\orcid{0000-0002-3723-3372}\inst{\ref{inst103},\ref{inst20}}\and
Carsten Kramer\orcid{0000-0002-4908-4925}\inst{\ref{inst33}}\and
Joana A. Kramer\orcid{0009-0003-3011-0454}\inst{\ref{inst12}}\and
Michael Kramer\orcid{0000-0002-4175-2271}\inst{\ref{inst12}}\and
Thomas P. Krichbaum\orcid{0000-0002-4892-9586}\inst{\ref{inst12}}\and
Cheng-Yu Kuo\orcid{0000-0001-6211-5581}\inst{\ref{inst104},\ref{inst20}}\and
Noemi La Bella\orcid{0000-0002-8116-9427}\inst{\ref{inst37}}\and
Sang-Sung Lee\orcid{0000-0002-6269-594X}\inst{\ref{inst48}}\and
Aviad Levis\orcid{0000-0001-7307-632X}\inst{\ref{inst30}}\and
Zhiyuan Li (\cntext{李志远})\orcid{0000-0003-0355-6437}\inst{\ref{inst105},\ref{inst106}}\and
Rocco Lico\orcid{0000-0001-7361-2460}\inst{\ref{inst6},\ref{inst15}}\and
Greg Lindahl\orcid{0000-0002-6100-4772}\inst{\ref{inst107}}\and
Michael Lindqvist\orcid{0000-0002-3669-0715}\inst{\ref{inst21}}\and
Mikhail Lisakov\orcid{0000-0001-6088-3819}\inst{\ref{inst108}}\and
Jun Liu (\cntext{刘俊})\orcid{0000-0002-7615-7499}\inst{\ref{inst12}}\and
Kuo Liu\orcid{0000-0002-2953-7376}\inst{\ref{inst46},\ref{inst47}}\and
Wen-Ping Lo\orcid{0000-0003-1869-2503}\inst{\ref{inst20},\ref{inst109}}\and
Andrei P. Lobanov\orcid{0000-0003-1622-1484}\inst{\ref{inst12}}\and
Laurent Loinard\orcid{0000-0002-5635-3345}\inst{\ref{inst110},\ref{inst14},\ref{inst111}}\and
Colin J. Lonsdale\orcid{0000-0003-4062-4654}\inst{\ref{inst4}}\and
Amy E. Lowitz\orcid{0000-0002-4747-4276}\inst{\ref{inst22}}\and
Ru-Sen Lu (\cntext{路如森})\orcid{0000-0002-7692-7967}\inst{\ref{inst46},\ref{inst47},\ref{inst12}}\and
Nicholas R. MacDonald\orcid{0000-0002-6684-8691}\inst{\ref{inst12}}\and
Jirong Mao (\cntext{毛基荣})\orcid{0000-0002-7077-7195}\inst{\ref{inst112},\ref{inst113},\ref{inst114}}\and
Sera Markoff\orcid{0000-0001-9564-0876}\inst{\ref{inst115},\ref{inst116}}\and
Daniel P. Marrone\orcid{0000-0002-2367-1080}\inst{\ref{inst22}}\and
Alan P. Marscher\orcid{0000-0001-7396-3332}\inst{\ref{inst95}}\and
Satoki Matsushita\orcid{0000-0002-2127-7880}\inst{\ref{inst20}}\and
Lia Medeiros\orcid{0000-0003-2342-6728}\inst{\ref{inst56},\ref{inst57}}\and
Karl M. Menten\orcid{0000-0001-6459-0669}\inst{\ref{inst12},\ref{inst117}}\and
Izumi Mizuno\orcid{0000-0002-7210-6264}\inst{\ref{inst28},\ref{inst29}}\and
Yosuke Mizuno\orcid{0000-0002-8131-6730}\inst{\ref{inst55},\ref{inst118},\ref{inst53}}\and
Joshua Montgomery\orcid{0000-0003-0345-8386}\inst{\ref{inst83},\ref{inst27}}\and
James M. Moran\orcid{0000-0002-3882-4414}\inst{\ref{inst14},\ref{inst19}}\and
Kotaro Moriyama\orcid{0000-0003-1364-3761}\inst{\ref{inst53},\ref{inst81}}\and
Monika Moscibrodzka\orcid{0000-0002-4661-6332}\inst{\ref{inst37}}\and
Wanga Mulaudzi\orcid{0000-0003-4514-625X}\inst{\ref{inst115}}\and
Cornelia Müller\orcid{0000-0002-2739-2994}\inst{\ref{inst12},\ref{inst37}}\and
Hendrik Müller\orcid{0000-0002-9250-0197}\inst{\ref{inst12}}\and
Gibwa Musoke\orcid{0000-0003-1984-189X}\inst{\ref{inst115},\ref{inst37}}\and
Ioannis Myserlis\orcid{0000-0003-3025-9497}\inst{\ref{inst119}}\and
Hiroshi Nagai\orcid{0000-0003-0292-3645}\inst{\ref{inst13},\ref{inst87}}\and
Neil M. Nagar\orcid{0000-0001-6920-662X}\inst{\ref{inst24}}\and
Dhanya G. Nair\orcid{0000-0001-5357-7805}\inst{\ref{inst24},\ref{inst12}}\and
Masanori Nakamura\orcid{0000-0001-6081-2420}\inst{\ref{inst120},\ref{inst20}}\and
Gopal Narayanan\orcid{0000-0002-4723-6569}\inst{\ref{inst38}}\and
Iniyan Natarajan\orcid{0000-0001-8242-4373}\inst{\ref{inst19},\ref{inst14}}\and
Antonios Nathanail\orcid{0000-0002-1655-9912}\inst{\ref{inst121},\ref{inst53}}\and
Santiago Navarro Fuentes\inst{\ref{inst119}}\and
Joey Neilsen\orcid{0000-0002-8247-786X}\inst{\ref{inst122}}\and
Chunchong Ni\orcid{0000-0003-1361-5699}\inst{\ref{inst35},\ref{inst36},\ref{inst34}}\and
Michael A. Nowak\orcid{0000-0001-6923-1315}\inst{\ref{inst123}}\and
Junghwan Oh\orcid{0000-0002-4991-9638}\inst{\ref{inst98}}\and
Hiroki Okino\orcid{0000-0003-3779-2016}\inst{\ref{inst81},\ref{inst88}}\and
Héctor Raúl Olivares Sánchez\orcid{0000-0001-6833-7580}\inst{\ref{inst124}}\and
Tomoaki Oyama\orcid{0000-0003-4046-2923}\inst{\ref{inst81}}\and
Feryal Özel\orcid{0000-0003-4413-1523}\inst{\ref{inst125}}\and
Daniel C. M. Palumbo\orcid{0000-0002-7179-3816}\inst{\ref{inst14},\ref{inst19}}\and
Georgios Filippos Paraschos\orcid{0000-0001-6757-3098}\inst{\ref{inst12}}\and
Jongho Park\orcid{0000-0001-6558-9053}\inst{\ref{inst126},\ref{inst20}}\and
Harriet Parsons\orcid{0000-0002-6327-3423}\inst{\ref{inst28},\ref{inst29}}\and
Nimesh Patel\orcid{0000-0002-6021-9421}\inst{\ref{inst19}}\and
Ue-Li Pen\orcid{0000-0003-2155-9578}\inst{\ref{inst20},\ref{inst34},\ref{inst127},\ref{inst128},\ref{inst129}}\and
Dominic W. Pesce\orcid{0000-0002-5278-9221}\inst{\ref{inst19},\ref{inst14}}\and
Vincent Piétu\inst{\ref{inst33}}\and
Aleksandar PopStefanija\inst{\ref{inst38}}\and
Oliver Porth\orcid{0000-0002-4584-2557}\inst{\ref{inst115},\ref{inst53}}\and
Ben Prather\orcid{0000-0002-0393-7734}\inst{\ref{inst25}}\and
Giacomo Principe\orcid{0000-0003-0406-7387}\inst{\ref{inst130},\ref{inst131},\ref{inst6}}\and
Dimitrios Psaltis\orcid{0000-0003-1035-3240}\inst{\ref{inst125}}\and
Hung-Yi Pu\orcid{0000-0001-9270-8812}\inst{\ref{inst132},\ref{inst133},\ref{inst20}}\and
Venkatessh Ramakrishnan\orcid{0000-0002-9248-086X}\inst{\ref{inst24},\ref{inst134},\ref{inst135}}\and
Ramprasad Rao\orcid{0000-0002-1407-7944}\inst{\ref{inst19}}\and
Mark G. Rawlings\orcid{0000-0002-6529-202X}\inst{\ref{inst136},\ref{inst28},\ref{inst29}}\and
Luciano Rezzolla\orcid{0000-0002-1330-7103}\inst{\ref{inst53},\ref{inst137},\ref{inst138}}\and
Angelo Ricarte\orcid{0000-0001-5287-0452}\inst{\ref{inst14},\ref{inst19}}\and
Bart Ripperda\orcid{0000-0002-7301-3908}\inst{\ref{inst127},\ref{inst139},\ref{inst128},\ref{inst34}}\and
Freek Roelofs\orcid{0000-0001-5461-3687}\inst{\ref{inst37}}\and
Cristina Romero-Cañizales\orcid{0000-0001-6301-9073}\inst{\ref{inst20}}\and
Eduardo Ros\orcid{0000-0001-9503-4892}\inst{\ref{inst12}}\and
Arash Roshanineshat\orcid{0000-0002-8280-9238}\inst{\ref{inst22}}\and
Helge Rottmann\inst{\ref{inst12}}\and
Alan L. Roy\orcid{0000-0002-1931-0135}\inst{\ref{inst12}}\and
Ignacio Ruiz\orcid{0000-0002-0965-5463}\inst{\ref{inst119}}\and
Chet Ruszczyk\orcid{0000-0001-7278-9707}\inst{\ref{inst4}}\and
Salvador Sánchez\orcid{0000-0002-8042-5951}\inst{\ref{inst119}}\and
David Sánchez-Argüelles\orcid{0000-0002-7344-9920}\inst{\ref{inst77},\ref{inst78}}\and
Miguel Sánchez-Portal\orcid{0000-0003-0981-9664}\inst{\ref{inst119}}\and
Mahito Sasada\orcid{0000-0001-5946-9960}\inst{\ref{inst140},\ref{inst81},\ref{inst141}}\and
Kaushik Satapathy\orcid{0000-0003-0433-3585}\inst{\ref{inst22}}\and
Saurabh\orcid{0000-0001-7156-4848}\inst{\ref{inst12}}\and
Tuomas Savolainen\orcid{0000-0001-6214-1085}\inst{\ref{inst142},\ref{inst135},\ref{inst12}}\and
F. Peter Schloerb\inst{\ref{inst38}}\and
Jonathan Schonfeld\orcid{0000-0002-8909-2401}\inst{\ref{inst19}}\and
Karl-Friedrich Schuster\orcid{0000-0003-2890-9454}\inst{\ref{inst33}}\and
Lijing Shao\orcid{0000-0002-1334-8853}\inst{\ref{inst86},\ref{inst12}}\and
Zhiqiang Shen (\cntext{沈志强})\orcid{0000-0003-3540-8746}\inst{\ref{inst46},\ref{inst47}}\and
Sasikumar Silpa\orcid{0000-0003-0667-7074}\inst{\ref{inst24}}\and
Des Small\orcid{0000-0003-3723-5404}\inst{\ref{inst98}}\and
Bong Won Sohn\orcid{0000-0002-4148-8378}\inst{\ref{inst48},\ref{inst96},\ref{inst49}}\and
Jason SooHoo\orcid{0000-0003-1938-0720}\inst{\ref{inst4}}\and
León D. S. Salas\orcid{0000-0003-1979-6363}\inst{\ref{inst115}}\and
Kamal Souccar\orcid{0000-0001-7915-5272}\inst{\ref{inst38}}\and
Joshua S. Stanway\orcid{0009-0003-7659-4642}\inst{\ref{inst143}}\and
He Sun (\cntext{孙赫})\orcid{0000-0003-1526-6787}\inst{\ref{inst144},\ref{inst145}}\and
Fumie Tazaki\orcid{0000-0003-0236-0600}\inst{\ref{inst146}}\and
Alexandra J. Tetarenko\orcid{0000-0003-3906-4354}\inst{\ref{inst147}}\and
Paul Tiede\orcid{0000-0003-3826-5648}\inst{\ref{inst19},\ref{inst14}}\and
Remo P. J. Tilanus\orcid{0000-0002-6514-553X}\inst{\ref{inst22},\ref{inst37},\ref{inst92},\ref{inst148}}\and
Michael Titus\orcid{0000-0001-9001-3275}\inst{\ref{inst4}}\and
Kenji Toma\orcid{0000-0002-7114-6010}\inst{\ref{inst149},\ref{inst150}}\and
Pablo Torne\orcid{0000-0001-8700-6058}\inst{\ref{inst119},\ref{inst12}}\and
Teresa Toscano\orcid{0000-0003-3658-7862}\inst{\ref{inst15}}\and
Efthalia Traianou\orcid{0000-0002-1209-6500}\inst{\ref{inst15},\ref{inst12}}\and
Tyler Trent\inst{\ref{inst22}}\and
Sascha Trippe\orcid{0000-0003-0465-1559}\inst{\ref{inst151}}\and
Matthew Turk\orcid{0000-0002-5294-0198}\inst{\ref{inst51}}\and
Ilse van Bemmel\orcid{0000-0001-5473-2950}\inst{\ref{inst61}}\and
Huib Jan van Langevelde\orcid{0000-0002-0230-5946}\inst{\ref{inst98},\ref{inst92},\ref{inst152}}\and
Daniel R. van Rossum\orcid{0000-0001-7772-6131}\inst{\ref{inst37}}\and
Jesse Vos\orcid{0000-0003-3349-7394}\inst{\ref{inst37}}\and
Jan Wagner\orcid{0000-0003-1105-6109}\inst{\ref{inst12}}\and
Derek Ward-Thompson\orcid{0000-0003-1140-2761}\inst{\ref{inst143}}\and
John Wardle\orcid{0000-0002-8960-2942}\inst{\ref{inst153}}\and
Jasmin E. Washington\orcid{0000-0002-7046-0470}\inst{\ref{inst22}}\and
Jonathan Weintroub\orcid{0000-0002-4603-5204}\inst{\ref{inst14},\ref{inst19}}\and
Robert Wharton\orcid{0000-0002-7416-5209}\inst{\ref{inst12}}\and
Maciek Wielgus\orcid{0000-0002-8635-4242}\inst{\ref{inst15}}\and
Kaj Wiik\orcid{0000-0002-0862-3398}\inst{\ref{inst154},\ref{inst134},\ref{inst135}}\and
Gunther Witzel\orcid{0000-0003-2618-797X}\inst{\ref{inst12}}\and
Michael F. Wondrak\orcid{0000-0002-6894-1072}\inst{\ref{inst37},\ref{inst155}}\and
George N. Wong\orcid{0000-0001-6952-2147}\inst{\ref{inst156},\ref{inst42}}\and
Qingwen Wu (\cntext{吴庆文})\orcid{0000-0003-4773-4987}\inst{\ref{inst157}}\and
Nitika Yadlapalli\orcid{0000-0003-3255-4617}\inst{\ref{inst30}}\and
Paul Yamaguchi\orcid{0000-0002-6017-8199}\inst{\ref{inst19}}\and
Aristomenis Yfantis\orcid{0000-0002-3244-7072}\inst{\ref{inst37}}\and
Doosoo Yoon\orcid{0000-0001-8694-8166}\inst{\ref{inst115}}\and
André Young\orcid{0000-0003-0000-2682}\inst{\ref{inst37}}\and
Ziri Younsi\orcid{0000-0001-9283-1191}\inst{\ref{inst158},\ref{inst53}}\and
Wei Yu (\cntext{于威})\orcid{0000-0002-5168-6052}\inst{\ref{inst19}}\and
Feng Yuan (\cntext{袁峰})\orcid{0000-0003-3564-6437}\inst{\ref{inst159}}\and
Ye-Fei Yuan (\cntext{袁业飞})\orcid{0000-0002-7330-4756}\inst{\ref{inst160}}\and
J. Anton Zensus\orcid{0000-0001-7470-3321}\inst{\ref{inst12}}\and
Shuo Zhang\orcid{0000-0002-2967-790X}\inst{\ref{inst161}}\and
Guang-Yao Zhao\orcid{0000-0002-4417-1659}\inst{\ref{inst15},\ref{inst12}}\and
Shan-Shan Zhao (\cntext{赵杉杉})\orcid{0000-0002-9774-3606}\inst{\ref{inst46}}
}
\institute{
Dipartimento di Fisica, Università degli Studi di Cagliari, SP Monserrato-Sestu km 0.7, I-09042 Monserrato (CA), Italy\label{inst1}\and
Instituto de Astronomia, Geofísica e Ciências Atmosféricas, Universidade de São Paulo, R. do Matão, 1226, São Paulo, SP 05508-090, Brazil\label{inst2}\and
INAF - Osservatorio Astronomico di Cagliari, via della Scienza 5, I-09047 Selargius (CA), Italy\label{inst3}\and
Massachusetts Institute of Technology Haystack Observatory, 99 Millstone Road, Westford, MA 01886, USA\label{inst4}\and
None\label{inst5}\and
INAF-Istituto di Radioastronomia, Via P. Gobetti 101, I-40129 Bologna, Italy\label{inst6}\and
Departament d'Astronomia i Astrofísica, Universitat de València, C. Dr. Moliner 50, E-46100 Burjassot, València, Spain\label{inst7}\and
Observatori Astronòmic, Universitat de València, C. Catedrático José Beltrán 2, E-46980 Paterna, València, Spain\label{inst8}\and
Dipartimento di Fisica ``E. Pancini'', Università di Napoli ``Federico II'', Compl. Univ. di Monte S. Angelo, Edificio G, Via Cinthia, I-80126, Napoli, Italy\label{inst9}\and
INFN Sez. di Napoli, Compl. Univ. di Monte S. Angelo, Edificio G, Via Cinthia, I-80126, Napoli, Italy\label{inst10}\and
INAF-Istituto di Radioastronomia \& Italian ALMA Regional Centre, Via P. Gobetti 101, I-40129 Bologna, Italy\label{inst11}\and
Max-Planck-Institut für Radioastronomie, Auf dem Hügel 69, D-53121 Bonn, Germany\label{inst12}\and
National Astronomical Observatory of Japan, 2-21-1 Osawa, Mitaka, Tokyo 181-8588, Japan\label{inst13}\and
Black Hole Initiative at Harvard University, 20 Garden Street, Cambridge, MA 02138, USA\label{inst14}\and
Instituto de Astrofísica de Andalucía-CSIC, Glorieta de la Astronomía s/n, E-18008 Granada, Spain\label{inst15}\and
Department of Physics, Faculty of Science, Universiti Malaya, 50603 Kuala Lumpur, Malaysia\label{inst16}\and
Department of Physics \& Astronomy, The University of Texas at San Antonio, One UTSA Circle, San Antonio, TX 78249, USA\label{inst17}\and
Physics \& Astronomy Department, Rice University, Houston, TX 77005-1827, USA\label{inst18}\and
Center for Astrophysics $|$ Harvard \& Smithsonian, 60 Garden Street, Cambridge, MA 02138, USA\label{inst19}\and
Institute of Astronomy and Astrophysics, Academia Sinica, 11F of Astronomy-Mathematics Building, AS/NTU No. 1, Sec. 4, Roosevelt Rd., Taipei 106216, Taiwan, R.O.C.\label{inst20}\and
Department of Space, Earth and Environment, Chalmers University of Technology, Onsala Space Observatory, SE-43992 Onsala, Sweden\label{inst21}\and
Steward Observatory and Department of Astronomy, University of Arizona, 933 N. Cherry Ave., Tucson, AZ 85721, USA\label{inst22}\and
Yale Center for Astronomy \& Astrophysics, Yale University, 52 Hillhouse Avenue, New Haven, CT 06511, USA\label{inst23}\and
Astronomy Department, Universidad de Concepción, Casilla 160-C, Concepción, Chile\label{inst24}\and
Department of Physics, University of Illinois, 1110 West Green Street, Urbana, IL 61801, USA\label{inst25}\and
Fermi National Accelerator Laboratory, MS209, P.O. Box 500, Batavia, IL 60510, USA\label{inst26}\and
Department of Astronomy and Astrophysics, University of Chicago, 5640 South Ellis Avenue, Chicago, IL 60637, USA\label{inst27}\and
East Asian Observatory, 660 N. A'ohoku Place, Hilo, HI 96720, USA\label{inst28}\and
James Clerk Maxwell Telescope (JCMT), 660 N. A'ohoku Place, Hilo, HI 96720, USA\label{inst29}\and
California Institute of Technology, 1200 East California Boulevard, Pasadena, CA 91125, USA\label{inst30}\and
Institute of Astronomy and Astrophysics, Academia Sinica, 645 N. A'ohoku Place, Hilo, HI 96720, USA\label{inst31}\and
Department of Physics and Astronomy, University of Hawaii at Manoa, 2505 Correa Road, Honolulu, HI 96822, USA\label{inst32}\and
Institut de Radioastronomie Millimétrique (IRAM), 300 rue de la Piscine, F-38406 Saint Martin d'Hères, France\label{inst33}\and
Perimeter Institute for Theoretical Physics, 31 Caroline Street North, Waterloo, ON N2L 2Y5, Canada\label{inst34}\and
Department of Physics and Astronomy, University of Waterloo, 200 University Avenue West, Waterloo, ON N2L 3G1, Canada\label{inst35}\and
Waterloo Centre for Astrophysics, University of Waterloo, Waterloo, ON N2L 3G1, Canada\label{inst36}\and
Department of Astrophysics, Institute for Mathematics, Astrophysics and Particle Physics (IMAPP), Radboud University, P.O. Box 9010, 6500 GL Nijmegen, The Netherlands\label{inst37}\and
Department of Astronomy, University of Massachusetts, Amherst, MA 01003, USA\label{inst38}\and
Kavli Institute for Cosmological Physics, University of Chicago, 5640 South Ellis Avenue, Chicago, IL 60637, USA\label{inst39}\and
Department of Physics, University of Chicago, 5720 South Ellis Avenue, Chicago, IL 60637, USA\label{inst40}\and
Enrico Fermi Institute, University of Chicago, 5640 South Ellis Avenue, Chicago, IL 60637, USA\label{inst41}\and
Princeton Gravity Initiative, Jadwin Hall, Princeton University, Princeton, NJ 08544, USA\label{inst42}\and
Data Science Institute, University of Arizona, 1230 N. Cherry Ave., Tucson, AZ 85721, USA\label{inst43}\and
Program in Applied Mathematics, University of Arizona, 617 N. Santa Rita, Tucson, AZ 85721, USA\label{inst44}\and
Cornell Center for Astrophysics and Planetary Science, Cornell University, Ithaca, NY 14853, USA\label{inst45}\and
Shanghai Astronomical Observatory, Chinese Academy of Sciences, 80 Nandan Road, Shanghai 200030, People's Republic of China\label{inst46}\and
Key Laboratory of Radio Astronomy and Technology, Chinese Academy of Sciences, A20 Datun Road, Chaoyang District, Beijing, 100101, People's Republic of China\label{inst47}\and
Korea Astronomy and Space Science Institute, Daedeok-daero 776, Yuseong-gu, Daejeon 34055, Republic of Korea\label{inst48}\and
Department of Astronomy, Yonsei University, Yonsei-ro 50, Seodaemun-gu, 03722 Seoul, Republic of Korea\label{inst49}\and
WattTime, 490 43rd Street, Unit 221, Oakland, CA 94609, USA\label{inst50}\and
Department of Astronomy, University of Illinois at Urbana-Champaign, 1002 West Green Street, Urbana, IL 61801, USA\label{inst51}\and
Instituto de Astronomía, Universidad Nacional Autónoma de México (UNAM), Apdo Postal 70-264, Ciudad de México, México\label{inst52}\and
Institut für Theoretische Physik, Goethe-Universität Frankfurt, Max-von-Laue-Straße 1, D-60438 Frankfurt am Main, Germany\label{inst53}\and
Research Center for Astronomical Computing, Zhejiang Laboratory, Hangzhou 311100, People's Republic of China\label{inst54}\and
Tsung-Dao Lee Institute, Shanghai Jiao Tong University, Shengrong Road 520, Shanghai, 201210, People's Republic of China\label{inst55}\and
Department of Astrophysical Sciences, Peyton Hall, Princeton University, Princeton, NJ 08544, USA\label{inst56}\and
NASA Hubble Fellowship Program, Einstein Fellow\label{inst57}\and
Wits Centre for Astrophysics, University of the Witwatersrand, 1 Jan Smuts Avenue, Braamfontein, Johannesburg 2050, South Africa\label{inst58}\and
Department of Physics, University of Pretoria, Hatfield, Pretoria 0028, South Africa\label{inst59}\and
Centre for Radio Astronomy Techniques and Technologies, Department of Physics and Electronics, Rhodes University, Makhanda 6140, South Africa\label{inst60}\and
ASTRON, Oude Hoogeveensedijk 4, 7991 PD Dwingeloo, The Netherlands\label{inst61}\and
LESIA, Observatoire de Paris, Université PSL, CNRS, Sorbonne Université, Université de Paris, 5 place Jules Janssen, F-92195 Meudon, France\label{inst62}\and
JILA and Department of Astrophysical and Planetary Sciences, University of Colorado, Boulder, CO 80309, USA\label{inst63}\and
National Astronomical Observatories, Chinese Academy of Sciences, 20A Datun Road, Chaoyang District, Beijing 100101, PR China\label{inst64}\and
Las Cumbres Observatory, 6740 Cortona Drive, Suite 102, Goleta, CA 93117-5575, USA\label{inst65}\and
Department of Physics, University of California, Santa Barbara, CA 93106-9530, USA\label{inst66}\and
National Radio Astronomy Observatory, 520 Edgemont Road, Charlottesville, VA 22903, USA\label{inst67}\and
Department of Electrical Engineering and Computer Science, Massachusetts Institute of Technology, 32-D476, 77 Massachusetts Ave., Cambridge, MA 02142, USA\label{inst68}\and
Google Research, 355 Main St., Cambridge, MA 02142, USA\label{inst69}\and
Institut für Theoretische Physik und Astrophysik, Universität Würzburg, Emil-Fischer-Str. 31, D-97074 Würzburg, Germany\label{inst70}\and
Department of History of Science, Harvard University, Cambridge, MA 02138, USA\label{inst71}\and
Department of Physics, Harvard University, Cambridge, MA 02138, USA\label{inst72}\and
NCSA, University of Illinois, 1205 W. Clark St., Urbana, IL 61801, USA\label{inst73}\and
Institute for Mathematics and Interdisciplinary Center for Scientific Computing, Heidelberg University, Im Neuenheimer Feld 205, Heidelberg 69120, Germany\label{inst74}\and
Institut f\"ur Theoretische Physik, Universit\"at Heidelberg, Philosophenweg 16, 69120 Heidelberg, Germany\label{inst75}\and
CP3-Origins, University of Southern Denmark, Campusvej 55, DK-5230 Odense, Denmark\label{inst76}\and
Instituto Nacional de Astrofísica, Óptica y Electrónica. Apartado Postal 51 y 216, 72000. Puebla Pue., México\label{inst77}\and
Consejo Nacional de Humanidades, Ciencia y Tecnología, Av. Insurgentes Sur 1582, 03940, Ciudad de México, México\label{inst78}\and
Key Laboratory for Research in Galaxies and Cosmology, Chinese Academy of Sciences, Shanghai 200030, People's Republic of China\label{inst79}\and
Graduate School of Science, Nagoya City University, Yamanohata 1, Mizuho-cho, Mizuho-ku, Nagoya, 467-8501, Aichi, Japan\label{inst80}\and
Mizusawa VLBI Observatory, National Astronomical Observatory of Japan, 2-12 Hoshigaoka, Mizusawa, Oshu, Iwate 023-0861, Japan\label{inst81}\and
Department of Physics, McGill University, 3600 rue University, Montréal, QC H3A 2T8, Canada\label{inst82}\and
Trottier Space Institute at McGill, 3550 rue University, Montréal,  QC H3A 2A7, Canada\label{inst83}\and
NOVA Sub-mm Instrumentation Group, Kapteyn Astronomical Institute, University of Groningen, Landleven 12, 9747 AD Groningen, The Netherlands\label{inst84}\and
Department of Astronomy, School of Physics, Peking University, Beijing 100871, People's Republic of China\label{inst85}\and
Kavli Institute for Astronomy and Astrophysics, Peking University, Beijing 100871, People's Republic of China\label{inst86}\and
Department of Astronomical Science, The Graduate University for Advanced Studies (SOKENDAI), 2-21-1 Osawa, Mitaka, Tokyo 181-8588, Japan\label{inst87}\and
Department of Astronomy, Graduate School of Science, The University of Tokyo, 7-3-1 Hongo, Bunkyo-ku, Tokyo 113-0033, Japan\label{inst88}\and
The Institute of Statistical Mathematics, 10-3 Midori-cho, Tachikawa, Tokyo, 190-8562, Japan\label{inst89}\and
Department of Statistical Science, The Graduate University for Advanced Studies (SOKENDAI), 10-3 Midori-cho, Tachikawa, Tokyo 190-8562, Japan\label{inst90}\and
Kavli Institute for the Physics and Mathematics of the Universe, The University of Tokyo, 5-1-5 Kashiwanoha, Kashiwa, 277-8583, Japan\label{inst91}\and
Leiden Observatory, Leiden University, Postbus 2300, 9513 RA Leiden, The Netherlands\label{inst92}\and
ASTRAVEO LLC, PO Box 1668, Gloucester, MA 01931, USA\label{inst93}\and
Applied Materials Inc., 35 Dory Road, Gloucester, MA 01930, USA\label{inst94}\and
Institute for Astrophysical Research, Boston University, 725 Commonwealth Ave., Boston, MA 02215, USA\label{inst95}\and
University of Science and Technology, Gajeong-ro 217, Yuseong-gu, Daejeon 34113, Republic of Korea\label{inst96}\and
Institute for Cosmic Ray Research, The University of Tokyo, 5-1-5 Kashiwanoha, Kashiwa, Chiba 277-8582, Japan\label{inst97}\and
Joint Institute for VLBI ERIC (JIVE), Oude Hoogeveensedijk 4, 7991 PD Dwingeloo, The Netherlands\label{inst98}\and
CSIRO, Space and Astronomy, PO Box 76, Epping, NSW 1710, Australia\label{inst99}\and
Department of Physics, Ulsan National Institute of Science and Technology (UNIST), Ulsan 44919, Republic of Korea\label{inst100}\and
Department of Physics, Korea Advanced Institute of Science and Technology (KAIST), 291 Daehak-ro, Yuseong-gu, Daejeon 34141, Republic of Korea\label{inst101}\and
Kogakuin University of Technology \& Engineering, Academic Support Center, 2665-1 Nakano, Hachioji, Tokyo 192-0015, Japan\label{inst102}\and
Graduate School of Science and Technology, Niigata University, 8050 Ikarashi 2-no-cho, Nishi-ku, Niigata 950-2181, Japan\label{inst103}\and
Physics Department, National Sun Yat-Sen University, No. 70, Lien-Hai Road, Kaosiung City 80424, Taiwan, R.O.C.\label{inst104}\and
School of Astronomy and Space Science, Nanjing University, Nanjing 210023, People's Republic of China\label{inst105}\and
Key Laboratory of Modern Astronomy and Astrophysics, Nanjing University, Nanjing 210023, People's Republic of China\label{inst106}\and
Common Crawl Foundation, 9663 Santa Monica Blvd. 425, Beverly Hills, CA 90210 USA\label{inst107}\and
Instituto de Física, Pontificia Universidad Católica de Valparaíso, Casilla 4059, Valparaíso, Chile\label{inst108}\and
Department of Physics, National Taiwan University, No. 1, Sec. 4, Roosevelt Rd., Taipei 106216, Taiwan, R.O.C\label{inst109}\and
Instituto de Radioastronomía y Astrofísica, Universidad Nacional Autónoma de México, Morelia 58089, México\label{inst110}\and
David Rockefeller Center for Latin American Studies, Harvard University, 1730 Cambridge Street, Cambridge, MA 02138, USA\label{inst111}\and
Yunnan Observatories, Chinese Academy of Sciences, 650011 Kunming, Yunnan Province, People's Republic of China\label{inst112}\and
Center for Astronomical Mega-Science, Chinese Academy of Sciences, 20A Datun Road, Chaoyang District, Beijing, 100012, People's Republic of China\label{inst113}\and
Key Laboratory for the Structure and Evolution of Celestial Objects, Chinese Academy of Sciences, 650011 Kunming, People's Republic of China\label{inst114}\and
Anton Pannekoek Institute for Astronomy, University of Amsterdam, Science Park 904, 1098 XH, Amsterdam, The Netherlands\label{inst115}\and
Gravitation and Astroparticle Physics Amsterdam (GRAPPA) Institute, University of Amsterdam, Science Park 904, 1098 XH Amsterdam, The Netherlands\label{inst116}\and
Deceased\label{inst117}\and
School of Physics and Astronomy, Shanghai Jiao Tong University, 800 Dongchuan Road, Shanghai, 200240, People's Republic of China\label{inst118}\and
Institut de Radioastronomie Millimétrique (IRAM), Avenida Divina Pastora 7, Local 20, E-18012, Granada, Spain\label{inst119}\and
National Institute of Technology, Hachinohe College, 16-1 Uwanotai, Tamonoki, Hachinohe City, Aomori 039-1192, Japan\label{inst120}\and
Research Center for Astronomy, Academy of Athens, Soranou Efessiou 4, 115 27 Athens, Greece\label{inst121}\and
Department of Physics, Villanova University, 800 Lancaster Avenue, Villanova, PA 19085, USA\label{inst122}\and
Physics Department, Washington University, CB 1105, St. Louis, MO 63130, USA\label{inst123}\and
Departamento de Matemática da Universidade de Aveiro and Centre for Research and Development in Mathematics and Applications (CIDMA), Campus de Santiago, 3810-193 Aveiro, Portugal\label{inst124}\and
School of Physics, Georgia Institute of Technology, 837 State St NW, Atlanta, GA 30332, USA\label{inst125}\and
School of Space Research, Kyung Hee University, 1732, Deogyeong-daero, Giheung-gu, Yongin-si, Gyeonggi-do 17104, Republic of Korea\label{inst126}\and
Canadian Institute for Theoretical Astrophysics, University of Toronto, 60 St. George Street, Toronto, ON M5S 3H8, Canada\label{inst127}\and
Dunlap Institute for Astronomy and Astrophysics, University of Toronto, 50 St. George Street, Toronto, ON M5S 3H4, Canada\label{inst128}\and
Canadian Institute for Advanced Research, 180 Dundas St West, Toronto, ON M5G 1Z8, Canada\label{inst129}\and
Dipartimento di Fisica, Università di Trieste, I-34127 Trieste, Italy\label{inst130}\and
INFN Sez. di Trieste, I-34127 Trieste, Italy\label{inst131}\and
Department of Physics, National Taiwan Normal University, No. 88, Sec. 4, Tingzhou Rd., Taipei 116, Taiwan, R.O.C.\label{inst132}\and
Center of Astronomy and Gravitation, National Taiwan Normal University, No. 88, Sec. 4, Tingzhou Road, Taipei 116, Taiwan, R.O.C.\label{inst133}\and
Finnish Centre for Astronomy with ESO, University of Turku, FI-20014 Turun Yliopisto, Finland\label{inst134}\and
Aalto University Metsähovi Radio Observatory, Metsähovintie 114, FI-02540 Kylmälä, Finland\label{inst135}\and
Gemini Observatory/NSF NOIRLab, 670 N. A'ohōkū Place, Hilo, HI 96720, USA\label{inst136}\and
Frankfurt Institute for Advanced Studies, Ruth-Moufang-Strasse 1, D-60438 Frankfurt, Germany\label{inst137}\and
School of Mathematics, Trinity College, Dublin 2, Ireland\label{inst138}\and
Department of Physics, University of Toronto, 60 St. George Street, Toronto, ON M5S 1A7, Canada\label{inst139}\and
Department of Physics, Tokyo Institute of Technology, 2-12-1 Ookayama, Meguro-ku, Tokyo 152-8551, Japan\label{inst140}\and
Hiroshima Astrophysical Science Center, Hiroshima University, 1-3-1 Kagamiyama, Higashi-Hiroshima, Hiroshima 739-8526, Japan\label{inst141}\and
Aalto University Department of Electronics and Nanoengineering, PL 15500, FI-00076 Aalto, Finland\label{inst142}\and
Jeremiah Horrocks Institute, University of Central Lancashire, Preston PR1 2HE, UK\label{inst143}\and
National Biomedical Imaging Center, Peking University, Beijing 100871, People's Republic of China\label{inst144}\and
College of Future Technology, Peking University, Beijing 100871, People's Republic of China\label{inst145}\and
Tokyo Electron Technology Solutions Limited, 52 Matsunagane, Iwayado, Esashi, Oshu, Iwate 023-1101, Japan\label{inst146}\and
Department of Physics and Astronomy, University of Lethbridge, Lethbridge, Alberta T1K 3M4, Canada\label{inst147}\and
Netherlands Organisation for Scientific Research (NWO), Postbus 93138, 2509 AC Den Haag, The Netherlands\label{inst148}\and
Frontier Research Institute for Interdisciplinary Sciences, Tohoku University, Sendai 980-8578, Japan\label{inst149}\and
Astronomical Institute, Tohoku University, Sendai 980-8578, Japan\label{inst150}\and
Department of Physics and Astronomy, Seoul National University, Gwanak-gu, Seoul 08826, Republic of Korea\label{inst151}\and
University of New Mexico, Department of Physics and Astronomy, Albuquerque, NM 87131, USA\label{inst152}\and
Physics Department, Brandeis University, 415 South Street, Waltham, MA 02453, USA\label{inst153}\and
Tuorla Observatory, Department of Physics and Astronomy, University of Turku, FI-20014 Turun Yliopisto, Finland\label{inst154}\and
Radboud Excellence Fellow of Radboud University, Nijmegen, The Netherlands\label{inst155}\and
School of Natural Sciences, Institute for Advanced Study, 1 Einstein Drive, Princeton, NJ 08540, USA\label{inst156}\and
School of Physics, Huazhong University of Science and Technology, Wuhan, Hubei, 430074, People's Republic of China\label{inst157}\and
Mullard Space Science Laboratory, University College London, Holmbury St. Mary, Dorking, Surrey, RH5 6NT, UK\label{inst158}\and
Center for Astronomy and Astrophysics and Department of Physics, Fudan University, Shanghai 200438, People's Republic of China\label{inst159}\and
Astronomy Department, University of Science and Technology of China, Hefei 230026, People's Republic of China\label{inst160}\and
Department of Physics and Astronomy, Michigan State University, 567 Wilson Rd, East Lansing, MI 48824, USA\label{inst161}
}

  \abstract
% Context heading (optional)  
{}  
% Aims heading (mandatory)  
{We investigated the polarization and Faraday properties of Messier 87 (M87) and seven other radio-loud active galactic nuclei (AGNs) at $\lambda$0.87~mm (345~GHz) using the Atacama Large Millimeter/submillimeter Array (ALMA). Our goal was to characterize the linear polarization (LP) fractions, measure Faraday rotation measures (RMs), and examine the magnetic field structures in the emission regions of these AGNs.}  
% Methods heading (mandatory)  
{We conducted full-polarization observations as part of the ALMA Band~7 very long baseline interferometry (VLBI) commissioning during the April 2021 Event Horizon Telescope (EHT) campaign. We analyzed the LP fractions and RMs to assess the nature of Faraday screens and magnetic fields in the submillimeter emission regions.}  
% Results heading (mandatory)  
{We find LP fractions between 1\% and 17\% and RMs exceeding $10^{5}$ \radmsq, which are 1–2 orders of magnitude higher than typically observed at longer wavelengths ($\lambda>$3 mm). This suggests denser Faraday screens or stronger magnetic fields. Additionally, we present the first submillimeter polarized images of the M87 jet and the observed AGNs, revealing RM gradients and sign reversals in the M87 jet indicative of a kiloparsec-scale helical magnetic field structure.}  
% Conclusions heading (optional)  
{Our results provide essential constraints for calibrating, analyzing, and interpreting VLBI data from the EHT at 345 GHz, representing a critical step toward submillimeter VLBI imaging.}  


   \maketitle

\section{Introduction} 
The development of a phased-array capability for the Atacama Large Millimeter/submillimeter Array (ALMA) has revolutionized very long baseline interferometry (VLBI) at millimeter wavelengths  \citep{APPPaper,Goddi2019}. 
By significantly enhancing the sensitivity of VLBI baselines, particularly at 230 GHz ($\lambda\approx 1.3$~mm), the inclusion of phased-ALMA arrays  into existing VLBI arrays made possible the first horizon-scale images of supermassive black holes, including M87* in the Messier 87 Galaxy \citep{EHTC2019_1,EHTC2019_2,EHTC2019_3,EHTC2019_4,EHTC2019_5,EHTC2019_6,EHT2021_1, EHT2021_2,EHT_2023,EHT2024}  and Sgr A* at the heart of the Milky Way \citep{EHTC2022_1,EHTC2022_2,EHTC2022_3,EHTC2022_4,EHTC2022_5,EHTC2022_6,EHT2024_1}.

In recent developments, ALMA’s phased array capabilities were extended to the submillimeter (ALMA Band~7; $\nu \approx 345$ GHz; $\lambda\approx 0.87$~mm), enabling a performance comparable to that of Band~6 ($\nu \approx 230$ GHz) under suitable observing conditions \citep{APP2B7}. The first test of ALMA’s Band 7 phasing capability occurred in October 2018 during a global VLBI campaign, marking the detection of the first VLBI fringes in the submillimeter regime between ALMA and three other Event Horizon Telescope (EHT) sites \citep{Raymond2024}. Following this success, a full end-to-end test of the Band 7 VLBI capability was conducted at the end of the ALMA-EHT science campaign in April 2021, with the aim of obtaining the first 345 GHz VLBI fringes toward the EHT key target M87* and selected radio-loud active galactic nuclei (AGNs), as well as assessing the feasibility of VLBI imaging in the submillimeter regime.

Observations at $\nu \approx 345$ GHz offer a 50\% improvement in angular resolution over 230~GHz and are expected to enhance $uv$ coverage through the combination with data from lower frequencies. Such a multifrequency synthesis will enable higher-fidelity imaging while minimizing interstellar scattering effects, which is particularly critical for imaging Sgr A* \citep{EHTC2022_3}.

To turn the ALMA array into a coherently phased aperture for millimeter VLBI and integrate it with other VLBI stations, the connected-element interferometric visibilities must first be calibrated \citep{QA2Paper}. This process inherently generates a full-polarization interferometric dataset as a by-product of VLBI observations with ALMA. For VLBI experiments, these interferometric datasets serve multiple purposes: they provide source-integrated parameters to refine and validate VLBI calibration and imaging workflows \citep{EHTC2019_2,EHTC2019_3,EHT2021_1,EHTC2022_2,EHTC2022_3,EHT2024_1} and provide critical observational constraints for theoretical models and physical interpretations \citep{EHT2021_2,EHTC2022_4,EHTC2022_5,EHTC2022_6}.
However, beyond their role in VLBI, these datasets have significant standalone scientific value, enabling the derivation of millimeter emission, polarization, and Faraday properties of VLBI targets on arcsecond scales. At $\lambda$0.87 mm, synchrotron emission originates from a more optically thin region closer to the supermassive black hole when compared to $\lambda>$1 mm, providing a unique window into the jet base and accretion flow.

In this paper we present the analysis of connected-element interferometric data from the ALMA Band 7 VLBI test conducted in April 2021\footnote{These ALMA commissioning data are made public with this publication.}. Results from the full 345 GHz VLBI campaign, including data from other stations, will be discussed in a follow-up paper (EHTC et al., in prep.; hereafter Paper II). 

Section~\ref{sec:data} details the observational setup (Sect. \ref{sec:obs}), observed targets (Sect. \ref{sec:targets}), calibration procedures (Sect. \ref{sec:datacal}), and full-polarization image deconvolution (Sect. \ref{sec:dataimg}). Section~\ref{sec:analysis} describes the data analysis, including the extraction of Stokes parameters, polarimetric and Faraday property estimation  (Sect. \ref{sec:polan}), and polarized image production (Sect. \ref{sec:imaging}). Section~\ref{sec:res} presents the polarimetric properties (Sect. \ref{sec:agnpol}) and spectral indices (Sect. \ref{sec:AGNpec}) of the observed AGN sources, with a focus on the submillimeter polarized emission in the M87 jet (Sect. \ref{sec:m87}). Section~\ref{sec:conclusions} summarizes our conclusions.

%==============================================================================
\section{Observations, data calibration, and imaging}
\label{sec:data}
%==============================================================================

ALMA Band~7 observations were performed on April 19, 2021, at the end of the EHT science campaign conducted in Band 6.   ALMA was operated as a phased array, and joined a global network of VLBI stations operating at this frequency for an end-to-end submillimeter VLBI commissioning observation.

Weather conditions at ALMA were excellent: typical opacity values were $\tau_{225} \sim 0.05$, while system temperatures ($T_{\rm sys}$) ranged from 95 to 276 K. The measured phasing efficiencies during the test varied between 81\% and 97\%, reflecting strong system performance under favorable conditions  \citep{APP2B7}. 

%%%%%%%%%%%%%%%%
%==============================================================================
\begin{table*}
\caption{VLBI sources observed on April 19, 2021, in Band\,7. }            
\label{table:sources}       
\begin{center}
\small
\begin{tabular}{ccllcr} 
\hline\hline                  
\noalign{\smallskip}
\multicolumn{2}{c}{Source}  & \multicolumn{2}{c}{Coordinates} & Calibration Intent$^{a}$  &   T$_{\rm on \ source}$  \\
Common Name& J2000 Name   & RA (J2000) &DEC (J2000) &&   (min)  \\
\noalign{\smallskip}
\hline
\noalign{\smallskip}  
3C~279&  J1256$-$0547 &12:56:11.167 &$-$05:47:21.52& Target 1  & 34.4 \\
3C~273  &J1229+0203&12:29:06.700  &+02:03:08.60& Flux, Bandpass & 25.8     \\
M87   & NGC 4486 &12:30:49.412 &+12:23:28.27 & Target 2 & 16.9  \\
4C 01.28&J1058+0133  & 10:58:29.604 &+01:33:58.83&  Polarization & 10.5  \\
S4 1144+40 &J1146+3958  &11:46:58.298 &+39:58:34.28 & \ldots  &7.5  \\ 
PKS 1510-089 &J1512$-$0905&15:12:50.533 &$-$09:05:59.83& \ldots &3.8 \\ 
PKS 1335-127 &J1337$-$1257  &13:37:39.783 &$-$12:57:24.70& \ldots & 3.8 \\ 
\hline
\multicolumn{6}{c}{non-VLBI scans$^{b}$}\\
4C 01.28&J1058+0133  & 10:58:29.604 &+01.33.58.83  &  Polarization & 40.3 \\
PKS 1243-072 &J1246$-$0730&12:46:04.232 &$-$07.30.46.575  & Phase  & 24.7  \\  3C~273&J1229+0203&12:56:11.167 &$-$05:47:21.52  & Flux, Bandpass & 10.1    \\
\noalign{\smallskip}
\hline    
\end{tabular}
\end{center}
%\table-comments
\tablefoottext{a}{In addition to the specified intent, amplitude and phase gains were estimated on each observed source.} 
\tablefoottext{b}{These sources were observed by ALMA for calibration purposes during gaps in the VLBI schedules. } 
\end{table*}
%==============================================================================
\begin{table*}
\caption{Frequency-averaged polarization properties of AGN targets (at a representative frequency of 343 GHz). 
} 
\begin{tabular}{cccccccc}
\hline\hline
Source & Day & I$^{a}$ & Spectral Index & LP & $\chi$ & $\chi_0$ & RM \\
 & (2021) & (Jy) & ($\alpha$) & ($\%$)  & (deg) & (deg) & ($10^5$ rad m$^{-2}$) \\
\hline
3C273 & Apr 19 & 4.17 $\pm$ 0.42 & -0.781 $\pm$ 0.008 & 1.67 $\pm$ 0.03 & -64.19 $\pm$ 0.52 & -38.9 $\pm$ 7.3 & -5.8 $\pm$ 1.7 \\
3C279 & Apr 19 & 8.73 $\pm$ 0.87 & -0.48 $\pm$ 0.04 & 10.63 $\pm$ 0.03 & -46.711 $\pm$ 0.081 & -41.2 $\pm$ 1.1 & -1.26 $\pm$ 0.26 \\
4C01.28 & Apr 19 & 1.50 $\pm$ 0.15 & -0.42 $\pm$ 0.03 & 8.67 $\pm$ 0.03 & -37.72 $\pm$ 0.11 & -37.4 $\pm$ 1.5 & -0.08 $\pm$ 0.35 \\
J1146+3958 & Apr 19 & 0.396 $\pm$ 0.040 & -0.4 $\pm$ 0.2 & 2.66 $\pm$ 0.07 & 87.27 $\pm$ 0.77 & 107 $\pm$ 11 & -4.4 $\pm$ 2.5 \\
J1337-1257 & Apr 19 & 3.13 $\pm$ 0.31 & -0.38 $\pm$ 0.05 & 16.68 $\pm$ 0.03 & -7.310 $\pm$ 0.055 & -8.15 $\pm$ 0.77 & 0.19 $\pm$ 0.18 \\
J1512-0905 & Apr 19 & 1.74 $\pm$ 0.17 & -0.42 $\pm$ 0.04 & 1.07 $\pm$ 0.04 & -65.72 $\pm$ 0.95 & -96 $\pm$ 13 & 6.9 $\pm$ 3.1 \\
M87 & Apr 19 & 0.970 $\pm$ 0.097 & -1.24 $\pm$ 0.02 & 2.80 $\pm$ 0.03 & 12.60 $\pm$ 0.35 & 8.5 $\pm$ 5.0 & 0.9 $\pm$ 1.1 \\
\hline
\multicolumn{8}{c}{non-VLBI scans}\\ 
3C273$^{b}$ & Apr 19 & 4.17 $\pm$ 0.42 & -0.68 $\pm$ 0.01 & ...& ...& ...& ...\\
4C01.28 & Apr 19 & 1.50 $\pm$ 0.15 & -0.36 $\pm$ 0.02 & 8.66 $\pm$ 0.03 & -37.68 $\pm$ 0.11 & -38.0 $\pm$ 1.6 & 0.07 $\pm$ 0.37 \\
J1246-0730 & Apr 19 & 0.280 $\pm$ 0.028 & -0.56 $\pm$ 0.04 & 0.6 $\pm$ 0.1 & -17.0 $\pm$ 6.6 & -119 $\pm$ 95 & 23 $\pm$ 21 \\
\hline\hline
\end{tabular}
\tablefoottext{a}{Errors in Stokes I include thermal and 10\% absolute flux systematic uncertainty summed in quadrature.}
\tablefoottext{b}{The short observation on 3C273 did not result in sufficiently good data to enable a meaningful polarization analysis.}
\label{tab:EHT_uvmf_RM}
\end{table*}

%%%%%%%%%%%%%%%%
%_______________________________________________________________
\subsection{Observational setup
\label{sec:obs}}
%_______________________________________________________________

\begin{figure}
\hspace{-5mm}
\includegraphics[width=0.5\textwidth]{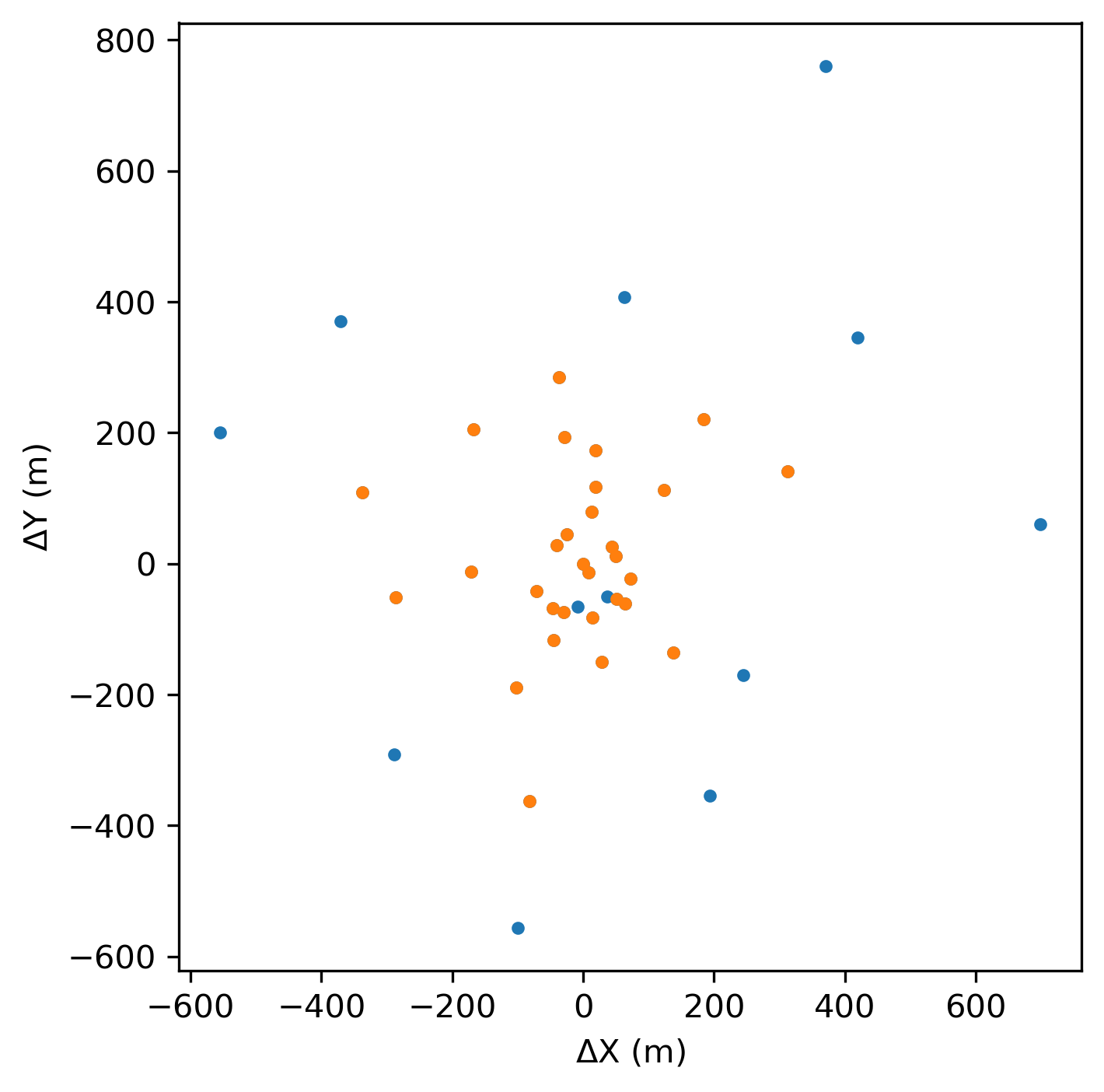} 
\caption{ALMA antenna locations for  the phased array (orange points) and the un-phased comparison antennas (blue points) during
  the Band~7 observations on April 19, 2021.  
Positions are { relative to the array reference antenna \citep{QA2Paper} and are} plotted with positive values of $X$ toward local east and positive values of $Y$ toward local north.
}
\label{fig:plotants}
\end{figure}

The observations at Band 7 spanned a continuous session of nearly 5 hours (from  01:16 to 06:08 UTC) and utilized 42 antennas. 
 Of these, 31 were configured within a 400-meter radius around the reference antenna to form the phased array (baselines to 0.8 km).  
Eleven additional antennas (baselines to 1.3 km) completed the array and  were used as un-phased comparison antennas for the determination of phasing efficiency estimates. 
The antenna locations on April 19 are plotted in Fig.~\ref{fig:plotants}. 

%%%%%%%%%%%%%%%%%%%%%%%%%%%%%%%%
%%%%%%%%%%%%%%%%%%%%%%%%%%%%%%%%

The setup incorporated four spectral windows (SPWs) in dual linear polarization (LP); two in the lower sideband and two in the upper sideband, with central frequencies of 335.600, 337.5411\footnote{The Local Oscillator setup at ALMA does not allow evenly spaced basebands in Band~7, unlike in VLBI's Band 6.}, 347.600, and  349.600  GHz.  
 Each SPW offered an effective bandwidth of 1875~MHz. The ALMA interferometric data were averaged by a factor of 8 in frequency, resulting in
 240 spectral channels per SPW with a channel spacing of 7.8125~MHz.

%_______________________________________________________________
\subsection{Observed targets}
\label{sec:targets}
%_______________________________________________________________

As these observations were designed as a comprehensive end-to-end commissioning test, the well-known quasar 3C279 was selected as the primary target. Given the novelty of VLBI at 345 GHz, there is currently limited guidance on suitable calibrators for this frequency. To address this, a selection of promising candidates was included for calibration and evaluation. This candidate sample comprises bright quasars and other compact extragalactic sources that had previously produced VLBI fringes in Band 6 observations. Notably, the sample includes M87*, a planned future science target for 345 GHz VLBI observations. 
{It should be noted that the observations include scans on three targets  during gaps in the VLBI schedule. These scans were observed by ALMA for calibration purposes. In total, eight targets were observed in VLBI mode, while one target was observed exclusively by ALMA.}
A detailed list of the observed sources and their adopted calibration roles is provided in Table~\ref{table:sources} (see also Sect. \ref{sec:datacal}).

%_______________________________________________________________
\subsection{Data calibration}
\label{sec:datacal}
%_______________________________________________________________

The ALMA data were calibrated using the Common Astronomy Software Applications ({\sc casa}) package, following the specialized “Quality Assurance Level 2” (QA2) procedures described in \citet[see also \citealt{APP2B7}]{QA2Paper}. {As detailed in those studies, the VLBI and non-VLBI scans  are calibrated independently, resulting in separate calibration solutions.}

3C273 was adopted as an absolute flux-density calibrator, 
a source routinely observed by ALMA as part of the flux-density monitoring program with the ALMA Compact Array (ACA; see Appendix \ref{app:amapola_comp}).
From the ALMA database, the flux density at 342 GHz was derived as $S_{342 GHz} = 4.164$~Jy  with a spectral index $\alpha$=--0.79.

Accurate calibration of ALMA VLBI science data requires a full polarization calibration of interferometric visibilities. 
{ALMA’s linearly polarized feeds simultaneously receive both orthogonal polarizations (X and Y), with all antennas in the phased array aligned to the reference antenna. Since the reference antenna’s phase is set to zero in both polarizations, a residual phase bandpass remains in the cross-hands of all baselines. Correcting this residual XY phase is essential for accurately combining cross- and parallel-hands to extract Stokes parameters.
Thus, the primary calibration requirement for VLBI correlation is determining the X-Y phase difference and delay at the reference antenna (see Sect. 5 of \citealt{QA2Paper} for details). This step enables the conversion of ALMA’s linearly polarized data into a circular polarization basis, ensuring consistency with other VLBI stations \citep{PCPaper,QA2Paper}.
J1058+0133 (4C 01.28) was chosen as the polarization calibrator due to its adequate parallactic angle coverage and the presence of a compact, strongly polarized core (at the $\sim$8\% level). This selection enabled the simultaneous determination of the source polarization model and an  estimate of the XY cross-phase at the reference antenna.} 
The Stokes parameters derived from the polarization model were determined as 
 {\it IQUV} = [1.50, 0.033, $-$0.126, 0.0]~Jy. 
%________________________________________________________________
\subsection{Full-Stokes imaging} 
\label{sec:dataimg}

All targets observed in Band 7 were imaged using the {\sc casa} task \texttt{tclean} in all Stokes parameters: $I$, $Q$, $U$, and $V$. A Briggs weighting scheme \citep{Briggs1995} was adopted with a robust parameter of 0.5 and a cleaning gain of 0.1. 

A first cleaning step (100 iterations across all Stokes parameters) was performed within the inner 4\arcsec. If significant emission ($>7\sigma$) remained in the residual maps (e.g., for M87), an automatic script updated the cleaning mask, and a second, deeper cleaning (2000 iterations across all Stokes parameters) was conducted down to 2$\sigma$. Both cleaning steps were run with the parameter \texttt{interactive=False}. A final interactive cleaning step (\texttt{interactive=True}) was performed to manually adjust the mask, capturing any real emission missed by the automatic process and cleaning deeper sources with complex structures or high-residual signals.

The array configuration during phased-array observations provided synthesized beams ranging from 
[0\pas36--0\pas50] $\times$ [0\pas29--0\pas34], depending on the target. We produced 576$\times$576 pixel maps with a pixel size of 0\pas06, resulting in a field of view of 34\pas6 $\times$ 34\pas6, which comfortably covered the primary beam of ALMA Band~7 (18\arcsec\ at 350 GHz).

Maps were produced for individual SPWs and by combining SPWs in each sideband (SPW=0,1 and SPW=2,3) using \texttt{deconvolver=`hogbom'} with \texttt{nterms=1}. Additionally, maps combining all four SPWs were created using \texttt{deconvolver=`mtmfs'} with \texttt{nterms=2}, achieving better sensitivity and producing higher-quality images\footnote{{The \texttt{deconvolver=`mtmfs'} configuration outperformed \texttt{deconvolver=`hogbom'} when combining all four SPWs, yielding 30--40\% better sensitivity on average, as expected for RMS $\propto 1/\sqrt{\Delta \nu}$. However, \texttt{deconvolver=`hogbom'} performed poorly when combining all four SPWs, especially for sources with steep spectral indices, resulting in up to 50\% worse RMS compared to \texttt{deconvolver=`mtmfs'}.}}. Thus, the combined SPW images were used for the imaging analysis presented in this paper, except in cases where a per-SPW analysis was necessary, such as for spectral index or Faraday rotation studies.

%====================================================
\section{Data analysis}
\label{sec:analysis} 
%====================================================

The data, calibrated and imaged as described in Sect. \ref{sec:obs}, were analyzed following the procedures outlined in \citet{Goddi2021}. The analysis consisted of two main components:
\begin{itemize}
    \item Extract the Stokes parameters in the compact cores of the observed targets and estimate their polarimetric and Faraday properties (Sect. \ref{sec:polan}).
    \item Produce polarized images for each target and determine the spatial distribution of the polarimetric quantities on arcsecond (kiloparsec) scales (Sect. \ref{sec:imaging}).
\end{itemize}
%_______________________________________________________________
\subsection{Polarization analysis of point sources}
\label{sec:polan}
%_______________________________________________________________
To extract flux values for Stokes $I$, $Q$, $U$, and $V$, we employed two alternative methods, one that utilizes the visibility data and the other full-Stokes images. 

In the $uv$-plane analysis, we used the external {\sc casa} library {\sc uvmultifit} \citep{UVMULTIFIT}. To optimize processing time, we first averaged all 240 frequency channels to produce eight-channel, four-SPW visibility $uv$ files. Assuming the emission is dominated by a central point source at the phase center, we fit a \texttt{delta} function to the visibilities, extracting Stokes $I$, $Q$, $U$, and $V$ parameters for each SPW individually. 

For the image-based approach, we summed the central $5\times5$ pixels of the \texttt{CLEAN} model component map. Summing only these central pixels isolates core emission from the surroundings, which is particularly important for sources with extended structure.

\citet{Goddi2021} performed a statistical comparison of flux-extraction methods in the $uv$ and image planes for a sample of VLBI targets observed in Bands 3 and 6. They found that the median absolute deviation of the Stokes parameters between methods is $<$1\% for both point and extended sources. This agreement holds for the current Band 7 observations as well.

Using the measured Stokes parameters, we determined polarization properties for all targets, including the fractional LP ($LP = \sqrt{Q^{2}+U^{2}}/I$), the electric vector position angle (EVPA; $2\chi=\mathrm{arctan}(U/Q)$), and its variation with frequency (Faraday rotation; see Sect. \ref{subsec:RM}). Uncertainties in LP include the thermal  errors of Stokes $Q$ and $U$ and a 1$\sigma$ systematic error (added in quadrature) associated with Stokes $I$ leakage into Stokes $Q$ and $U$ (0.03\% of Stokes $I$). This analysis results in LP uncertainties $<0.1\%$, consistent with previous studies \citep{Nagai2016,Bower2018,Goddi2021}.

The polarization quantities, averaged across the four SPWs, are reported in Table~\ref{tab:EHT_uvmf_RM}. Table~\ref{tab:EHT_uvmf_spw} provides the polarimetric quantities for each SPW.
{ We note that while the polarization parameters of 4C 01.28 from non-VLBI scans closely match those from VLBI scans (indicating consistent independent polarization calibrations), the short observation of 3C 273 did not yield data of sufficient quality for meaningful polarization analysis. Consequently, non-VLBI scans of 3C 273 were excluded, and the polarization analysis is based solely on VLBI scans.}

  \subsubsection{Comparison with the AMAPOLA survey} 
\label{sec:amapola_comp}

{For the purposes of absolute flux calibration, ALMA regularly monitors the flux density of bright sources (mainly blazars or quasi-stellar objects) distributed across the entire right ascension range ("the Grid"). These observations are conducted together with solar system objects as part of the Grid Survey (GS) program, which operates on a cadence of approximately 10 days. The observations are executed with the Atacama Compact Array (ACA) in Bands~3, 6, and 7. Since the full-polarization mode is employed, it is possible to extract polarimetric information from the GS sources.
This polarimetric analysis is performed using AMAPOLA\footnote{\url{http://www.alma.cl/~skameno/AMAPOLA/}}, a set of CASA-compatible Python scripts designed to reduce the full-Stokes polarimetry of GS observations. 
 While the AMAPOLA values are primarily used for observation planning and the ACA and ALMA-VLBI arrays cover different $uv$ ranges, comparing our data with the AMAPOLA database helps identify any systematic effects or clear inconsistencies due to variability within a week-long time frame.}

{The GS includes multiple measurements in Band 3 and Band 7 from April 2021 for all our targets, except M87.
Our analysis shows that the polarimetric measurements are generally consistent with the historical trends reported for the Grid sources.}
Additional details, including comparison plots, are provided in Appendix~\ref{app:amapola_comp}.
%--------------------------------------------------------------------------------------
\subsubsection{Rotation measure}
\label{subsec:RM}
%--------------------------------------------------------------------------------------

Since we measured the EVPA at four distinct frequencies (one for each SPW) spanning a 16~GHz frequency range (334.6--350.6 GHz), we could estimate
 the Faraday  rotation measure (RM) in the 0.87~mm band. Assuming that the Faraday rotation arises from a single, external, homogeneous Faraday screen (i.e., the rotation occurs outside the plasma responsible for the polarized emission), a linear dependence between the EVPA and the wavelength squared is expected. 

We modeled this relationship by fitting the RM and the mean wavelength EVPA ($\bar{\chi}$) using the standard linear  relation:
\begin{equation}
\chi = \bar{\chi} + RM (\lambda^2 - \bar{\lambda}^2) \,,
\label{eq:rm}
\end{equation}
where $\chi$ is the observed EVPA at wavelength $\lambda$, and $\bar{\chi}$ is the EVPA at the mean wavelength $\bar{\lambda}$ (corresponding to the band center). Additionally, the EVPA extrapolated to zero wavelength (assuming the $\lambda^2$ dependence holds) is given by%
\begin{equation}
\chi_{0} = \bar{\chi} - RM \bar{\lambda}^2 \,.
\label{eq:chi0}
\end{equation}
The RM fitting is performed using a weighted least-squares method applied to $\chi$ as a function of $\lambda^2$. The values of $\bar{\chi}$, $\chi_{0}$, and the fitted RM are reported in the sixth, seventh, and eighth columns of Table~\ref{tab:EHT_uvmf_RM}. 
{We determine EVPA uncertainties ranging from 0.06\dg\ to 1\dg\ (excluding J1246-0730, where LP$<1$\%), corresponding to RM propagated errors between 0.08 and $3 \times 10^5$ \radmsq. Despite these relatively high uncertainties and the limited frequency coverage at these high frequencies, we achieve $3\sigma$ RM detections in 3C273 and 3C279. Details on the calculation of EVPA and RM uncertainties, as well as an evaluation of the robustness of RM fits derived from relatively narrow frequency coverage (16–18 GHz) at submillimeter wavelengths, are provided in \citet{Goddi2021}.}

%--------------------------------------
\begin{figure*}
%\centering
\includegraphics[width=0.515\textwidth]{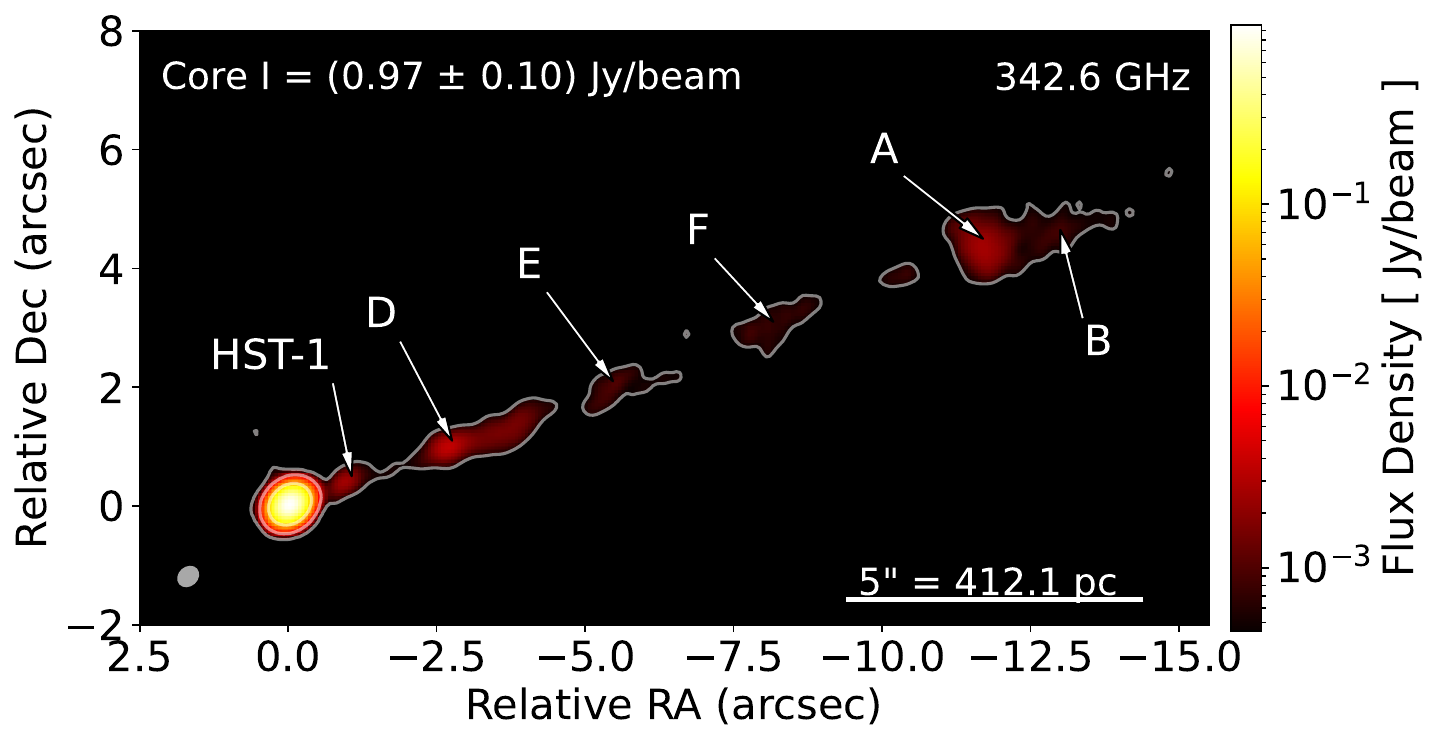} 
\includegraphics[width=0.5\textwidth]{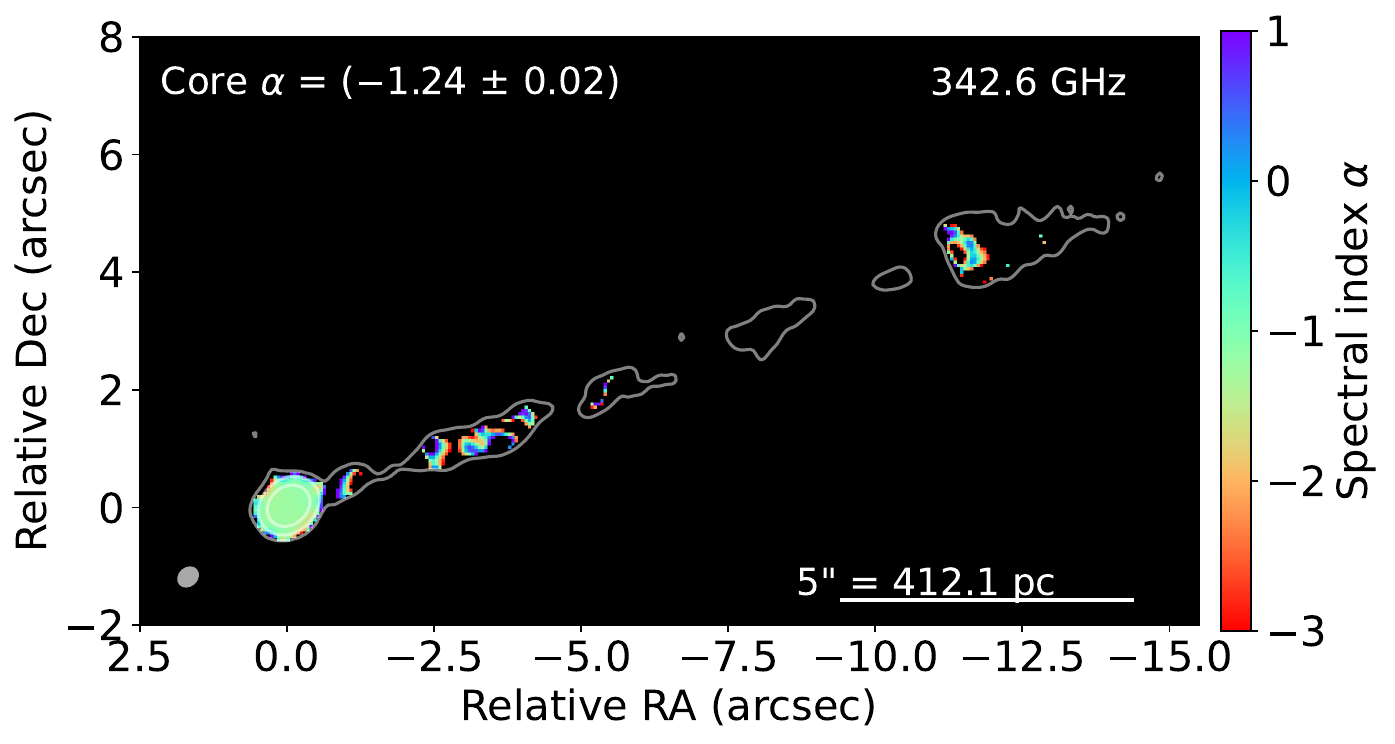} 
\includegraphics[width=0.5\textwidth]{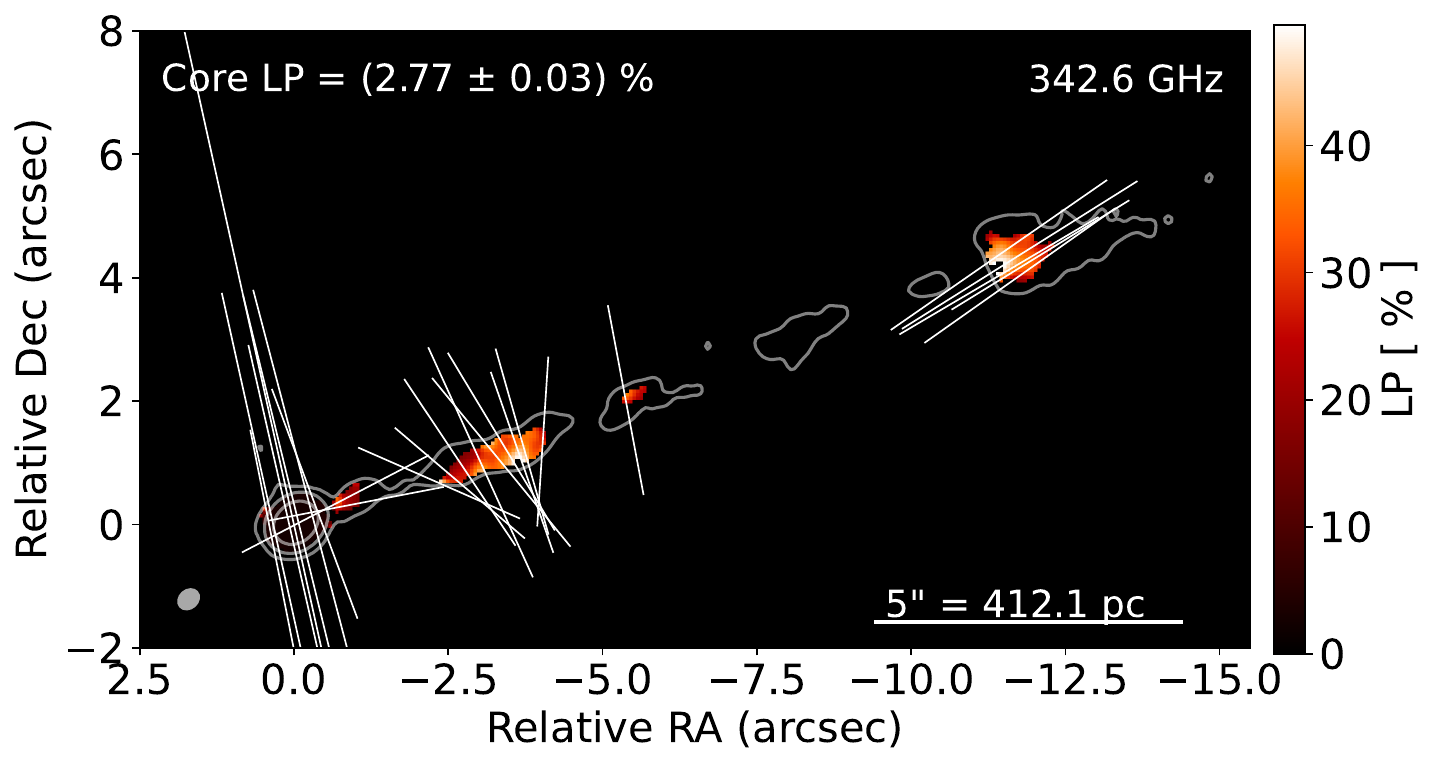}  \hspace{1.7mm}
\includegraphics[width=0.515\textwidth]{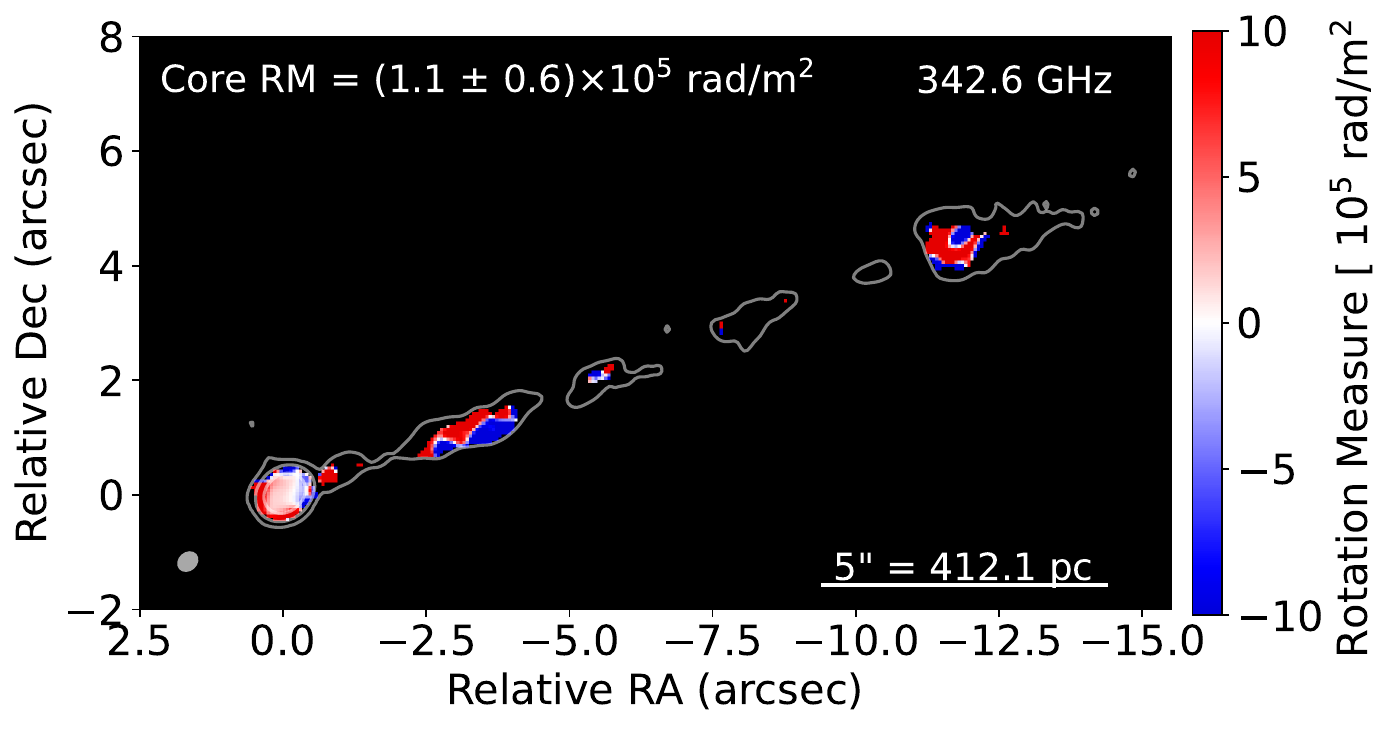}  
\caption{
Polarization images of M87 at $\lambda$0.87~mm observed on April 19, 2021. 
{The raster images in each panel cover an area of $\approx 1.5 \times 0.8$~kpc and display the following}: total intensity, spectral index, fractional LP, and Faraday RM (from the top left to bottom right). 
White vectors overlaid in the LP panel (bottom left) represent the orientation of the EVPAs, with vector lengths linearly proportional to the polarized intensity. 
In each panel, the white contour corresponds to the $4\sigma_I$ level, where $\sigma_I=0.11$  mJy/beam is the RMS noise in the Stokes $I$ map. The total intensity brightness is plotted using a logarithmic scale starting at the $3\sigma$ level. 
For the spectral index map, we applied a threshold of $\rm 5 \times \sigma$ in Stokes $I$. For the LP fraction and RM maps, thresholds are defined as $\rm 3 \times \sigma_I$ for Stokes $I$ and $\rm 2 \times \sigma_{Ip}$ for the polarized flux density (here the $\sigma_{Ip} = 0.08$ mJy/beam includes the thermal noise and the systematic error from Stokes $I$ leakage into Stokes $Q$ and $U$, combined in quadrature). 
The total intensity, spectral index, LP fraction, and RM values at the peak of the compact core are annotated in the upper-left corner of each panel. EVPAs are sampled every six pixels for clarity. 
The synthesized beam, represented as an ellipse in the lower-left corner of each panel, measures 0\pas40~$\times$~0\pas32 at a position angle of $-48^\circ$. Note that no primary beam correction is applied to these maps. 
}
\label{fig:m87_polimage}
\end{figure*}
%_______________________________________________________________
\subsection{Polarization images}
\label{sec:imaging}
%_______________________________________________________________

We used the full-Stokes images (produced as described in Sect. \ref{sec:dataimg}) to 
determine the spatial distribution of the polarimetric quantities on arcsecond scales. 
Specifically, we executed custom Python scripts in \casa\ that take as input the Stokes $I$, $Q$, and $U$ images and output images of the linear polarized intensity, the fractional LP, the EVPA, and the Faraday RM. All these polarization quantities were calculated as described in Sects. \ref{sec:polan} and \ref{subsec:RM} on a pixel-by-pixel basis after convolving all SPW images to the same synthesized beam. 
In generating the final images, we applied a threshold defined as $\rm 3 \times \sigma$ (rms noise level) for Stokes $I$ and $\rm 2 \times \sigma$ for the polarized flux density\footnote{The $\sigma$ error on the polarized flux includes the thermal error (residual image rms noise) and the systematic error (added in quadrature) associated with Stokes $I$ leakage onto Stokes $Q$ and $U$ (0.03\% of Stokes~$I$).}. 

\begin{figure*}[ht!]
\centering
\includegraphics[width=0.33\textwidth]{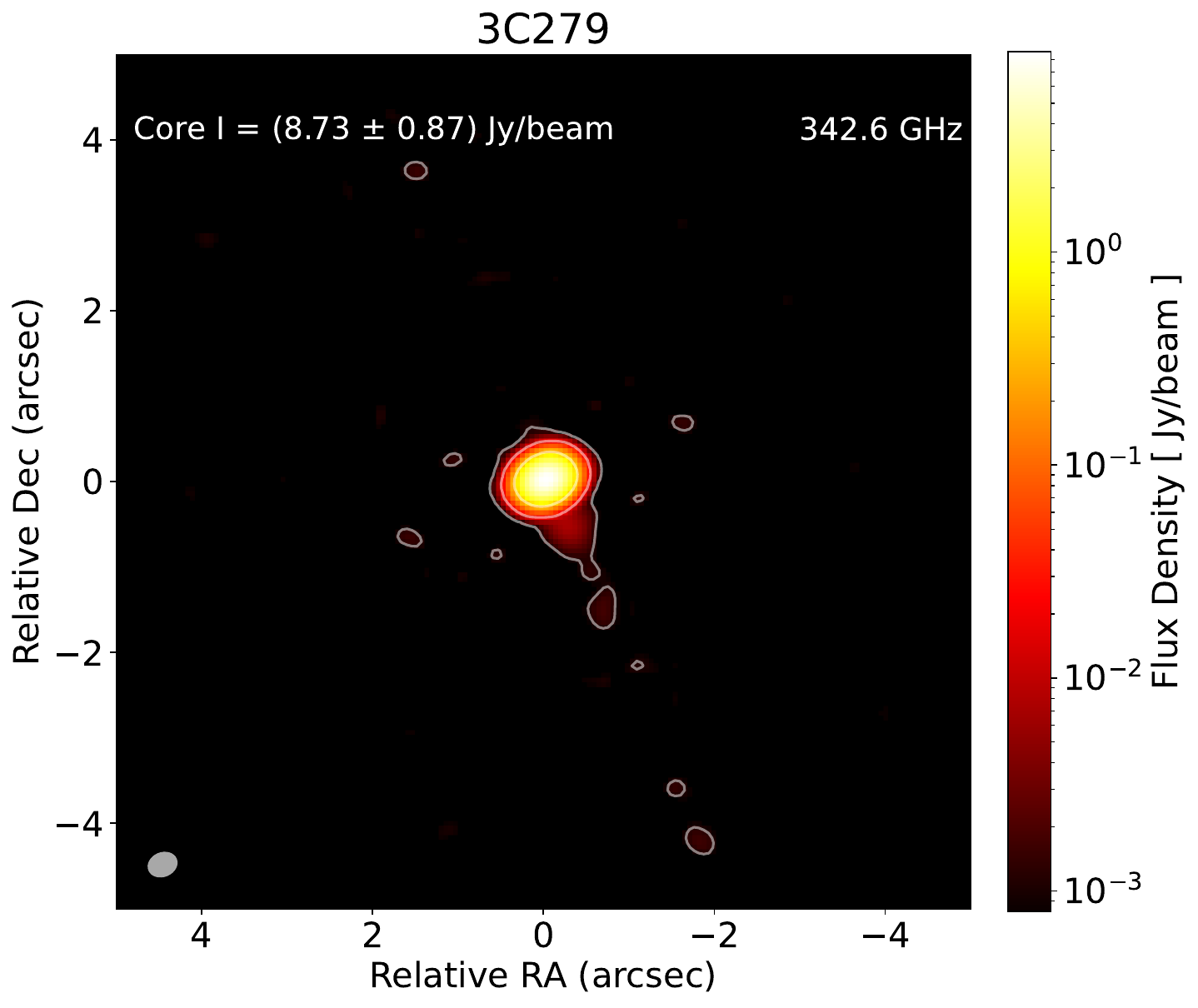} \hspace{-0.3cm}
\includegraphics[width=0.32\textwidth]{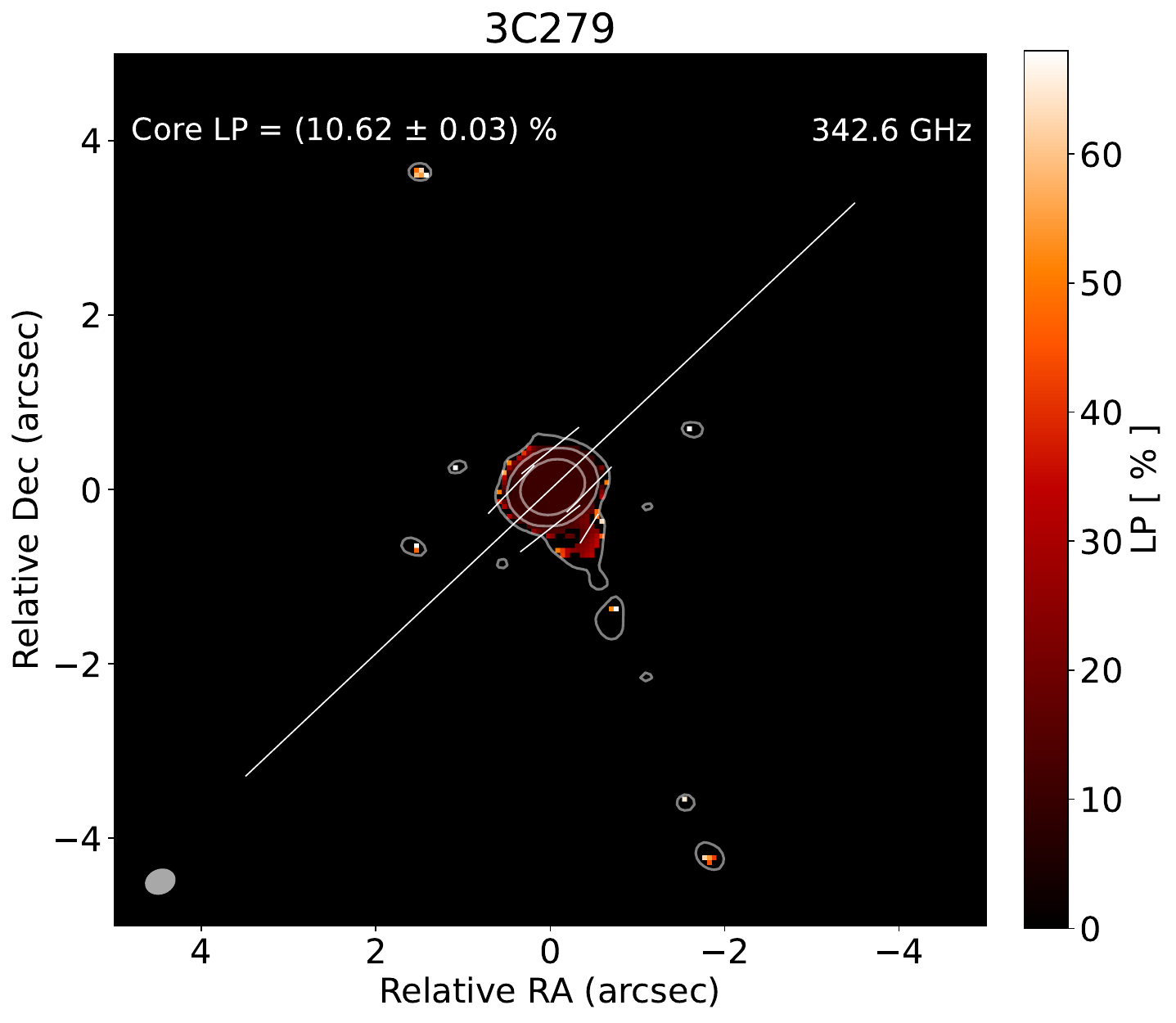}  \hspace{-0.3cm}
\includegraphics[width=0.34\textwidth]{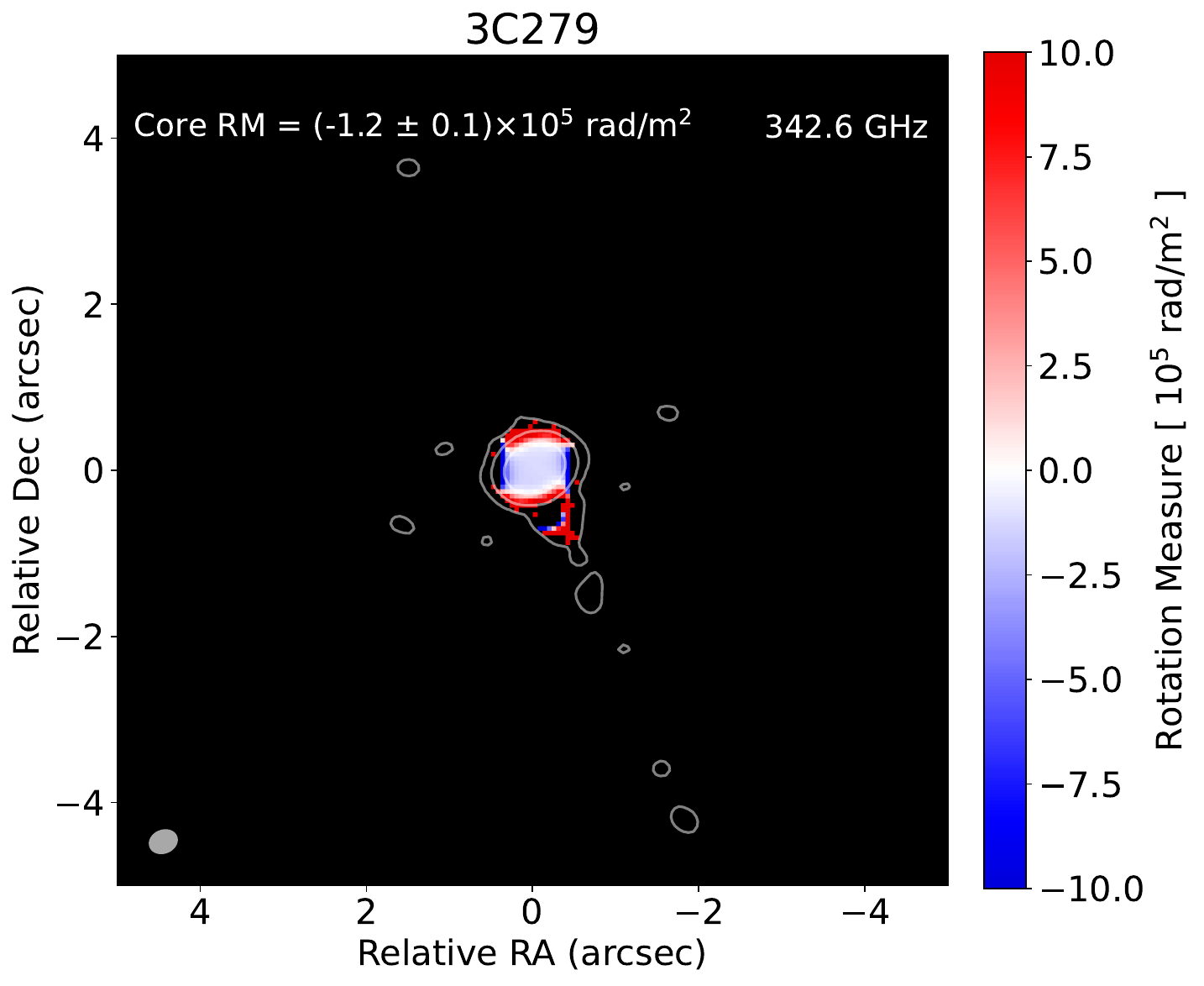}  \hspace{-0.3cm}

\includegraphics[width=0.33\textwidth]{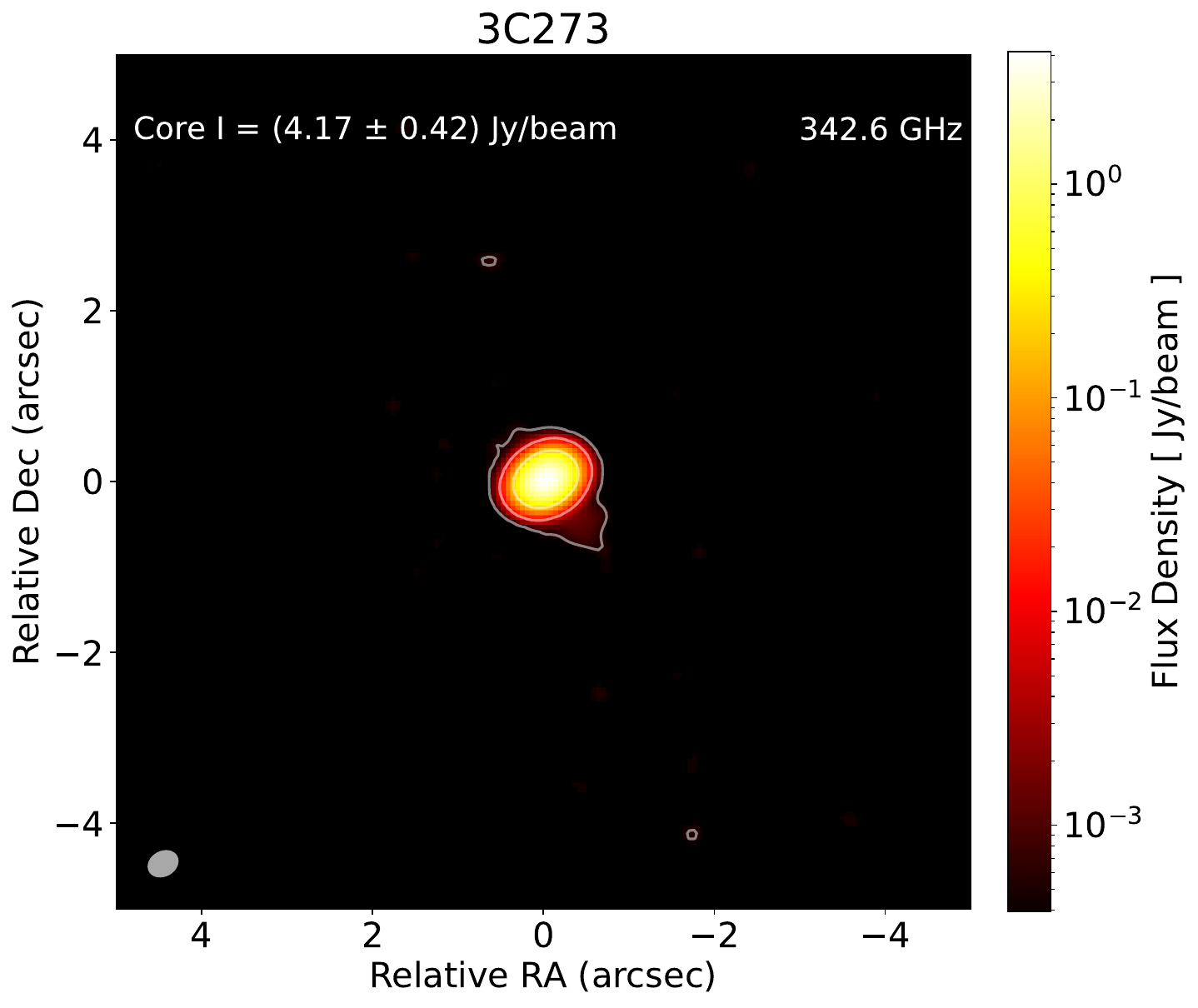} \hspace{-0.3cm}
\includegraphics[width=0.32\textwidth]{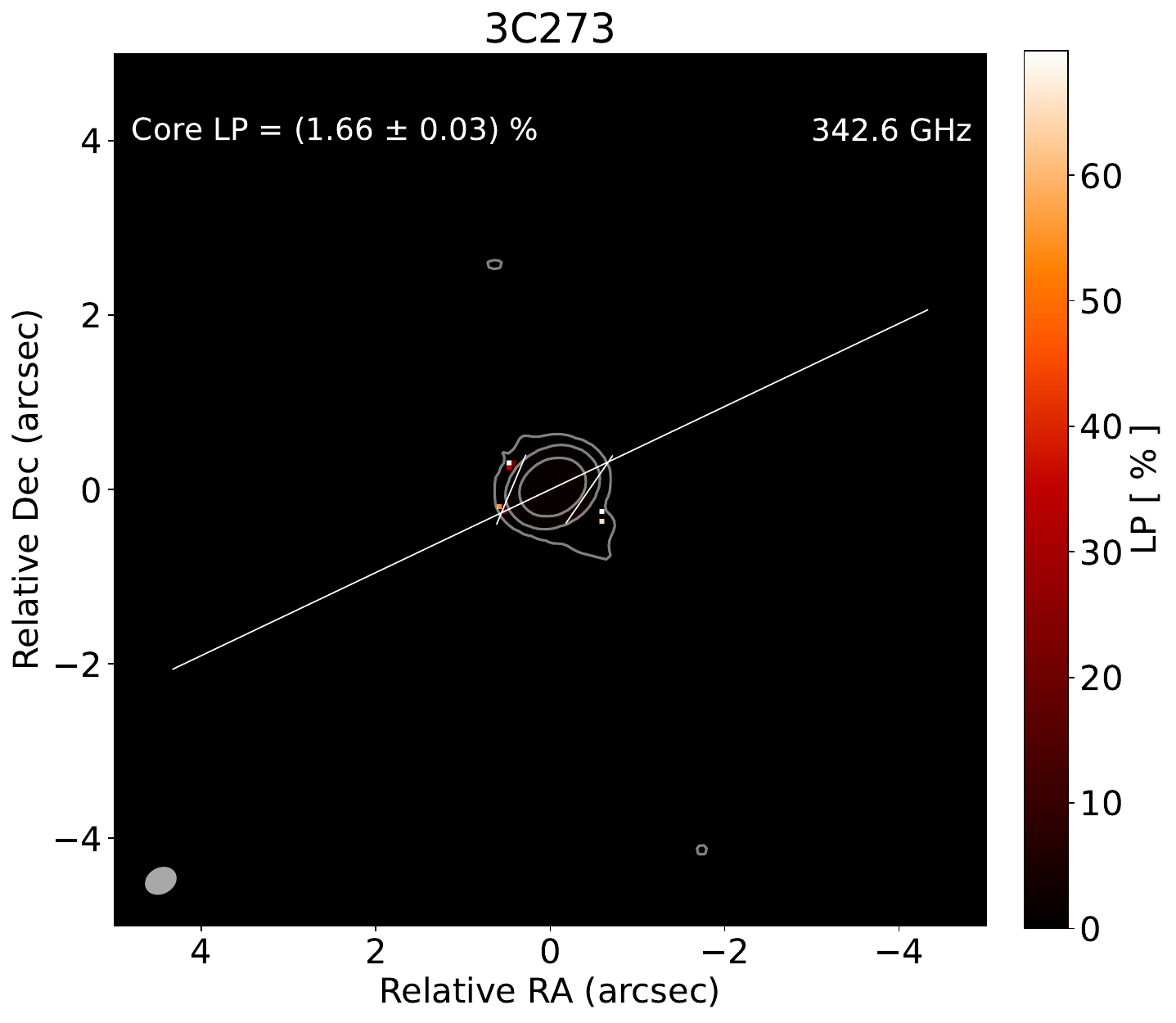}  \hspace{-0.3cm}
\includegraphics[width=0.34\textwidth]{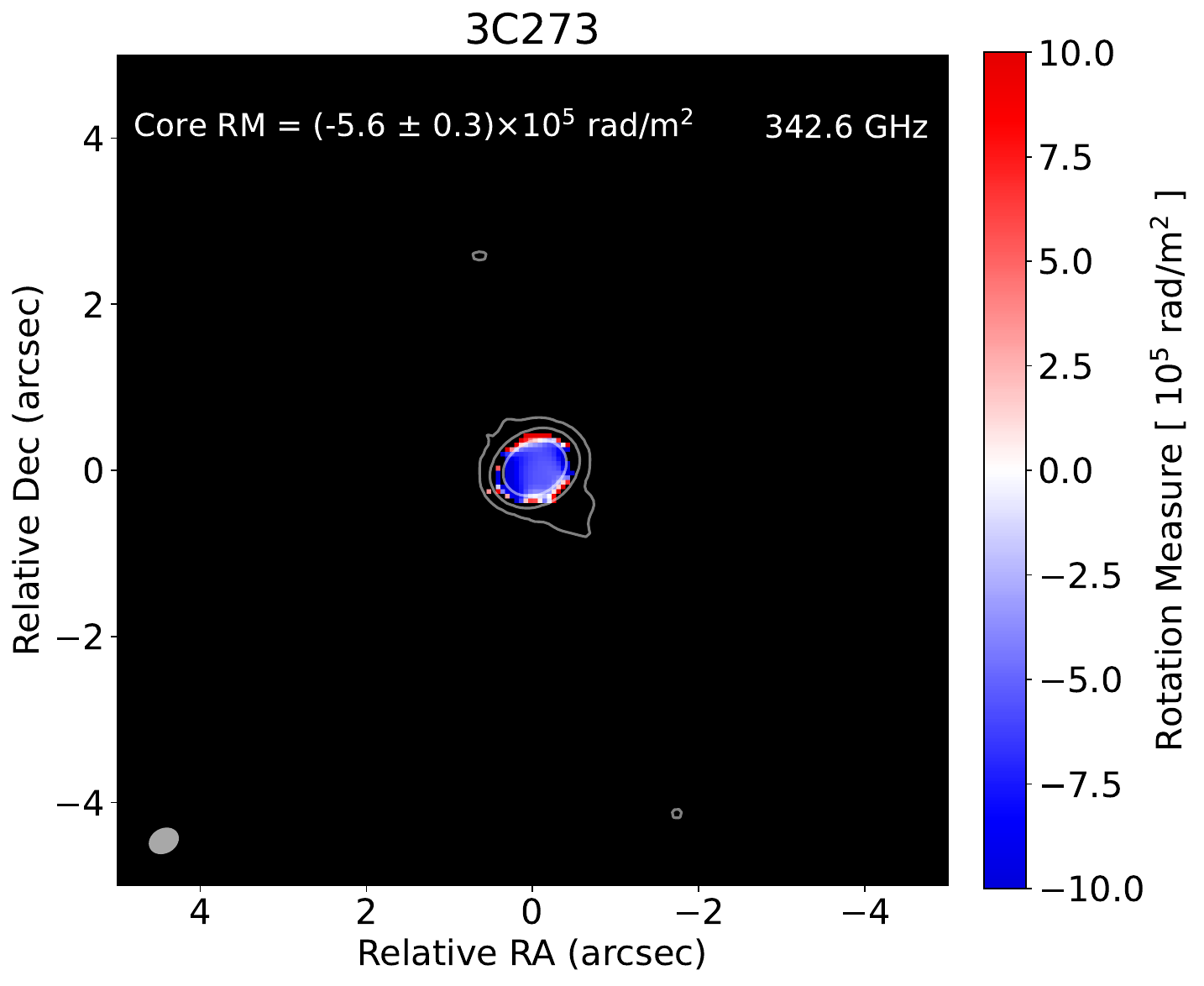}  \hspace{-0.3cm}
\includegraphics[width=0.33\textwidth]{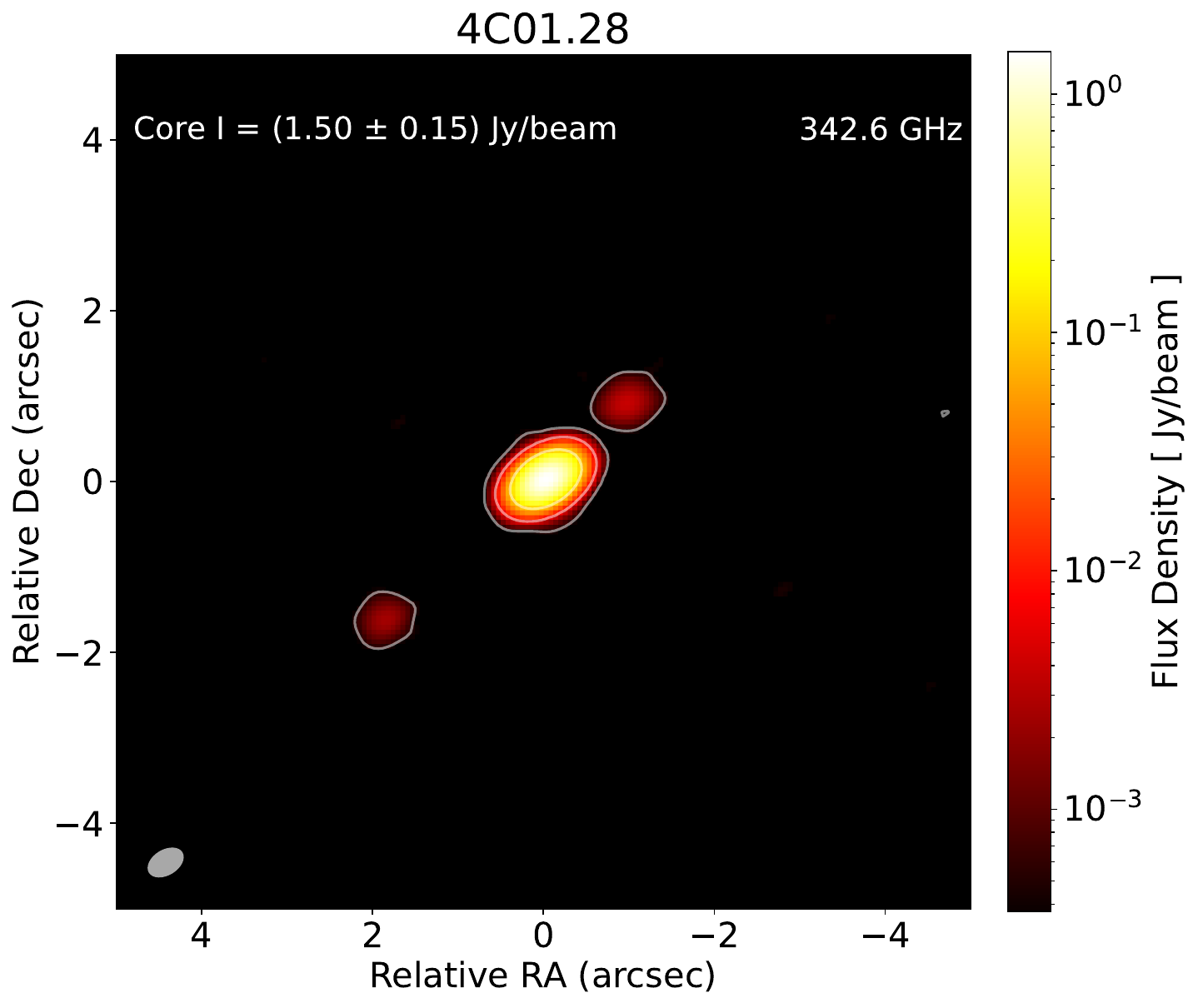} \hspace{-0.3cm}
\includegraphics[width=0.32\textwidth]{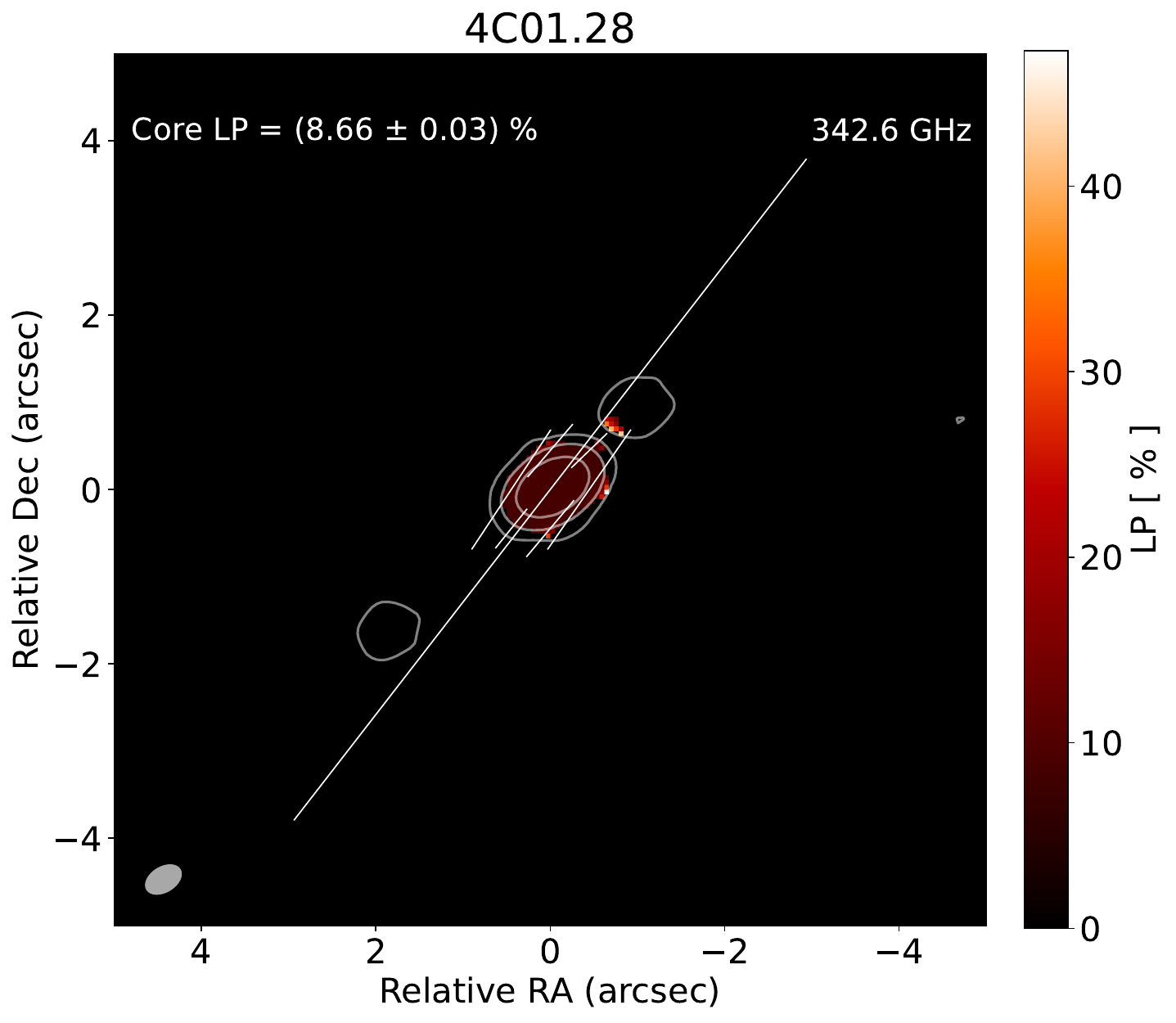}  \hspace{-0.3cm}
\includegraphics[width=0.34\textwidth]{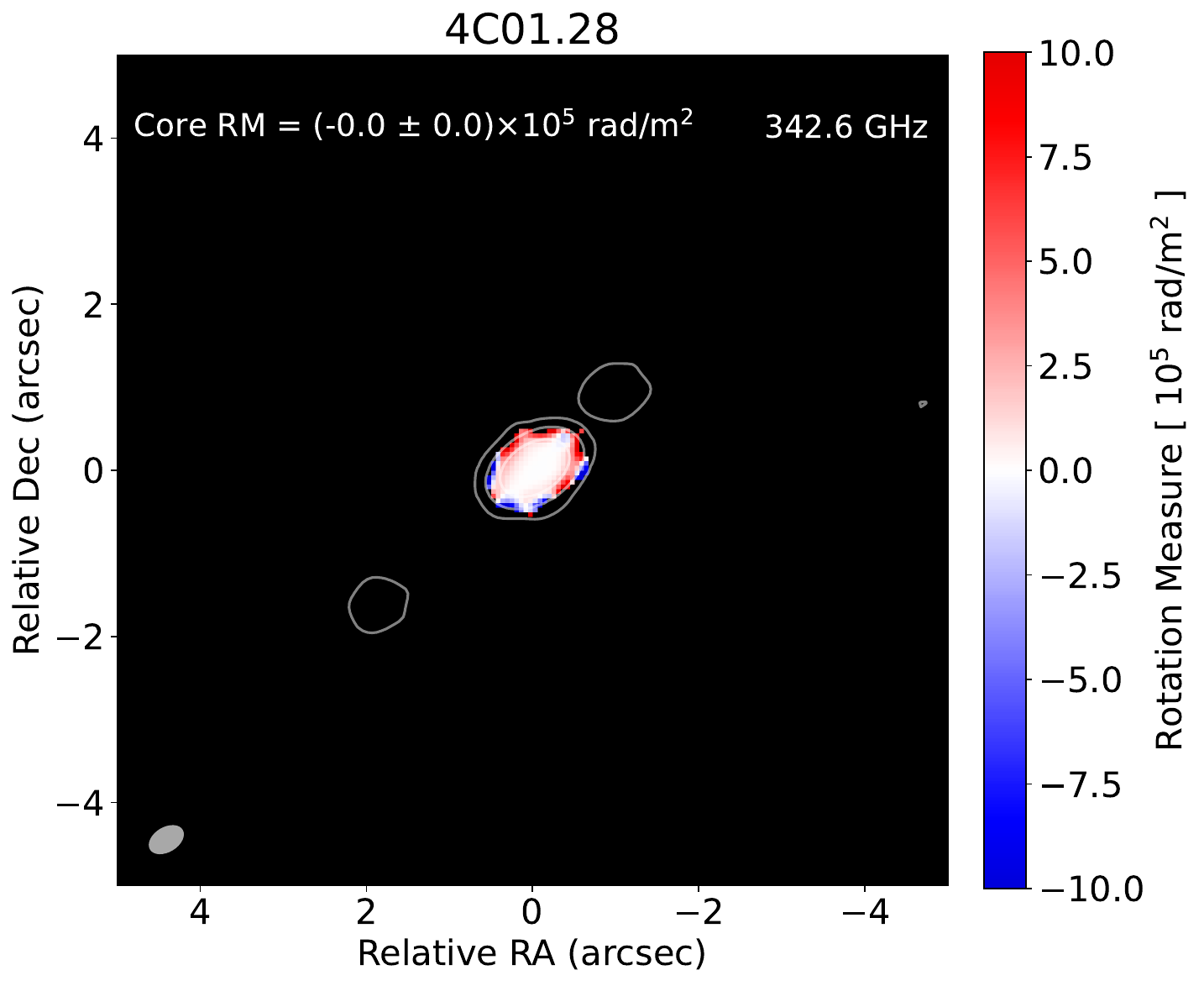} 
\caption{Polarization images of selected AGNs observed with ALMA at 0.87~mm on April 19, 2021 (see Fig.~\ref{fig:m87_polimage} for a description of the plotted quantities). 
The synthesized beams  (represented as an ellipse in the
lower-left corner of each panel) have the following sizes (and position angles): 0\pas36 $\times$ 0\pas29 ($-67.5^\circ$) for 3C279, 0\pas38 $\times$ 0\pas30 ($-60.8^\circ$) for 3C273, and 0\pas46 $\times$ 0\pas30 ($-58.7^\circ$) for 4C01.28. 
{Note that the EVPAs are not Faraday-corrected and that the magnetic field vectors should be rotated by 90\dg,  ignoring Lorentz transformation and light aberration. 
}}
\label{fig:polimages}
\end{figure*}

{ Representative images are presented in Fig.\ref{fig:m87_polimage}, showcasing M87, and Fig.\ref{fig:polimages}, featuring 3C 279, 3C 273, and 4C 01.28}\footnote{ Images for other targets are not included, as they remain unresolved on arcsecond scales and do not provide additional information beyond what is already summarized in Table~\ref {tab:EHT_uvmf_RM}.}.  
The raster image in each panel displays the total intensity, spectral index (for M87), LP fraction, and RM. White vectors overlaid on these images represent the orientation of the EVPAs, with their lengths linearly proportional to the polarized flux. It should be noted that these EVPAs are not corrected for Faraday rotation and that the magnetic field vectors should be rotated by $90\deg$.

%--------------------------------------------------------------------------------------

%--------------------------------------------------------------------------------------

%====================================================
\section{Results and discussion}
\label{sec:res}
%====================================================

We derived  polarimetric properties { and produced polarized images}  of eight AGNs observed in full-polarization mode with ALMA in the 0.87 mm band for the first time. In Sect. \ref{sec:agnpol} we analyze the polarization characteristics of the AGN; in Sect. \ref{sec:AGNpec} we discuss their spectral indices; and in Sect. \ref{sec:m87} we focus on the polarized submillimeter emission from the kiloparsec-scale jet in M87.

%_______________________________________________________________
\subsection{AGN polarization properties}
\label{sec:agnpol}
%_______________________________________________________________

The polarimetric quantities for our AGN targets, derived as described in Sect. \ref{sec:polan}, are summarized in Table~\ref{tab:EHT_uvmf_RM}.  
The table also includes the measured LP fractions, EVPAs, and RMs, which can provide key insights into the magnetic field structures and plasma conditions within the AGN jets. 
The LP fractions in the AGN central cores span a wide range, from $\lesssim 1$\% for weakly polarized targets (e.g., PKS1243-072 and PKS1510-089) to 10–17\% for strongly polarized sources like 3C279 and PKS1335-127, consistent with previous measurements (Appendix~\ref{app:amapola_comp}). 

This is the first time Faraday RMs have been measured in the submillimeter. 
In 3C273, we measure RM = $(-5.8 \pm 1.6) \times 10^5$ \ \radmsq   \ at 0.87mm, which is consistent with previous ALMA observations at 1.3~mm \citep[December 2016: RM = $(5.0 \pm 0.3) \times 10^5$~ \ \radmsq  ; April 2017: RM = $(2.5 \pm 0.3) \times 10^5$  \ \radmsq  ;][]{Hovatta2019,Goddi2021} and higher than 3mm observations \citep[RM = $(-0.60 \pm 0.14) \times 10^4$  \ \radmsq  ;][]{Goddi2021}. Interestingly, the sign of the RM at 0.87mm differs from previous 1.3~mm observations, which we interpret as a result of time variability rather than a frequency-dependent effect (Carlos et al., in prep.).

In 3C279, earlier measurements showed RM values ranging from 1800 to 2700 \ \radmsq   \ at 3.5mm \citep[e.g.,][]{Lee2015,Goddi2021} and an upper limit of 5000 \ \radmsq   \ at 1.3~mm \citep[][]{Goddi2021}. Our new observations suggest a significant increase in RM at shorter wavelengths, supporting the idea that higher frequencies probe the innermost regions with stronger magnetic fields and denser plasma.  
Given the time variability of these sources, simultaneous multi-band observations are required to confirm this trend and establish the dependence of RM on wavelength.
Some targets, including M87, 3C279, 3C273, and 4C~01.28, were observed  at lower frequencies (Bands 3 and 6; $\nu\approx$86~GHz and 230~GHz, respectively) during the same week. A comparative analysis of polarization properties across ALMA bands is planned for a future study (Carlos et al., in prep.).

Our findings of RM exceeding $10^5$ \radmsq \ in AGN cores at $\lambda \sim$0.87mm align with previous studies at millimeter wavelengths \citep[e.g.,][]{Plambeck2014,IMV2015,Hovatta2019,Goddi2021}. These high RM values, which are 1–2 orders of magnitude greater than those typically reported for AGNs at $\lambda >$ 3~mm \citep[e.g.,][]{Gabuzda2017,Peng2024}, point to a denser Faraday screen or stronger magnetic fields in the submillimeter emission region.

%_______________________________________________________________
\subsection{AGN spectral indices }
\label{sec:AGNpec} 
%_______________________________________________________________  

In addition to polarization parameters, we derived the total intensity spectral index $\alpha$ for all sources, where $\alpha$ is defined such that $I(\nu) \propto \nu^{\alpha}$. The spectral index was computed “in-band” using a weighted least-squares fit across the four flux-density measurements  
in each SPW. The AGN cores systematically show negative spectral indices in the range $\alpha = -1.3$ to  $-0.4$, consistent with previous findings at millimeter wavelengths \citep[e.g.,][]{Agudo2018,Goddi2021}. This contrasts with the flat spectral indices ($\alpha \approx 0$) typically observed at longer centimeter wavelengths, further supporting the idea that AGN cores become progressively more optically thin at shorter wavelengths.

For M87, the compact core exhibits a spectral index of $\alpha = -1.25$, consistent with previous measurements at 1.3 mm \citep[][]{Goddi2021} but contrasting with the flatter spectra observed at 3 mm \citep[e.g.,][]{Doi2013} and at centimeter wavelengths \citep[e.g.,][]{Kravchenko2020}. This steep spectral index suggests a spectral break between 3 mm and 1.3 mm, transitioning to a consistent power law from 1.3 mm to 0.87 mm. Such a break is likely due to the inclusion of contributions from both the compact core and the inner jet within the ALMA beam. While the compact (VLBI) core typically displays a flat spectrum, the jet component has a steeper spectral index, which dominates at higher frequencies due to decreased opacity at the jet base, as predicted by the standard jet model \citep{BlandfordKonigl1979}.

%_______________________________________________________________
\subsection{Polarization properties of the M87 jet at 345 GHz}
\label{sec:m87} 
%_______________________________________________________________
High-resolution polarization imaging of the relativistic jet in M87 at millimeter wavelengths has been achieved through ALMA observations at $\lambda$3mm with a resolution $\sim$2.5\arcsec  \citep{Peng2024} and $\lambda$1.3mm with a resolution $\sim$1\arcsec  \citep{Goddi2021}. These studies revealed the narrow, straight kiloparsec-scale jet extending over $\sim$25\arcsec\ from the nucleus, including several prominent knots (D, F, A, B, C) previously identified in optical and radio images. At $\lambda$3~mm, the jet-inflated radio lobes are also visible, with features imaged in greater detail at lower frequencies (e.g., the NRAO 20 cm Very Large Array image{\footnote{\url{https://www.nrao.edu/archives/items/show/33382}}}).

Our $\lambda$0.87mm ALMA observations show a similar structure to the $\lambda$1.3mm image  \citep[e.g., see Fig. 2 in][]{Goddi2021} but with improved angular resolution ($\sim$0\pas3), allowing us to resolve HST-1 from the core.
HST-1 is a bright, knot-like feature discovered with the \textit{Hubble} Space Telescope and located  approximately 0.85 arcseconds (about 60–70 parsecs) downstream from the central  black hole \citep{Biretta1999}.
HST-1 has exhibited remarkable properties over the years, including rapid variability, superluminal motion, and significant flaring activity \citep{Cheung2007,Giroletti2012},  making it 
a critical site for understanding particle acceleration, jet collimation, and magnetic field dynamics in the M87 jet and in AGNs in general.

 The radio core dominates the Stokes I emission with a peak brightness of $\sim$1Jy beam$^{-1}$ (Fig.\ref{fig:m87_polimage}, top left). 
In contrast, the jet knots become more prominent in the linearly polarized intensity, relative to the total intensity, as the fractional polarization increases from the core outward. 
The LP image (Fig.\ref{fig:m87_polimage}, bottom left) shows the lowest fractional polarization ($\lesssim$3\%) at the core, which rises to $\sim$20\% toward HST-1 and peaks at $\sim$55\%, 42\%, and 40\% in between knots D, E, and A, respectively. 
The high degree of polarization observed in the knots is indicative of a well-ordered magnetic field structure within these regions, likely resulting from shock compression or shear flows in the jet plasma \citep[e.g.,][]{Laing1980}.

The EVPA distribution observed at $\lambda$0.87mm closely matches prior results at $\lambda$1.3mm \citep{Goddi2021} and $\lambda$3~mm \citep{ Peng2024} and agrees with centimeter-wave polarization measurements from the Very Large Array \citep[VLA;][]{Algaba2016, Pasetto2021}. This consistency across multiple epochs suggests a stable magnetic field configuration. 
The EVPA distribution is generally perpendicular to the jet axis, except in the regions HST-1 and Knot A. Ignoring Lorentz transformation and light aberration, rotating the EVPA by 90° (without Faraday correction) indicates that the magnetic field is mostly parallel to the jet axis, except in HST-1 and Knot A, where it becomes nearly perpendicular. 
These deviations are likely caused by recollimation or standing shocks, which can alter the helicity of the magnetic field locally due to variations in the radial profiles of the poloidal and toroidal components \citep{Mizuno2015}.

 The RM, derived from EVPA measurements between 334.6GHz and 350.6GHz (Fig.~\ref{fig:m87_polimage}, bottom right), reveals both gradients  and sign reversals along the jet. Near the core, the RM exhibits an east-to-west gradient, ranging from $(2.5\pm3.9)\times10^{5}$  \radmsq \ at 0.3\arcsec\ east to $(-1.2\pm0.5)\times10^{5}$ \radmsq \ at 0.3\arcsec\ west. Downstream in HST-1, the RM reaches $(44\pm18)\times10^{5}$ \radmsq \ at 0.96\arcsec\ from the core, decreasing with distance. Knot D exhibits RM values ranging from $(47\pm23)\times10^{5}$ \radmsq \ to $(-29\pm11)\times10^{5}$  \radmsq, while knot A displays an RM of $(20.3\pm4.1)\times10^{5}$  \radmsq\  that varies significantly across the region.
 
The observed reversals in RM sign along the jet indicate changes in the line-of-sight magnetic field direction, while the RM gradients across the jet width reveal oppositely directed line-of-sight magnetic fields at the jet edges. These results are consistent with previous lower-frequency studies, which reported similar RM gradients observed with ALMA at 3 mm \citep{Peng2024} and the VLA at 1.7–7.5 cm \citep{Pasetto2021}. 
Such RM gradients and sign reversals are  evidence of a helical magnetic field threading the jet, potentially persisting up to kiloparsec scales \citep{Pasetto2021}. This configuration aligns with theoretical predictions of dynamically significant poloidal magnetic fields being twisted into a helix by the rotation of the black hole–accretion disk system \citep{Tchekhovskoy2011}, and  possibly with the "cosmic battery" model \citep[e.g.,][]{Myserlis2021,Contopoulos2022}. Independent support for this interpretation is provided by EHT observations of polarized emission near the M87 black hole \citep{EHT2021_2}, further corroborating the presence of a helical magnetic field structure.

A helical magnetic field could potentially also explain another feature observed in the M87 core, its high RM variability  \citep{Goddi2021}. 
Several scenarios may account for this variability, including turbulence in the accretion flow causing internal Faraday rotation, a dynamically changing external Faraday screen, or a rapidly varying source at horizon scales with a static external screen. {Alternatively, a helical magnetic field could introduce RM variability through beam-averaging effects. Variations in beam size across different observations may sample regions with oppositely directed magnetic fields along the line of sight, leading to distinct RM measurements.}

Our submillimeter observations, however, are limited by angular resolution, sensitivity, and frequency coverage, preventing a definitive differentiation between internal and external Faraday rotation and a precise characterization of the helical magnetic field  in the M87 jet. {While spatial RM variations along the jet axis are evident, the current imaging sensitivity is insufficient to continuously recover the polarized emission structure on kiloparsec scales. This limitation prevents us from confirming whether consistent RM and LP transverse gradients persist along the full extent of the jet}. Moreover, the limited resolution impedes the separation of emissions originating from the jet’s edges and central axis, leaving open the possibility that the observed RM gradients are due to external material rather than the jet’s intrinsic magnetic field.
Simultaneous observations with the EHT (which will be discussed in Paper II) are expected to shed light on the properties of the Faraday rotation medium in the core region and provide insights into the magnetic field structure at the base of the jet. 

The observed time variability in RM \citep{Goddi2021} adds another layer of complexity, as it precludes using nonsimultaneous datasets to reliably confirm RM-wavelength dependences. Addressing these challenges requires simultaneous, beam-matched, and multifrequency ALMA observations.
To this end, we plan to analyze a comprehensive dataset of ALMA observations, covering multiple frequency bands (Bands 3, 6, and 7) and spanning several years, with data obtained on multiple days. This systematic approach will enable a detailed investigation of time- and frequency-dependent effects and thus a more robust determination of the RM’s frequency dependence. Such an analysis will provide critical insights into whether the Faraday rotation is internal or external and refine our understanding of the helical magnetic field structure in the M87 jet. The findings from this extended analysis will be presented in an upcoming publication.

%==============================================================================
\section{Summary and conclusions}
\label{sec:conclusions}
 
 We have presented the first submillimeter full-polarization study of radio-loud AGNs with ALMA, analyzing their polarization and Faraday properties. We find  LP fractions ranging from 1\% to 17\% and RMs exceeding $10^{5}$ \radmsq, consistent with earlier studies at millimeter wavelengths \citep[e.g.,][]{Plambeck2014,IMV2015,Hovatta2019,Goddi2021}. These RM values are 1–2 orders of magnitude higher than those observed in AGNs at $\lambda>$3 mm \citep[e.g.,][]{Gabuzda2017,Peng2024}, indicating a denser Faraday screen or stronger magnetic fields in the submillimeter emission regions.

We produced the highest-frequency polarized images ever of these AGNs, which included M87 and its jet. For M87, we observe RM gradients and sign reversals both along and across the jet axis, potentially reflecting reversals in the magnetic field direction relative to the line of sight. If confirmed, this would support a helical magnetic field configuration on kiloparsec scales, as suggested by lower-frequency VLA studies \citep[e.g.,][]{Pasetto2021}. 
In future work we will analyze multifrequency, multi-epoch ALMA data and explore the time- and frequency-dependent properties of Faraday rotation to better constrain the magnetic field structure in M87.

The ALMA data were obtained in the 0.87 \ \text{mm} band  during VLBI commissioning tests conducted in collaboration with the EHT Collaboration. These data provided essential calibration and interpretation for simultaneous VLBI observations with the EHT. Notably, phased ALMA observations on April 19, 2021, enabled the first detection of VLBI fringes at 345~GHz for M87 and selected AGNs \citep[][Paper II]{app2_report_2024}. This milestone represents a key step in evaluating VLBI imaging feasibility in the submillimeter band.

Very long baseline interferometry  observations at this frequency offer a 50\% improvement in resolution compared to 230 GHz and substantially reduce interstellar scattering, which is especially relevant for imaging Sgr A*.
The combination of 230 GHz and 345 GHz data enhances \textit{uv} coverage, enabling high-fidelity imaging with multifrequency synthesis. 
These advancements will open new pathways for studying black hole shadows in M87 and Sgr A*, as well as accretion and jet formation in nearby AGNs. Reduced opacity at 345 GHz allows jet-launching regions closer to the black hole to be probed, offering critical insights into jet formation, collimation, acceleration, and the phenomenon of limb-brightening in inner jets, and thus deepening our understanding of AGN physics. 

\begin{acknowledgements}
The Event Horizon Telescope Collaboration thanks the following
organizations and programs: the Academia Sinica; the Academy
of Finland (projects 274477, 284495, 312496, 315721); the Agencia Nacional de Investigaci\'{o}n 
y Desarrollo (ANID), Chile via NCN$19\_058$ (TITANs), Fondecyt 1221421 and BASAL FB210003; the Alexander
von Humboldt Stiftung; an Alfred P. Sloan Research Fellowship;
Allegro, the European ALMA Regional Centre node in the Netherlands, the NL astronomy
research network NOVA and the astronomy institutes of the University of Amsterdam, Leiden University, and Radboud University;
the ALMA North America Development Fund; the Astrophysics and High Energy Physics programme by MCIN (with funding from European Union NextGenerationEU, PRTR-C17I1); the Black Hole Initiative, which is funded by grants from the John Templeton Foundation (60477, 61497, 62286) and the Gordon and Betty Moore Foundation (Grant GBMF-8273) - although the opinions expressed in this work are those of the author and do not necessarily reflect the views of these Foundations; 
the Brinson Foundation; ``la Caixa'' Foundation (ID 100010434) through fellowship codes LCF/BQ/DI22/11940027 and LCF/BQ/DI22/11940030; the Canada Research Chairs (CRC) program; Chandra DD7-18089X and TM6-17006X; the China Scholarship
Council; the China Postdoctoral Science Foundation fellowships (2020M671266, 2022M712084); Conicyt through Fondecyt Postdoctorado (project 3220195); Consejo Nacional de Humanidades, Ciencia y Tecnología (CONAHCYT, Mexico, projects U0004-246083, U0004-259839, F0003-272050, M0037-279006, F0003-281692, 104497, 275201, 263356, CBF2023-2024-1102, 257435); the Colfuturo Scholarship; 
the Consejer\'{i}a de Econom\'{i}a, Conocimiento, 
Empresas y Universidad 
of the Junta de Andaluc\'{i}a (grant P18-FR-1769), the Consejo Superior de Investigaciones 
Cient\'{i}ficas (grant 2019AEP112);
the Delaney Family via the Delaney Family John A.
Wheeler Chair at Perimeter Institute; Dirección General de Asuntos del Personal Académico-Universidad Nacional Autónoma de México (DGAPA-UNAM, projects IN112820 and IN108324); the Dutch Research Council (NWO) for the VICI award (grant 639.043.513), the grant OCENW.KLEIN.113, and the Dutch Black Hole Consortium (with project No. NWA 1292.19.202) of the research programme the National Science Agenda; the Dutch National Supercomputers, Cartesius and Snellius  (NWO grant 2021.013); 
the EACOA Fellowship awarded by the East Asia Core
Observatories Association, which consists of the Academia Sinica Institute of Astronomy and Astrophysics, the National Astronomical Observatory of Japan, Center for Astronomical Mega-Science,
Chinese Academy of Sciences, and the Korea Astronomy and Space Science Institute; 
the European Research Council (ERC) Synergy Grant ``BlackHoleCam: Imaging the Event Horizon of Black Holes'' (grant 610058) and Synergy Grant ``BlackHolistic:  Colour Movies of Black Holes:
Understanding Black Hole Astrophysics from the Event Horizon to Galactic Scales'' (grant 10107164); 
the European Union Horizon 2020
research and innovation programme under grant agreements
RadioNet (No. 730562), 
M2FINDERS (No. 101018682) and FunFiCO (No. 777740); the European Research Council for advanced grant ``JETSET: Launching, propagation and 
emission of relativistic jets from binary mergers and across mass scales'' (grant No. 884631); the European Horizon Europe staff exchange (SE) programme HORIZON-MSCA-2021-SE-01 grant NewFunFiCO (No. 10108625); the Horizon ERC Grants 2021 programme under grant agreement No. 101040021; the FAPESP (Funda\c{c}\~ao de Amparo \'a Pesquisa do Estado de S\~ao Paulo) under grant 2021/01183-8; the Fondes de Recherche Nature et Technologies (FRQNT); the Fondo CAS-ANID folio CAS220010; the Generalitat Valenciana (grants APOSTD/2018/177 and  ASFAE/2022/018) and
GenT Program (project CIDEGENT/2018/021); the Gordon and Betty Moore Foundation (GBMF-3561, GBMF-5278, GBMF-10423);   
the Institute for Advanced Study; the ICSC – Centro Nazionale di Ricerca in High Performance Computing, Big Data and Quantum Computing, funded by European Union – NextGenerationEU; the Istituto Nazionale di Fisica
Nucleare (INFN) sezione di Napoli, iniziative specifiche
TEONGRAV; 
the International Max Planck Research
School for Astronomy and Astrophysics at the
Universities of Bonn and Cologne; 
the European Union NextGenerationEU" RRF M4C2 1.1 project n. 2022YAPMJH;
DFG research grant ``Jet physics on horizon scales and beyond'' (grant No. 443220636);
Joint Columbia/Flatiron Postdoctoral Fellowship (research at the Flatiron Institute is supported by the Simons Foundation); 
the Japan Ministry of Education, Culture, Sports, Science and Technology (MEXT; grant JPMXP1020200109);  
the Japan Society for the Promotion of Science (JSPS) Grant-in-Aid for JSPS
Research Fellowship (JP17J08829); the Joint Institute for Computational Fundamental Science, Japan; the Key Research
Program of Frontier Sciences, Chinese Academy of
Sciences (CAS, grants QYZDJ-SSW-SLH057, QYZDJSSW-SYS008, ZDBS-LY-SLH011); 
the Leverhulme Trust Early Career Research
Fellowship; the Max-Planck-Gesellschaft (MPG);
the Max Planck Partner Group of the MPG and the
CAS; the MEXT/JSPS KAKENHI (grants 18KK0090, JP21H01137,
JP18H03721, JP18K13594, 18K03709, JP19K14761, 18H01245, 25120007, 19H01943, 21H01137, 21H04488, 22H00157, 23K03453); the MICINN Research Projects PID2019-108995GB-C22, PID2022-140888NB-C22; the MIT International Science
and Technology Initiatives (MISTI) Funds; 
the Ministry of Science and Technology (MOST) of Taiwan (103-2119-M-001-010-MY2, 105-2112-M-001-025-MY3, 105-2119-M-001-042, 106-2112-M-001-011, 106-2119-M-001-013, 106-2119-M-001-027, 106-2923-M-001-005, 107-2119-M-001-017, 107-2119-M-001-020, 107-2119-M-001-041, 107-2119-M-110-005, 107-2923-M-001-009, 108-2112-M-001-048, 108-2112-M-001-051, 108-2923-M-001-002, 109-2112-M-001-025, 109-2124-M-001-005, 109-2923-M-001-001,  
110-2112-M-001-033, 110-2124-M-001-007 and 110-2923-M-001-001); the National Science and Technology Council (NSTC) of Taiwan
(111-2124-M-001-005, 112-2124-M-001-014 and  112-2112-M-003-010-MY3);
the Ministry of Education (MoE) of Taiwan Yushan Young Scholar Program;
the Physics Division, National Center for Theoretical Sciences of Taiwan;
the National Aeronautics and
Space Administration (NASA, Fermi Guest Investigator
grant  
80NSSC23K1508, NASA Astrophysics Theory Program grant 80NSSC20K0527, NASA NuSTAR award 
80NSSC20K0645); NASA Hubble Fellowship Program Einstein Fellowship;
NASA Hubble Fellowship 
grants HST-HF2-51431.001-A, HST-HF2-51482.001-A, HST-HF2-51539.001-A, HST-HF2-51552.001A awarded 
by the Space Telescope Science Institute, which is operated by the Association of Universities for 
Research in Astronomy, Inc., for NASA, under contract NAS5-26555; 
the National Institute of Natural Sciences (NINS) of Japan; the National
Key Research and Development Program of China
(grant 2016YFA0400704, 2017YFA0402703, 2016YFA0400702); the National Science and Technology Council (NSTC, grants NSTC 111-2112-M-001 -041, NSTC 111-2124-M-001-005, NSTC 112-2124-M-001-014); the US National
Science Foundation (NSF, grants AST-0096454,
AST-0352953, AST-0521233, AST-0705062, AST-0905844, AST-0922984, AST-1126433, OIA-1126433, AST-1140030,
DGE-1144085, AST-1207704, AST-1207730, AST-1207752, MRI-1228509, OPP-1248097, AST-1310896, AST-1440254, 
AST-1555365, AST-1614868, AST-1615796, AST-1715061, AST-1716327,  AST-1726637, %AST-1716536, 
OISE-1743747, AST-1743747, AST-1816420, AST-1935980, AST-1952099, AST-2034306,  AST-2205908, AST-2307887); 
NSF Astronomy and Astrophysics Postdoctoral Fellowship (AST-1903847); 
the Natural Science Foundation of China (grants 11650110427, 10625314, 11721303, 11725312, 11873028, 11933007, 11991052, 11991053, 12192220, 12192223, 12273022, 12325302, 12303021); 
the Natural Sciences and Engineering Research Council of
Canada (NSERC); %the National Youth Thousand Talents Program of China; 
the National Research Foundation of Korea (the Global PhD Fellowship Grant: grants NRF-2015H1A2A1033752; the Korea Research Fellowship Program: NRF-2015H1D3A1066561; Brain Pool Program: RS-2024-00407499;  Basic Research Support Grant 2019R1F1A1059721, 2021R1A6A3A01086420, 2022R1C1C1005255, 2022R1F1A1075115); 
Netherlands Research School for Astronomy (NOVA) Virtual Institute of Accretion (VIA) postdoctoral fellowships; NOIRLab, which is managed by the Association of Universities for Research in Astronomy (AURA) under a cooperative agreement with the National Science Foundation; 
Onsala Space Observatory (OSO) national infrastructure, for the provisioning
of its facilities/observational support (OSO receives funding through the Swedish Research Council under grant 2017-00648);  the Perimeter Institute for Theoretical Physics (research at Perimeter Institute is supported by the Government of Canada through the Department of Innovation, Science and Economic Development and by the Province of Ontario through the Ministry of Research, Innovation and Science); the Portuguese Foundation for Science and Technology (FCT) grants (Individual CEEC program - 5th edition, \url{https://doi.org/10.54499/UIDB/04106/2020}, \url{https://doi.org/10.54499/UIDP/04106/2020}, PTDC/FIS-AST/3041/2020, CERN/FIS-PAR/0024/2021, 2022.04560.PTDC); the Princeton Gravity Initiative; the Spanish Ministerio de Ciencia e Innovaci\'{o}n (grants PGC2018-098915-B-C21, AYA2016-80889-P,
PID2019-108995GB-C21, PID2020-117404GB-C21, RYC2023-042988-I); 
the University of Pretoria for financial aid in the provision of the new 
Cluster Server nodes and SuperMicro (USA) for a SEEDING GRANT approved toward these 
nodes in 2020; the Shanghai Municipality orientation program of basic research for international scientists (grant no. 22JC1410600); 
the Shanghai Pilot Program for Basic Research, Chinese Academy of Science, 
Shanghai Branch (JCYJ-SHFY-2021-013); the Simons Foundation (grant 00001470);
the State Agency for Research of the Spanish MCIU through
the ``Center of Excellence Severo Ochoa'' award for
the Instituto de Astrof\'{i}sica de Andaluc\'{i}a (SEV-2017-
0709); the Spanish Ministry for Science and Innovation grant CEX2021-001131-S funded by MCIN/AEI/10.13039/501100011033; the Spinoza Prize SPI 78-409; the South African Research Chairs Initiative, through the 
South African Radio Astronomy Observatory (SARAO, grant ID 77948),  which is a facility of the National 
Research Foundation (NRF), an agency of the Department of Science and Innovation (DSI) of South Africa; the Swedish Research Council (VR); the Taplin Fellowship; the Toray Science Foundation; the UK Science and Technology Facilities Council (grant no. ST/X508329/1); the US Department of Energy (USDOE) through the Los Alamos National
Laboratory (operated by Triad National Security,
LLC, for the National Nuclear Security Administration
of the USDOE, contract 89233218CNA000001); and the YCAA Prize Postdoctoral Fellowship. This work was also supported by the National Research Foundation of Korea (NRF) grant funded by the Korea government(MSIT) (RS-2024-00449206). We acknowledge support from the Coordenação de Aperfeiçoamento de Pessoal de Nível Superior (CAPES) of Brazil through PROEX grant number 88887.845378/2023-00. We acknowledge financial support from Millenium Nucleus NCN23\_002 (TITANs) and Comité Mixto ESO-Chile.

We thank
the staff at the participating observatories, correlation
centers, and institutions for their enthusiastic support.
This paper makes use of the following ALMA data: ADS/JAO.ALMA\#2011.0.00013.E.
ALMA is a partnership of ESO (representing its member states), NSF (USA) and NINS (Japan), together with NRC (Canada), NSTC and ASIAA (Taiwan), and KASI (Republic of Korea), in cooperation with the Republic of Chile. The Joint ALMA Observatory is operated by ESO, AUI/NRAO and NAOJ.
The NRAO
is a facility of the NSF operated under cooperative agreement
by AUI.
This research used resources of the Oak Ridge Leadership Computing Facility at the Oak Ridge National
Laboratory, which is supported by the Office of Science of the U.S. Department of Energy under contract
No. DE-AC05-00OR22725; the ASTROVIVES FEDER infrastructure, with project code IDIFEDER-2021-086; the computing cluster of Shanghai VLBI correlator supported by the Special Fund 
for Astronomy from the Ministry of Finance in China;  
We also thank the Center for Computational Astrophysics, National Astronomical Observatory of Japan. This work was supported by FAPESP (Fundacao de Amparo a Pesquisa do Estado de Sao Paulo) under grant 2021/01183-8.

APEX is a collaboration between the
Max-Planck-Institut f{\"u}r Radioastronomie (Germany),
ESO, and the Onsala Space Observatory (Sweden). The
SMA is a joint project between the SAO and ASIAA
and is funded by the Smithsonian Institution and the
Academia Sinica. The JCMT is operated by the East
Asian Observatory on behalf of the NAOJ, ASIAA, and
KASI, as well as the Ministry of Finance of China, Chinese
Academy of Sciences, and the National Key Research and Development
Program (No. 2017YFA0402700) of China
and Natural Science Foundation of China grant 11873028.
Additional funding support for the JCMT is provided by the Science
and Technologies Facility Council (UK) and participating
universities in the UK and Canada. 
The LMT is a project operated by the Instituto Nacional
de Astr\'{o}fisica, \'{O}ptica, y Electr\'{o}nica (Mexico) and the
University of Massachusetts at Amherst (USA). The
IRAM 30-m telescope on Pico Veleta, Spain is operated
by IRAM and supported by CNRS (Centre National de
la Recherche Scientifique, France), MPG (Max-Planck-Gesellschaft, Germany), 
and IGN (Instituto Geogr\'{a}fico
Nacional, Spain). The SMT is operated by the Arizona
Radio Observatory, a part of the Steward Observatory
of the University of Arizona, with financial support of
operations from the State of Arizona and financial support
for instrumentation development from the NSF.
Support for SPT participation in the EHT is provided by the National Science Foundation through award OPP-1852617 
to the University of Chicago. Partial support is also 
provided by the Kavli Institute of Cosmological Physics at the University of Chicago. The SPT hydrogen maser was 
provided on loan from the GLT, courtesy of ASIAA.

This work used the
Extreme Science and Engineering Discovery Environment
(XSEDE), supported by NSF grant ACI-1548562,
and CyVerse, supported by NSF grants DBI-0735191,
DBI-1265383, and DBI-1743442. XSEDE Stampede2 resource
at TACC was allocated through TG-AST170024
and TG-AST080026N. XSEDE JetStream resource at
PTI and TACC was allocated through AST170028.
This research is part of the Frontera computing project at the Texas Advanced 
Computing Center through the Frontera Large-Scale Community Partnerships allocation
AST20023. Frontera is made possible by National Science Foundation award OAC-1818253.
This research was done using services provided by the OSG Consortium~\citep{osg07,osg09}, which is supported by the National Science Foundation award Nos. 2030508 and 1836650.
Additional work used ABACUS2.0, which is part of the eScience center at Southern Denmark University, and the Kultrun Astronomy Hybrid Cluster (projects Conicyt Programa de Astronomia Fondo Quimal QUIMAL170001, Conicyt PIA ACT172033, Fondecyt Iniciacion 11170268, Quimal 220002). 
Simulations were also performed on the SuperMUC cluster at the LRZ in Garching, 
on the LOEWE cluster in CSC in Frankfurt, on the HazelHen cluster at the HLRS in Stuttgart, 
and on the Pi2.0 and Siyuan Mark-I at Shanghai Jiao Tong University.
The computer resources of the Finnish IT Center for Science (CSC) and the Finnish Computing 
Competence Infrastructure (FCCI) project are acknowledged. This
research was enabled in part by support provided
by Compute Ontario (http://computeontario.ca), Calcul
Quebec (http://www.calculquebec.ca), and the Digital Research Alliance of Canada (https://alliancecan.ca/en).

The EHTC has
received generous donations of FPGA chips from Xilinx
Inc., under the Xilinx University Program. The EHTC
has benefited from technology shared under open-source
license by the Collaboration for Astronomy Signal Processing
and Electronics Research (CASPER). The EHT
project is grateful to T4Science and Microsemi for their
assistance with hydrogen masers. This research has
made use of NASA's Astrophysics Data System. We
gratefully acknowledge the support provided by the extended
staff of the ALMA, from the inception of
the ALMA Phasing Project through the observational
campaigns of 2017 and 2018. We would like to thank
A. Deller and W. Brisken for EHT-specific support with
the use of DiFX. We thank Martin Shepherd for the addition of extra features in the Difmap software 
that were used for the CLEAN imaging results presented in this paper.
We acknowledge the significance that
Maunakea, where the SMA and JCMT EHT stations
are located, has for the indigenous Hawaiian people.

\end{acknowledgements}

 \bibliographystyle{aa} % style aa.bst
\bibliography{2021_band7_m87}{}

\begin{thebibliography}{51}
\expandafter\ifx\csname natexlab\endcsname\relax\def\natexlab#1{#1}\fi

\bibitem[{{Agudo} {et~al.}(2018){Agudo}, {Thum}, {Ramakrishnan}, {Molina},
  {Casadio}, \& {G{\'o}mez}}]{Agudo2018}
{Agudo}, I., {Thum}, C., {Ramakrishnan}, V., {et~al.} 2018, \mnras, 473, 1850

\bibitem[{{Algaba} {et~al.}(2016){Algaba}, {Asada}, \& {Nakamura}}]{Algaba2016}
{Algaba}, J.~C., {Asada}, K., \& {Nakamura}, M. 2016, \apj, 823, 86

\bibitem[{{Biretta} {et~al.}(1999){Biretta}, {Sparks}, \&
  {Macchetto}}]{Biretta1999}
{Biretta}, J.~A., {Sparks}, W.~B., \& {Macchetto}, F. 1999, \apj, 520, 621

\bibitem[{{Blandford} \& {K{\"o}nigl}(1979)}]{BlandfordKonigl1979}
{Blandford}, R.~D. \& {K{\"o}nigl}, A. 1979, \apj, 232, 34

\bibitem[{{Bower} {et~al.}(2018){Bower}, {Broderick}, {Dexter}, {Doeleman},
  {Falcke}, {Fish}, {Johnson}, {Marrone}, {Moran}, {Moscibrodzka}, {Peck},
  {Plambeck}, \& {Rao}}]{Bower2018}
{Bower}, G.~C., {Broderick}, A., {Dexter}, J., {et~al.} 2018, \apj, 868, 101

\bibitem[{{Briggs}(1995)}]{Briggs1995}
{Briggs}, D.~S. 1995, in American Astronomical Society Meeting Abstracts, Vol.
  187, American Astronomical Society Meeting Abstracts, 112.02

\bibitem[{{Cheung} {et~al.}(2007){Cheung}, {Harris}, \& {Stawarz}}]{Cheung2007}
{Cheung}, C.~C., {Harris}, D.~E., \& {Stawarz}, {\L}. 2007, \apjl, 663, L65

\bibitem[{{Contopoulos} {et~al.}(2022){Contopoulos}, {Myserlis}, {Kazanas}, \&
  {Nathanail}}]{Contopoulos2022}
{Contopoulos}, I., {Myserlis}, I., {Kazanas}, D., \& {Nathanail}, A. 2022,
  Galaxies, 10, 80

\bibitem[{{Crew} {et~al.}(2023){Crew}, {Goddi}, {Matthews}, {Rottmann}, {Saez},
  \& {Mart{\'\i}-Vidal}}]{APP2B7}
{Crew}, G.~B., {Goddi}, C., {Matthews}, L.~D., {et~al.} 2023, \pasp, 135,
  025002

\bibitem[{{Doi} {et~al.}(2013){Doi}, {Hada}, {Nagai}, {Kino}, {Honma},
  {Akiyama}, {Oyama}, \& {Kono}}]{Doi2013}
{Doi}, A., {Hada}, K., {Nagai}, H., {et~al.} 2013, in European Physical Journal
  Web of Conferences, Vol.~61, European Physical Journal Web of Conferences,
  08008

\bibitem[{{Event Horizon Telescope Collaboration}
  {et~al.}(2024{\natexlab{a}}){Event Horizon Telescope Collaboration},
  {Akiyama}, {Alberdi}, {Alef}, {Algaba}, {Anantua}, {Asada}, {Azulay}, {Bach},
  {Baczko}, {Ball}, {Balokovic}, {Bandyopadhyay}, {Barrett}, {Baub{\"o}ck},
  {Benson}, {Bintley}, {Blackburn}, {Blundell}, {Bouman}, {Bower}, {Boyce},
  {Bremer}, {Brinkerink}, {Brissenden}, {Britzen}, {Broderick}, {Broguiere},
  {Bronzwaer}, {Bustamante}, {Byun}, {Carlstrom}, {Ceccobello}, {Chael},
  {Chan}, {Chang}, {Chatterjee}, {Chatterjee}, {Chen}, {Chen}, {Cheng}, {Cho},
  {Christian}, {Conroy}, {Conway}, {Cordes}, {Crawford}, {Crew}, {Cruz-Osorio},
  {Cui}, {Dahale}, {Davelaar}, {De Laurentis}, {Deane}, {Dempsey}, {Desvignes},
  {Dexter}, {Dhruv}, {Dihingia}, {Doeleman}, {Dougal}, {Dzib}, {Eatough},
  {Emami}, {Falcke}, {Farah}, {Fish}, {Fomalont}, {Ford}, {Foschi},
  {Fraga-Encinas}, {Freeman}, {Friberg}, {Fromm}, {Fuentes}, {Galison},
  {Gammie}, {Garc{\'\i}a}, {Gentaz}, {Georgiev}, {Goddi}, {Gold},
  {G{\'o}mez-Ruiz}, {G{\'o}mez}, {Gu}, {Gurwell}, {Hada}, {Haggard}, {Haworth},
  {Hecht}, {Hesper}, {Heumann}, {Ho}, {Ho}, {Honma}, {Huang}, {Huang},
  {Hughes}, {Ikeda}, {Impellizzeri}, {Inoue}, {Issaoun}, {James}, {Jannuzi},
  {Janssen}, {Jeter}, {Jiang}, {Jim{\'e}nez-Rosales}, {Johnson}, {Jorstad},
  {Joshi}, {Jung}, {Karami}, {Karuppusamy}, {Kawashima}, {Keating}, {Kettenis},
  {Kim}, {Kim}, {Kim}, {Kim}, {Kino}, {Koay}, {Kocherlakota}, {Kofuji}, {Koch},
  {Koyama}, {Kramer}, {Kramer}, {Kramer}, {Krichbaum}, {Kuo}, {La Bella},
  {Lauer}, {Lee}, {Lee}, {Leung}, {Levis}, {Li}, {Lico}, {Lindahl},
  {Lindqvist}, {Lisakov}, {Liu}, {Liu}, {Liuzzo}, {Lo}, {Lobanov}, {Loinard},
  {Lonsdale}, {Lowitz}, {Lu}, {MacDonald}, {Mao}, {Marchili}, {Markoff},
  {Marrone}, {Marscher}, {Mart{\'\i}-Vidal}, {Matsushita}, {Matthews},
  {Medeiros}, {Menten}, {Michalik}, {Mizuno}, {Mizuno}, {Moran}, {Moriyama},
  {Moscibrodzka}, {Mulaudzi}, {M{\"u}ller}, {M{\"u}ller}, {Mus}, {Musoke},
  {Myserlis}, {Nadolski}, {Nagai}, {Nagar}, {Nakamura}, {Narayanan},
  {Natarajan}, {Nathanail}, {Fuentes}, {Neilsen}, {Neri}, {Ni}, {Noutsos},
  {Nowak}, {Oh}, {Okino}, {Olivares}, {Ortiz-Le{\'o}n}, {Oyama}, {{\"O}zel},
  {Palumbo}, {Paraschos}, {Park}, {Parsons}, {Patel}, {Pen}, {Pesce},
  {Pi{\'e}tu}, {Plambeck}, {PopStefanija}, {Porth}, {P{\"o}tzl}, {Prather},
  {Preciado-L{\'o}pez}, {Psaltis}, {Pu}, {Ramakrishnan}, {Rao}, {Rawlings},
  {Raymond}, {Rezzolla}, {Ricarte}, {Ripperda}, {Roelofs}, {Rogers},
  {Romero-Ca{\~n}izales}, {Ros}, {Roshanineshat}, {Rottmann}, {Roy}, {Ruiz},
  {Ruszczyk}, {Rygl}, {S{\'a}nchez}, {S{\'a}nchez-Arg{\"u}elles},
  {S{\'a}nchez-Portal}, {Sasada}, {Satapathy}, {Savolainen}, {Schloerb},
  {Schonfeld}, {Schuster}, {Shao}, {Shen}, {Small}, {Sohn}, {SooHoo},
  {Sosapanta Salas}, {Souccar}, {Stanway}, {Sun}, {Tazaki}, {Tetarenko},
  {Tiede}, {Tilanus}, {Titus}, {Torne}, {Toscano}, {Traianou}, {Trent},
  {Trippe}, {Turk}, {van Bemmel}, {van Langevelde}, {van Rossum}, {Vos},
  {Wagner}, {Ward-Thompson}, {Wardle}, {Washington}, {Weintroub}, {Wharton},
  {Wielgus}, {Wiik}, {Witzel}, {Wondrak}, {Wong}, {Wu}, {Yadlapalli},
  {Yamaguchi}, {Yfantis}, {Yoon}, {Young}, {Young}, {Younsi}, {Yu}, {Yuan},
  {Yuan}, {Zensus}, {Zhang}, {Zhao}, \& {Zhao}}]{EHT2024_1}
{Event Horizon Telescope Collaboration}, {Akiyama}, K., {Alberdi}, A., {et~al.}
  2024{\natexlab{a}}, \apjl, 964, L25

\bibitem[{{Event Horizon Telescope Collaboration}
  {et~al.}(2024{\natexlab{b}}){Event Horizon Telescope Collaboration},
  {Akiyama}, {Alberdi}, {Alef}, {Algaba}, {Anantua}, {Asada}, {Azulay}, {Bach},
  {Baczko}, {Ball}, {Balokovi{\'c}}, {Bandyopadhyay}, {Barrett}, {Baub{\"o}ck},
  {Benson}, {Bintley}, {Blackburn}, {Blundell}, {Bouman}, {Bower}, {Boyce},
  {Bremer}, {Brissenden}, {Britzen}, {Broderick}, {Broguiere}, {Bronzwaer},
  {Bustamante}, {Carlstrom}, {Chael}, {Chan}, {Chang}, {Chatterjee},
  {Chatterjee}, {Chen}, {Chen}, {Cheng}, {Cho}, {Christian}, {Conroy},
  {Conway}, {Crawford}, {Crew}, {Cruz-Osorio}, {Cui}, {Dahale}, {Davelaar}, {De
  Laurentis}, {Deane}, {Dempsey}, {Desvignes}, {Dexter}, {Dhruv}, {Dihingia},
  {Doeleman}, {Dzib}, {Eatough}, {Emami}, {Falcke}, {Farah}, {Fish},
  {Fomalont}, {Ford}, {Foschi}, {Fraga-Encinas}, {Freeman}, {Friberg}, {Fromm},
  {Fuentes}, {Galison}, {Gammie}, {Garc{\'\i}a}, {Gentaz}, {Georgiev}, {Goddi},
  {Gold}, {G{\'o}mez-Ruiz}, {G{\'o}mez}, {Gu}, {Gurwell}, {Hada}, {Haggard},
  {Hesper}, {Heumann}, {Ho}, {Ho}, {Honma}, {Huang}, {Huang}, {Hughes},
  {Ikeda}, {Violette Impellizzeri}, {Inoue}, {Issaoun}, {James}, {Jannuzi},
  {Janssen}, {Jeter}, {Jiang}, {Jim{\'e}nez-Rosales}, {Johnson}, {Jorstad},
  {Jones}, {Joshi}, {Jung}, {Karuppusamy}, {Kawashima}, {Keating}, {Kettenis},
  {Kim}, {Kim}, {Kim}, {Kim}, {Kino}, {Koay}, {Kocherlakota}, {Kofuji}, {Koch},
  {Koyama}, {Kramer}, {Kramer}, {Kramer}, {Krichbaum}, {Kuo}, {La Bella},
  {Lee}, {Levis}, {Li}, {Lico}, {Lindahl}, {Lindqvist}, {Lisakov}, {Liu},
  {Liu}, {Liuzzo}, {Lo}, {Lobanov}, {Loinard}, {Lonsdale}, {Lowitz}, {Lu},
  {MacDonald}, {Mao}, {Marchili}, {Markoff}, {Marrone}, {Marscher},
  {Mart{\'\i}-Vidal}, {Matsushita}, {Matthews}, {Medeiros}, {Menten}, {Mizuno},
  {Mizuno}, {Montgomery}, {Moran}, {Moriyama}, {Moscibrodzka}, {Mulaudzi},
  {M{\"u}ller}, {M{\"u}ller}, {Mus}, {Musoke}, {Myserlis}, {Nagai}, {Nagar},
  {Nakamura}, {Narayanan}, {Natarajan}, {Nathanail}, {Fuentes}, {Neilsen},
  {Ni}, {Nowak}, {Oh}, {Okino}, {Olivares}, {Oyama}, {{\"O}zel}, {Palumbo},
  {Paraschos}, {Park}, {Parsons}, {Patel}, {Pen}, {Pesce}, {Pi{\'e}tu},
  {PopStefanija}, {Porth}, {Prather}, {Psaltis}, {Pu}, {Ramakrishnan}, {Rao},
  {Rawlings}, {Raymond}, {Rezzolla}, {Ricarte}, {Ripperda}, {Roelofs},
  {Romero-Ca{\~n}izales}, {Ros}, {Roshanineshat}, {Rottmann}, {Roy}, {Ruiz},
  {Ruszczyk}, {Rygl}, {S{\'a}nchez}, {S{\'a}nchez-Arg{\"u}elles},
  {S{\'a}nchez-Portal}, {Sasada}, {Satapathy}, {Savolainen}, {Schloerb},
  {Schonfeld}, {Schuster}, {Shao}, {Shen}, {Small}, {Sohn}, {SooHoo}, {Salas},
  {Souccar}, {Stanway}, {Sun}, {Tazaki}, {Tetarenko}, {Tiede}, {Tilanus},
  {Titus}, {Toma}, {Torne}, {Toscano}, {Traianou}, {Trent}, {Trippe}, {Turk},
  {van Bemmel}, {van Langevelde}, {van Rossum}, {Vos}, {Wagner},
  {Ward-Thompson}, {Wardle}, {Washington}, {Weintroub}, {Wharton}, {Wielgus},
  {Wiik}, {Witzel}, {Wondrak}, {Wong}, {Wu}, {Yadlapalli}, {Yamaguchi},
  {Yfantis}, {Yoon}, {Young}, {Younsi}, {Yu}, {Yuan}, {Yuan}, {Anton Zensus},
  {Zhang}, {Zhao}, {Zhao}, {Allardi}, {Chang}, {Chang}, {Chang}, {Chen},
  {Chilson}, {Faber}, {Gale}, {Han}, {Han}, {Hasegawa},
  {Hern{\'a}ndez-Rebollar}, {Huang}, {Jiang}, {Jinchi}, {Kimura}, {Kubo}, {Li},
  {Lin}, {Liu}, {Liu}, {Lu}, {Martin-Cocher}, {Meyer-Zhao}, {Monta{\~n}a},
  {Moraghan}, {Moreno-Nolasco}, {Nishioka}, {Norton}, {Nystrom}, {Ogawa},
  {Oshiro}, {Pradel}, {Principe}, {Raffin}, {Rodr{\'\i}guez-Montoya}, {Shaw},
  {Snow}, {Sridharan}, {Srinivasan}, {Wei}, \& {Yu}}]{EHT2024}
{Event Horizon Telescope Collaboration}, {Akiyama}, K., {Alberdi}, A., {et~al.}
  2024{\natexlab{b}}, \aap, 681, A79

\bibitem[{{Event Horizon Telescope Collaboration} {et~al.}(2023){Event Horizon
  Telescope Collaboration}, {Akiyama}, {Alberdi}, {Alef}, {Algaba}, {Anantua},
  {Asada}, {Azulay}, {Bach}, {Baczko}, {Ball}, {Balokovi{\'c}}, {Barrett},
  {Baub{\"o}ck}, {Benson}, {Bintley}, {Blackburn}, {Blundell}, {Bouman},
  {Bower}, {Boyce}, {Bremer}, {Brinkerink}, {Brissenden}, {Britzen},
  {Broderick}, {Broguiere}, {Bronzwaer}, {Bustamante}, {Byun}, {Carlstrom},
  {Ceccobello}, {Chael}, {Chan}, {Chang}, {Chatterjee}, {Chatterjee}, {Chen},
  {Chen}, {Cheng}, {Cho}, {Christian}, {Conroy}, {Conway}, {Cordes},
  {Crawford}, {Crew}, {Cruz-Osorio}, {Cui}, {Dahale}, {Davelaar}, {De
  Laurentis}, {Deane}, {Dempsey}, {Desvignes}, {Dexter}, {Dhruv}, {Doeleman},
  {Dougal}, {Dzib}, {Eatough}, {Emami}, {Falcke}, {Farah}, {Fish}, {Fomalont},
  {Ford}, {Foschi}, {Fraga-Encinas}, {Freeman}, {Friberg}, {Fromm}, {Fuentes},
  {Galison}, {Gammie}, {Garc{\'\i}a}, {Gentaz}, {Georgiev}, {Goddi}, {Gold},
  {G{\'o}mez-Ruiz}, {G{\'o}mez}, {Gu}, {Gurwell}, {Hada}, {Haggard}, {Haworth},
  {Hecht}, {Hesper}, {Heumann}, {Ho}, {Ho}, {Honma}, {Huang}, {Huang},
  {Hughes}, {Ikeda}, {Impellizzeri}, {Inoue}, {Issaoun}, {James}, {Jannuzi},
  {Janssen}, {Jeter}, {Jiang}, {Jim{\'e}nez-Rosales}, {Johnson}, {Jorstad},
  {Joshi}, {Jung}, {Karami}, {Karuppusamy}, {Kawashima}, {Keating}, {Kettenis},
  {Kim}, {Kim}, {Kim}, {Kim}, {Kino}, {Koay}, {Kocherlakota}, {Kofuji}, {Koch},
  {Koyama}, {Kramer}, {Kramer}, {Kramer}, {Krichbaum}, {Kuo}, {La Bella},
  {Lauer}, {Lee}, {Lee}, {Leung}, {Levis}, {Li}, {Lico}, {Lindahl},
  {Lindqvist}, {Lisakov}, {Liu}, {Liu}, {Liuzzo}, {Lo}, {Lobanov}, {Loinard},
  {Lonsdale}, {Lowitz}, {Lu}, {MacDonald}, {Mao}, {Marchili}, {Markoff},
  {Marrone}, {Marscher}, {Mart{\'\i}-Vidal}, {Matsushita}, {Matthews},
  {Medeiros}, {Menten}, {Michalik}, {Mizuno}, {Mizuno}, {Moran}, {Moriyama},
  {Moscibrodzka}, {Mulaudzi}, {M{\"u}ller}, {M{\"u}ller}, {Mus}, {Musoke},
  {Myserlis}, {Nadolski}, {Nagai}, {Nagar}, {Nakamura}, {Narayan}, {Narayanan},
  {Natarajan}, {Nathanail}, {Fuentes}, {Neilsen}, {Neri}, {Ni}, {Noutsos},
  {Nowak}, {Oh}, {Okino}, {Olivares}, {Ortiz-Le{\'o}n}, {Oyama}, {{\"O}zel},
  {Palumbo}, {Paraschos}, {Park}, {Parsons}, {Patel}, {Pen}, {Pesce},
  {Pi{\'e}tu}, {Plambeck}, {PopStefanija}, {Porth}, {P{\"o}tzl}, {Prather},
  {Preciado-L{\'o}pez}, {Psaltis}, {Pu}, {Ramakrishnan}, {Rao}, {Rawlings},
  {Raymond}, {Rezzolla}, {Ricarte}, {Ripperda}, {Roelofs}, {Rogers},
  {Romero-Ca{\~n}izales}, {Ros}, {Roshanineshat}, {Rottmann}, {Roy}, {Ruiz},
  {Ruszczyk}, {Rygl}, {S{\'a}nchez}, {S{\'a}nchez-Arg{\"u}elles},
  {S{\'a}nchez-Portal}, {Sasada}, {Satapathy}, {Savolainen}, {Schloerb},
  {Schonfeld}, {Schuster}, {Shao}, {Shen}, {Small}, {Sohn}, {SooHoo},
  {Sosapanta Salas}, {Souccar}, {Sun}, {Tazaki}, {Tetarenko}, {Tiede},
  {Tilanus}, {Titus}, {Torne}, {Toscano}, {Traianou}, {Trent}, {Trippe},
  {Turk}, {van Bemmel}, {van Langevelde}, {van Rossum}, {Vos}, {Wagner},
  {Ward-Thompson}, {Wardle}, {Washington}, {Weintroub}, {Wharton}, {Wielgus},
  {Wiik}, {Witzel}, {Wondrak}, {Wong}, {Wu}, {Yadlapalli}, {Yamaguchi},
  {Yfantis}, {Yoon}, {Young}, {Young}, {Younsi}, {Yu}, {Yuan}, {Yuan},
  {Zensus}, {Zhang}, {Zhao}, \& {Zhao}}]{EHT_2023}
{Event Horizon Telescope Collaboration}, {Akiyama}, K., {Alberdi}, A., {et~al.}
  2023, \apjl, 957, L20

\bibitem[{{Event Horizon Telescope Collaboration}
  {et~al.}(2022{\natexlab{a}}){Event Horizon Telescope Collaboration},
  {Akiyama}, {Alberdi}, {Alef}, {Algaba}, {Anantua}, {Asada}, {Azulay}, {Bach},
  {Baczko}, {Ball}, {Balokovi{\'c}}, {Barrett}, {Baub{\"o}ck}, {Benson},
  {Bintley}, {Blackburn}, {Blundell}, {Bouman}, {Bower}, {Boyce}, {Bremer},
  {Brinkerink}, {Brissenden}, {Britzen}, {Broderick}, {Broguiere}, {Bronzwaer},
  {Bustamante}, {Byun}, {Carlstrom}, {Ceccobello}, {Chael}, {Chan},
  {Chatterjee}, {Chatterjee}, {Chen}, {Chen}, {Cheng}, {Cho}, {Christian},
  {Conroy}, {Conway}, {Cordes}, {Crawford}, {Crew}, {Cruz-Osorio}, {Cui},
  {Davelaar}, {Laurentis}, {Deane}, {Dempsey}, {Desvignes}, {Dexter}, {Dhruv},
  {Doeleman}, {Dougal}, {Dzib}, {Eatough}, {Emami}, {Falcke}, {Farah}, {Fish},
  {Fomalont}, {Ford}, {Fraga-Encinas}, {Freeman}, {Friberg}, {Fromm},
  {Fuentes}, {Galison}, {Gammie}, {Garc{\'\i}a}, {Gentaz}, {Georgiev}, {Goddi},
  {Gold}, {G{\'o}mez-Ruiz}, {G{\'o}mez}, {Gu}, {Gurwell}, {Hada}, {Haggard},
  {Haworth}, {Hecht}, {Hesper}, {Heumann}, {Ho}, {Ho}, {Honma}, {Huang},
  {Huang}, {Hughes}, {Ikeda}, {Impellizzeri}, {Inoue}, {Issaoun}, {James},
  {Jannuzi}, {Janssen}, {Jeter}, {Jiang}, {Jim{\'e}nez-Rosales}, {Johnson},
  {Jorstad}, {Joshi}, {Jung}, {Karami}, {Karuppusamy}, {Kawashima}, {Keating},
  {Kettenis}, {Kim}, {Kim}, {Kim}, {Kim}, {Kino}, {Koay}, {Kocherlakota},
  {Kofuji}, {Koch}, {Koyama}, {Kramer}, {Kramer}, {Krichbaum}, {Kuo}, {Bella},
  {Lauer}, {Lee}, {Lee}, {Leung}, {Levis}, {Li}, {Lico}, {Lindahl},
  {Lindqvist}, {Lisakov}, {Liu}, {Liu}, {Liuzzo}, {Lo}, {Lobanov}, {Loinard},
  {Lonsdale}, {Lu}, {Mao}, {Marchili}, {Markoff}, {Marrone}, {Marscher},
  {Mart{\'\i}-Vidal}, {Matsushita}, {Matthews}, {Medeiros}, {Menten},
  {Michalik}, {Mizuno}, {Mizuno}, {Moran}, {Moriyama}, {Moscibrodzka},
  {M{\"u}ller}, {Mus}, {Musoke}, {Myserlis}, {Nadolski}, {Nagai}, {Nagar},
  {Nakamura}, {Narayan}, {Narayanan}, {Natarajan}, {Nathanail}, {Fuentes},
  {Neilsen}, {Neri}, {Ni}, {Noutsos}, {Nowak}, {Oh}, {Okino}, {Olivares},
  {Ortiz-Le{\'o}n}, {Oyama}, {{\"O}zel}, {Palumbo}, {Paraschos}, {Park},
  {Parsons}, {Patel}, {Pen}, {Pesce}, {Pi{\'e}tu}, {Plambeck}, {PopStefanija},
  {Porth}, {P{\"o}tzl}, {Prather}, {Preciado-L{\'o}pez}, {Psaltis}, {Pu},
  {Ramakrishnan}, {Rao}, {Rawlings}, {Raymond}, {Rezzolla}, {Ricarte},
  {Ripperda}, {Roelofs}, {Rogers}, {Ros}, {Romero-Ca{\~n}izales},
  {Roshanineshat}, {Rottmann}, {Roy}, {Ruiz}, {Ruszczyk}, {Rygl},
  {S{\'a}nchez}, {S{\'a}nchez-Arg{\"u}elles}, {S{\'a}nchez-Portal}, {Sasada},
  {Satapathy}, {Savolainen}, {Schloerb}, {Schonfeld}, {Schuster}, {Shao},
  {Shen}, {Small}, {Sohn}, {SooHoo}, {Souccar}, {Sun}, {Tazaki}, {Tetarenko},
  {Tiede}, {Tilanus}, {Titus}, {Torne}, {Traianou}, {Trent}, {Trippe}, {Turk},
  {van Bemmel}, {van Langevelde}, {van Rossum}, {Vos}, {Wagner},
  {Ward-Thompson}, {Wardle}, {Weintroub}, {Wex}, {Wharton}, {Wielgus}, {Wiik},
  {Witzel}, {Wondrak}, {Wong}, {Wu}, {Yamaguchi}, {Yoon}, {Young}, {Young},
  {Younsi}, {Yuan}, {Yuan}, {Zensus}, {Zhang}, {Zhao}, {Zhao}, {Agurto},
  {Allardi}, {Amestica}, {Araneda}, {Arriagada}, {Berghuis}, {Bertarini},
  {Berthold}, {Blanchard}, {Brown}, {C{\'a}rdenas}, {Cantzler}, {Caro},
  {Castillo-Dom{\'\i}nguez}, {Chan}, {Chang}, {Chang}, {Chang}, {Chang},
  {Chen}, {Chilson}, {Chuter}, {Ciechanowicz}, {Colin-Beltran}, {Coulson},
  {Crowley}, {Degenaar}, {Dornbusch}, {Dur{\'a}n}, {Everett}, {Faber},
  {Forster}, {Fuchs}, {Gale}, {Geertsema}, {Gonz{\'a}lez}, {Graham}, {Gueth},
  {Halverson}, {Han}, {Han}, {Hasegawa}, {Hern{\'a}ndez-Rebollar}, {Herrera},
  {Herrero-Illana}, {Heyminck}, {Hirota}, {Hoge}, {Hostler Schimpf}, {Howie},
  {Huang}, {Jiang}, {Jinchi}, {John}, {Kimura}, {Klein}, {Kubo}, {Kuroda},
  {Kwon}, {Lacasse}, {Laing}, {Leitch}, {Li}, {Liu}, {Liu}, {Lin}, {Lu},
  {Mac-Auliffe}, {Martin-Cocher}, {Matulonis}, {Maute}, {Messias},
  {Meyer-Zhao}, {Monta{\~n}a}, {Montenegro-Montes}, {Montgomerie}, {Moreno
  Nolasco}, {Muders}, {Nishioka}, {Norton}, {Nystrom}, {Ogawa}, {Olivares},
  {Oshiro}, {P{\'e}rez-Beaupuits}, {Parra}, {Phillips}, {Poirier}, {Pradel},
  {Qiu}, {Raffin}, {Rahlin}, {Ram{\'\i}rez}, {Ressler}, {Reynolds},
  {Rodr{\'\i}guez-Montoya}, {Saez-Madain}, {Santana}, {Shaw}, {Shirkey},
  {Silva}, {Snow}, {Sousa}, {Sridharan}, {Stahm}, {Stark}, {Test},
  {Torstensson}, {Venegas}, {Walther}, {Wei}, {White}, {Wieching}, {Wijnands},
  {Wouterloot}, {Yu}, {Yu (于威)}, \& {Zeballos}}]{EHTC2022_1}
{Event Horizon Telescope Collaboration}, {Akiyama}, K., {Alberdi}, A., {et~al.}
  2022{\natexlab{a}}, \apjl, 930, L12

\bibitem[{{Event Horizon Telescope Collaboration}
  {et~al.}(2022{\natexlab{b}}){Event Horizon Telescope Collaboration},
  {Akiyama}, {Alberdi}, {Alef}, {Algaba}, {Anantua}, {Asada}, {Azulay}, {Bach},
  {Baczko}, {Ball}, {Balokovi{\'c}}, {Barrett}, {Baub{\"o}ck}, {Benson},
  {Bintley}, {Blackburn}, {Blundell}, {Bouman}, {Bower}, {Boyce}, {Bremer},
  {Brinkerink}, {Brissenden}, {Britzen}, {Broderick}, {Broguiere}, {Bronzwaer},
  {Bustamante}, {Byun}, {Carlstrom}, {Ceccobello}, {Chael}, {Chan},
  {Chatterjee}, {Chatterjee}, {Chen}, {Chen}, {Cheng}, {Cho}, {Christian},
  {Conroy}, {Conway}, {Cordes}, {Crawford}, {Crew}, {Cruz-Osorio}, {Cui},
  {Davelaar}, {De Laurentis}, {Deane}, {Dempsey}, {Desvignes}, {Dexter},
  {Dhruv}, {Doeleman}, {Dougal}, {Dzib}, {Eatough}, {Emami}, {Falcke}, {Farah},
  {Fish}, {Fomalont}, {Ford}, {Fraga-Encinas}, {Freeman}, {Friberg}, {Fromm},
  {Fuentes}, {Galison}, {Gammie}, {Garc{\'\i}a}, {Gentaz}, {Georgiev}, {Goddi},
  {Gold}, {G{\'o}mez-Ruiz}, {G{\'o}mez}, {Gu}, {Gurwell}, {Hada}, {Haggard},
  {Haworth}, {Hecht}, {Hesper}, {Heumann}, {Ho}, {Ho}, {Honma}, {Huang},
  {Huang}, {Hughes}, {Ikeda}, {Impellizzeri}, {Inoue}, {Issaoun}, {James},
  {Jannuzi}, {Janssen}, {Jeter}, {Jiang}, {Jim{\'e}nez-Rosales}, {Johnson},
  {Jorstad}, {Joshi}, {Jung}, {Karami}, {Karuppusamy}, {Kawashima}, {Keating},
  {Kettenis}, {Kim}, {Kim}, {Kim}, {Kim}, {Kino}, {Koay}, {Kocherlakota},
  {Kofuji}, {Koch}, {Koyama}, {Kramer}, {Kramer}, {Krichbaum}, {Kuo}, {Bella},
  {Lauer}, {Lee}, {Lee}, {Leung}, {Levis}, {Li}, {Lico}, {Lindahl},
  {Lindqvist}, {Lisakov}, {Liu}, {Liu}, {Liuzzo}, {Lo}, {Lobanov}, {Loinard},
  {Lonsdale}, {Lu}, {Mao}, {Marchili}, {Markoff}, {Marrone}, {Marscher},
  {Mart{\'\i}-Vidal}, {Matsushita}, {Matthews}, {Medeiros}, {Menten},
  {Michalik}, {Mizuno}, {Mizuno}, {Moran}, {Moriyama}, {Moscibrodzka},
  {M{\"u}ller}, {Mus}, {Musoke}, {Myserlis}, {Nadolski}, {Nagai}, {Nagar},
  {Nakamura}, {Narayan}, {Narayanan}, {Natarajan}, {Nathanail}, {Fuentes},
  {Neilsen}, {Neri}, {Ni}, {Noutsos}, {Nowak}, {Oh}, {Okino}, {Olivares},
  {Ortiz-Le{\'o}n}, {Oyama}, {{\"O}zel}, {Palumbo}, {Paraschos}, {Park},
  {Parsons}, {Patel}, {Pen}, {Pesce}, {Pi{\'e}tu}, {Plambeck}, {PopStefanija},
  {Porth}, {P{\"o}tzl}, {Prather}, {Preciado-L{\'o}pez}, {Psaltis}, {Pu},
  {Ramakrishnan}, {Rao}, {Rawlings}, {Raymond}, {Rezzolla}, {Ricarte},
  {Ripperda}, {Roelofs}, {Rogers}, {Ros}, {Romero-Ca{\~n}izales},
  {Roshanineshat}, {Rottmann}, {Roy}, {Ruiz}, {Ruszczyk}, {Rygl},
  {S{\'a}nchez}, {S{\'a}nchez-Arg{\"u}elles}, {S{\'a}nchez-Portal}, {Sasada},
  {Satapathy}, {Savolainen}, {Schloerb}, {Schonfeld}, {Schuster}, {Shao},
  {Shen}, {Small}, {Sohn}, {SooHoo}, {Souccar}, {Sun}, {Tazaki}, {Tetarenko},
  {Tiede}, {Tilanus}, {Titus}, {Torne}, {Traianou}, {Trent}, {Trippe}, {Turk},
  {van Bemmel}, {van Langevelde}, {van Rossum}, {Vos}, {Wagner},
  {Ward-Thompson}, {Wardle}, {Weintroub}, {Wex}, {Wharton}, {Wielgus}, {Wiik},
  {Witzel}, {Wondrak}, {Wong}, {Wu}, {Yamaguchi}, {Yoon}, {Young}, {Young},
  {Younsi}, {Yuan}, {Yuan}, {Zensus}, {Zhang}, {Zhao}, {Zhao}, {Agurto},
  {Araneda}, {Arriagada}, {Bertarini}, {Berthold}, {Blanchard}, {Brown},
  {C{\'a}rdenas}, {Cantzler}, {Caro}, {Chuter}, {Ciechanowicz}, {Coulson},
  {Crowley}, {Degenaar}, {Dornbusch}, {Dur{\'a}n}, {Forster}, {Geertsema},
  {Gonz{\'a}lez}, {Graham}, {Gueth}, {Han}, {Herrera}, {Herrero-Illana},
  {Heyminck}, {Hoge}, {Huang}, {Jiang}, {John}, {Klein}, {Kubo}, {Kuroda},
  {Kwon}, {Laing}, {Liu}, {Liu}, {Mac-Auliffe}, {Martin-Cocher}, {Matulonis},
  {Messias}, {Meyer-Zhao}, {Montenegro-Montes}, {Montgomerie}, {Muders},
  {Nishioka}, {Norton}, {Olivares}, {P{\'e}rez-Beaupuits}, {Parra}, {Poirier},
  {Pradel}, {Raffin}, {Ram{\'\i}rez}, {Reynolds}, {Saez-Madain}, {Santana},
  {Silva}, {Sousa}, {Stahm}, {Torstensson}, {Venegas}, {Walther}, {Wieching},
  {Wijnands}, \& {Wouterloot}}]{EHTC2022_2}
{Event Horizon Telescope Collaboration}, {Akiyama}, K., {Alberdi}, A., {et~al.}
  2022{\natexlab{b}}, \apjl, 930, L13

\bibitem[{{Event Horizon Telescope Collaboration}
  {et~al.}(2022{\natexlab{c}}){Event Horizon Telescope Collaboration},
  {Akiyama}, {Alberdi}, {Alef}, {Algaba}, {Anantua}, {Asada}, {Azulay}, {Bach},
  {Baczko}, {Ball}, {Balokovi{\'c}}, {Barrett}, {Baub{\"o}ck}, {Benson},
  {Bintley}, {Blackburn}, {Blundell}, {Bouman}, {Bower}, {Boyce}, {Bremer},
  {Brinkerink}, {Brissenden}, {Britzen}, {Broderick}, {Broguiere}, {Bronzwaer},
  {Bustamante}, {Byun}, {Carlstrom}, {Ceccobello}, {Chael}, {Chan},
  {Chatterjee}, {Chatterjee}, {Chen}, {Chen}, {Cheng}, {Cho}, {Christian},
  {Conroy}, {Conway}, {Cordes}, {Crawford}, {Crew}, {Cruz-Osorio}, {Cui},
  {Davelaar}, {De Laurentis}, {Deane}, {Dempsey}, {Desvignes}, {Dexter},
  {Dhruv}, {Doeleman}, {Dougal}, {Dzib}, {Eatough}, {Emami}, {Falcke}, {Farah},
  {Fish}, {Fomalont}, {Ford}, {Fraga-Encinas}, {Freeman}, {Friberg}, {Fromm},
  {Fuentes}, {Galison}, {Gammie}, {Garc{\'\i}a}, {Gentaz}, {Georgiev}, {Goddi},
  {Gold}, {G{\'o}mez-Ruiz}, {G{\'o}mez}, {Gu}, {Gurwell}, {Hada}, {Haggard},
  {Haworth}, {Hecht}, {Hesper}, {Heumann}, {Ho}, {Ho}, {Honma}, {Huang},
  {Huang}, {Hughes}, {Ikeda}, {Impellizzeri}, {Inoue}, {Issaoun}, {James},
  {Jannuzi}, {Janssen}, {Jeter}, {Jiang}, {Jim{\'e}nez-Rosales}, {Johnson},
  {Jorstad}, {Joshi}, {Jung}, {Karami}, {Karuppusamy}, {Kawashima}, {Keating},
  {Kettenis}, {Kim}, {Kim}, {Kim}, {Kim}, {Kino}, {Koay}, {Kocherlakota},
  {Kofuji}, {Koch}, {Koyama}, {Kramer}, {Kramer}, {Krichbaum}, {Kuo}, {La
  Bella}, {Lauer}, {Lee}, {Lee}, {Leung}, {Levis}, {Li}, {Lico}, {Lindahl},
  {Lindqvist}, {Lisakov}, {Liu}, {Liu}, {Liuzzo}, {Lo}, {Lobanov}, {Loinard},
  {Lonsdale}, {Lu}, {Mao}, {Marchili}, {Markoff}, {Marrone}, {Marscher},
  {Mart{\'\i}-Vidal}, {Matsushita}, {Matthews}, {Medeiros}, {Menten},
  {Michalik}, {Mizuno}, {Mizuno}, {Moran}, {Moriyama}, {Moscibrodzka},
  {M{\"u}ller}, {Mus}, {Musoke}, {Myserlis}, {Nadolski}, {Nagai}, {Nagar},
  {Nakamura}, {Narayan}, {Narayanan}, {Natarajan}, {Nathanail}, {Navarro
  Fuentes}, {Neilsen}, {Neri}, {Ni}, {Noutsos}, {Nowak}, {Oh}, {Okino},
  {Olivares}, {Ortiz-Le{\'o}n}, {Oyama}, {{\"O}zel}, {Palumbo}, {Paraschos},
  {Park}, {Parsons}, {Patel}, {Pen}, {Pesce}, {Pi{\'e}tu}, {Plambeck},
  {PopStefanija}, {Porth}, {P{\"o}tzl}, {Prather}, {Preciado-L{\'o}pez},
  {Psaltis}, {Pu}, {Ramakrishnan}, {Rao}, {Rawlings}, {Raymond}, {Rezzolla},
  {Ricarte}, {Ripperda}, {Roelofs}, {Rogers}, {Ros}, {Romero-Ca{\~n}izales},
  {Roshanineshat}, {Rottmann}, {Roy}, {Ruiz}, {Ruszczyk}, {Rygl},
  {S{\'a}nchez}, {S{\'a}nchez-Arg{\"u}elles}, {S{\'a}nchez-Portal}, {Sasada},
  {Satapathy}, {Savolainen}, {Schloerb}, {Schonfeld}, {Schuster}, {Shao},
  {Shen}, {Small}, {Sohn}, {SooHoo}, {Souccar}, {Sun}, {Tazaki}, {Tetarenko},
  {Tiede}, {Tilanus}, {Titus}, {Torne}, {Traianou}, {Trent}, {Trippe}, {Turk},
  {van Bemmel}, {van Langevelde}, {van Rossum}, {Vos}, {Wagner},
  {Ward-Thompson}, {Wardle}, {Weintroub}, {Wex}, {Wharton}, {Wielgus}, {Wiik},
  {Witzel}, {Wondrak}, {Wong}, {Wu}, {Yamaguchi}, {Yoon}, {Young}, {Young},
  {Younsi}, {Yuan}, {Yuan}, {Zensus}, {Zhang}, {Zhao}, {Zhao}, \& {EHT
  Collaboration}}]{EHTC2022_3}
{Event Horizon Telescope Collaboration}, {Akiyama}, K., {Alberdi}, A., {et~al.}
  2022{\natexlab{c}}, \apjl, 930, L14

\bibitem[{{Event Horizon Telescope Collaboration}
  {et~al.}(2022{\natexlab{d}}){Event Horizon Telescope Collaboration},
  {Akiyama}, {Alberdi}, {Alef}, {Algaba}, {Anantua}, {Asada}, {Azulay}, {Bach},
  {Baczko}, {Ball}, {Balokovi{\'c}}, {Barrett}, {Baub{\"o}ck}, {Benson},
  {Bintley}, {Blackburn}, {Blundell}, {Bouman}, {Bower}, {Boyce}, {Bremer},
  {Brinkerink}, {Brissenden}, {Britzen}, {Broderick}, {Broguiere}, {Bronzwaer},
  {Bustamante}, {Byun}, {Carlstrom}, {Ceccobello}, {Chael}, {Chan},
  {Chatterjee}, {Chatterjee}, {Chen}, {Chen}, {Cheng}, {Cho}, {Christian},
  {Conroy}, {Conway}, {Cordes}, {Crawford}, {Crew}, {Cruz-Osorio}, {Cui},
  {Davelaar}, {Laurentis}, {Deane}, {Dempsey}, {Desvignes}, {Dexter}, {Dhruv},
  {Doeleman}, {Dougal}, {Dzib}, {Eatough}, {Emami}, {Falcke}, {Farah}, {Fish},
  {Fomalont}, {Ford}, {Fraga-Encinas}, {Freeman}, {Friberg}, {Fromm},
  {Fuentes}, {Galison}, {Gammie}, {Garc{\'\i}a}, {Gentaz}, {Georgiev}, {Goddi},
  {Gold}, {G{\'o}mez-Ruiz}, {G{\'o}mez}, {Gu}, {Gurwell}, {Hada}, {Haggard},
  {Haworth}, {Hecht}, {Hesper}, {Heumann}, {Ho}, {Ho}, {Honma}, {Huang},
  {Huang}, {Hughes}, {Ikeda}, {Impellizzeri}, {Inoue}, {Issaoun}, {James},
  {Jannuzi}, {Janssen}, {Jeter}, {Jiang}, {Jim{\'e}nez-Rosales}, {Johnson},
  {Jorstad}, {Joshi}, {Jung}, {Karami}, {Karuppusamy}, {Kawashima}, {Keating},
  {Kettenis}, {Kim}, {Kim}, {Kim}, {Kim}, {Kino}, {Koay}, {Kocherlakota},
  {Kofuji}, {Koch}, {Koyama}, {Kramer}, {Kramer}, {Krichbaum}, {Kuo}, {Bella},
  {Lauer}, {Lee}, {Lee}, {Leung}, {Levis}, {Li}, {Lico}, {Lindahl},
  {Lindqvist}, {Lisakov}, {Liu}, {Liu}, {Liuzzo}, {Lo}, {Lobanov}, {Loinard},
  {Lonsdale}, {Lu}, {Mao}, {Marchili}, {Markoff}, {Marrone}, {Marscher},
  {Mart{\'\i}-Vidal}, {Matsushita}, {Matthews}, {Medeiros}, {Menten},
  {Michalik}, {Mizuno}, {Mizuno}, {Moran}, {Moriyama}, {Moscibrodzka},
  {M{\"u}ller}, {Mus}, {Musoke}, {Myserlis}, {Nadolski}, {Nagai}, {Nagar},
  {Nakamura}, {Narayan}, {Narayanan}, {Natarajan}, {Nathanail}, {Fuentes},
  {Neilsen}, {Neri}, {Ni}, {Noutsos}, {Nowak}, {Oh}, {Okino}, {Olivares},
  {Ortiz-Le{\'o}n}, {Oyama}, {Palumbo}, {Paraschos}, {Park}, {Parsons},
  {Patel}, {Pen}, {Pesce}, {Pi{\'e}tu}, {Plambeck}, {PopStefanija}, {Porth},
  {P{\"o}tzl}, {Prather}, {Preciado-L{\'o}pez}, {Pu}, {Ramakrishnan}, {Rao},
  {Rawlings}, {Raymond}, {Rezzolla}, {Ricarte}, {Ripperda}, {Roelofs},
  {Rogers}, {Ros}, {Romero-Ca{\~n}izales}, {Roshanineshat}, {Rottmann}, {Roy},
  {Ruiz}, {Ruszczyk}, {Rygl}, {S{\'a}nchez}, {S{\'a}nchez-Arg{\"u}elles},
  {S{\'a}nchez-Portal}, {Sasada}, {Satapathy}, {Savolainen}, {Schloerb},
  {Schonfeld}, {Schuster}, {Shao}, {Shen}, {Small}, {Sohn}, {SooHoo},
  {Souccar}, {Sun}, {Tazaki}, {Tetarenko}, {Tiede}, {Tilanus}, {Titus},
  {Torne}, {Traianou}, {Trent}, {Trippe}, {Turk}, {van Bemmel}, {van
  Langevelde}, {van Rossum}, {Vos}, {Wagner}, {Ward-Thompson}, {Wardle},
  {Weintroub}, {Wex}, {Wharton}, {Wielgus}, {Wiik}, {Witzel}, {Wondrak},
  {Wong}, {Wu}, {Yamaguchi}, {Yoon}, {Young}, {Young}, {Younsi}, {Yuan},
  {Yuan}, {Zensus}, {Zhang}, {Zhao}, {Zhao}, \& {Chang}}]{EHTC2022_4}
{Event Horizon Telescope Collaboration}, {Akiyama}, K., {Alberdi}, A., {et~al.}
  2022{\natexlab{d}}, \apjl, 930, L15

\bibitem[{{Event Horizon Telescope Collaboration}
  {et~al.}(2022{\natexlab{e}}){Event Horizon Telescope Collaboration},
  {Akiyama}, {Alberdi}, {Alef}, {Algaba}, {Anantua}, {Asada}, {Azulay}, {Bach},
  {Baczko}, {Ball}, {Balokovi{\'c}}, {Barrett}, {Baub{\"o}ck}, {Benson},
  {Bintley}, {Blackburn}, {Blundell}, {Bouman}, {Bower}, {Boyce}, {Bremer},
  {Brinkerink}, {Brissenden}, {Britzen}, {Broderick}, {Broguiere}, {Bronzwaer},
  {Bustamante}, {Byun}, {Carlstrom}, {Ceccobello}, {Chael}, {Chan},
  {Chatterjee}, {Chatterjee}, {Chen}, {Chen}, {Cheng}, {Cho}, {Christian},
  {Conroy}, {Conway}, {Cordes}, {Crawford}, {Crew}, {Cruz-Osorio}, {Cui},
  {Davelaar}, {De Laurentis}, {Deane}, {Dempsey}, {Desvignes}, {Dexter},
  {Dhruv}, {Doeleman}, {Dougal}, {Dzib}, {Eatough}, {Emami}, {Falcke}, {Farah},
  {Fish}, {Fomalont}, {Ford}, {Fraga-Encinas}, {Freeman}, {Friberg}, {Fromm},
  {Fuentes}, {Galison}, {Gammie}, {Garc{\'\i}a}, {Gentaz}, {Georgiev}, {Goddi},
  {Gold}, {G{\'o}mez-Ruiz}, {G{\'o}mez}, {Gu}, {Gurwell}, {Hada}, {Haggard},
  {Haworth}, {Hecht}, {Hesper}, {Heumann}, {Ho}, {Ho}, {Honma}, {Huang},
  {Huang}, {Hughes}, {Ikeda}, {Impellizzeri}, {Inoue}, {Issaoun}, {James},
  {Jannuzi}, {Janssen}, {Jeter}, {Jiang}, {Jim{\'e}nez-Rosales}, {Johnson},
  {Jorstad}, {Joshi}, {Jung}, {Karami}, {Karuppusamy}, {Kawashima}, {Keating},
  {Kettenis}, {Kim}, {Kim}, {Kim}, {Kim}, {Kino}, {Koay}, {Kocherlakota},
  {Kofuji}, {Koch}, {Koyama}, {Kramer}, {Kramer}, {Krichbaum}, {Kuo}, {Bella},
  {Lauer}, {Lee}, {Lee}, {Leung}, {Levis}, {Li}, {Lico}, {Lindahl},
  {Lindqvist}, {Lisakov}, {Liu}, {Liu}, {Liuzzo}, {Lo}, {Lobanov}, {Loinard},
  {Lonsdale}, {Lu}, {Mao}, {Marchili}, {Markoff}, {Marrone}, {Marscher},
  {Mart{\'\i}-Vidal}, {Matsushita}, {Matthews}, {Medeiros}, {Menten},
  {Michalik}, {Mizuno}, {Mizuno}, {Moran}, {Moriyama}, {Moscibrodzka},
  {M{\"u}ller}, {Mus}, {Musoke}, {Myserlis}, {Nadolski}, {Nagai}, {Nagar},
  {Nakamura}, {Narayan}, {Narayanan}, {Natarajan}, {Nathanail}, {Fuentes},
  {Neilsen}, {Neri}, {Ni}, {Noutsos}, {Nowak}, {Oh}, {Okino}, {Olivares},
  {Ortiz-Le{\'o}n}, {Oyama}, {{\"O}zel}, {Palumbo}, {Paraschos}, {Park},
  {Parsons}, {Patel}, {Pen}, {Pesce}, {Pi{\'e}tu}, {Plambeck}, {PopStefanija},
  {Porth}, {P{\"o}tzl}, {Prather}, {Preciado-L{\'o}pez}, {Psaltis}, {Pu},
  {Ramakrishnan}, {Rao}, {Rawlings}, {Raymond}, {Rezzolla}, {Ricarte},
  {Ripperda}, {Roelofs}, {Rogers}, {Ros}, {Romero-Ca{\~n}izales},
  {Roshanineshat}, {Rottmann}, {Roy}, {Ruiz}, {Ruszczyk}, {Rygl},
  {S{\'a}nchez}, {S{\'a}nchez-Arg{\"u}elles}, {S{\'a}nchez-Portal}, {Sasada},
  {Satapathy}, {Savolainen}, {Schloerb}, {Schonfeld}, {Schuster}, {Shao},
  {Shen}, {Small}, {Sohn}, {SooHoo}, {Souccar}, {Sun}, {Tazaki}, {Tetarenko},
  {Tiede}, {Tilanus}, {Titus}, {Torne}, {Traianou}, {Trent}, {Trippe}, {Turk},
  {van Bemmel}, {van Langevelde}, {van Rossum}, {Vos}, {Wagner},
  {Ward-Thompson}, {Wardle}, {Weintroub}, {Wex}, {Wharton}, {Wielgus}, {Wiik},
  {Witzel}, {Wondrak}, {Wong}, {Wu}, {Yamaguchi}, {Yoon}, {Young}, {Young},
  {Younsi}, {Yuan}, {Yuan}, {Zensus}, {Zhang}, {Zhao}, \& {Zhao}}]{EHTC2022_6}
{Event Horizon Telescope Collaboration}, {Akiyama}, K., {Alberdi}, A., {et~al.}
  2022{\natexlab{e}}, \apjl, 930, L17

\bibitem[{{Event Horizon Telescope Collaboration}
  {et~al.}(2019{\natexlab{a}}){Event Horizon Telescope Collaboration},
  {Akiyama}, {Alberdi}, {Alef}, {Asada}, {Azulay}, {Baczko}, {Ball},
  {Balokovi{\'c}}, {Barrett}, {Bintley}, {Blackburn}, {Boland}, {Bouman},
  {Bower}, {Bremer}, {Brinkerink}, {Brissenden}, {Britzen}, {Broderick},
  {Broguiere}, {Bronzwaer}, {Byun}, {Carlstrom}, {Chael}, {Chan}, {Chatterjee},
  {Chatterjee}, {Chen}, {Chen}, {Cho}, {Christian}, {Conway}, {Cordes}, {Crew},
  {Cui}, {Davelaar}, {De Laurentis}, {Deane}, {Dempsey}, {Desvignes}, {Dexter},
  {Doeleman}, {Eatough}, {Falcke}, {Fish}, {Fomalont}, {Fraga-Encinas},
  {Freeman}, {Friberg}, {Fromm}, {G{\'o}mez}, {Galison}, {Gammie},
  {Garc{\'\i}a}, {Gentaz}, {Georgiev}, {Goddi}, {Gold}, {Gu}, {Gurwell},
  {Hada}, {Hecht}, {Hesper}, {Ho}, {Ho}, {Honma}, {Huang}, {Huang}, {Hughes},
  {Ikeda}, {Inoue}, {Issaoun}, {James}, {Jannuzi}, {Janssen}, {Jeter}, {Jiang},
  {Johnson}, {Jorstad}, {Jung}, {Karami}, {Karuppusamy}, {Kawashima},
  {Keating}, {Kettenis}, {Kim}, {Kim}, {Kim}, {Kino}, {Koay}, {Koch}, {Koyama},
  {Kramer}, {Kramer}, {Krichbaum}, {Kuo}, {Lauer}, {Lee}, {Li}, {Li},
  {Lindqvist}, {Liu}, {Liuzzo}, {Lo}, {Lobanov}, {Loinard}, {Lonsdale}, {Lu},
  {MacDonald}, {Mao}, {Markoff}, {Marrone}, {Marscher}, {Mart{\'\i}-Vidal},
  {Matsushita}, {Matthews}, {Medeiros}, {Menten}, {Mizuno}, {Mizuno}, {Moran},
  {Moriyama}, {Moscibrodzka}, {M{\"u}ller}, {Nagai}, {Nagar}, {Nakamura},
  {Narayan}, {Narayanan}, {Natarajan}, {Neri}, {Ni}, {Noutsos}, {Okino},
  {Olivares}, {Ortiz-Le{\'o}n}, {Oyama}, {{\"O}zel}, {Palumbo}, {Patel}, {Pen},
  {Pesce}, {Pi{\'e}tu}, {Plambeck}, {PopStefanija}, {Porth}, {Prather},
  {Preciado-L{\'o}pez}, {Psaltis}, {Pu}, {Ramakrishnan}, {Rao}, {Rawlings},
  {Raymond}, {Rezzolla}, {Ripperda}, {Roelofs}, {Rogers}, {Ros}, {Rose},
  {Roshanineshat}, {Rottmann}, {Roy}, {Ruszczyk}, {Ryan}, {Rygl},
  {S{\'a}nchez}, {S{\'a}nchez-Arguelles}, {Sasada}, {Savolainen}, {Schloerb},
  {Schuster}, {Shao}, {Shen}, {Small}, {Sohn}, {SooHoo}, {Tazaki}, {Tiede},
  {Tilanus}, {Titus}, {Toma}, {Torne}, {Trent}, {Trippe}, {Tsuda}, {van
  Bemmel}, {van Langevelde}, {van Rossum}, {Wagner}, {Wardle}, {Weintroub},
  {Wex}, {Wharton}, {Wielgus}, {Wong}, {Wu}, {Young}, {Young}, {Younsi},
  {Yuan}, {Yuan}, {Zensus}, {Zhao}, {Zhao}, {Zhu}, {Algaba}, {Allardi},
  {Amestica}, {Anczarski}, {Bach}, {Baganoff}, {Beaudoin}, {Benson},
  {Berthold}, {Blanchard}, {Blundell}, {Bustamente}, {Cappallo},
  {Castillo-Dom{\'\i}nguez}, {Chang}, {Chang}, {Chang}, {Chen}, {Chilson},
  {Chuter}, {C{\'o}rdova Rosado}, {Coulson}, {Crawford}, {Crowley}, {David},
  {Derome}, {Dexter}, {Dornbusch}, {Dudevoir}, {Dzib}, {Eckart}, {Eckert},
  {Erickson}, {Everett}, {Faber}, {Farah}, {Fath}, {Folkers}, {Forbes},
  {Freund}, {G{\'o}mez-Ruiz}, {Gale}, {Gao}, {Geertsema}, {Graham}, {Greer},
  {Grosslein}, {Gueth}, {Haggard}, {Halverson}, {Han}, {Han}, {Hao},
  {Hasegawa}, {Henning}, {Hern{\'a}ndez-G{\'o}mez}, {Herrero-Illana},
  {Heyminck}, {Hirota}, {Hoge}, {Huang}, {Impellizzeri}, {Jiang}, {Kamble},
  {Keisler}, {Kimura}, {Kono}, {Kubo}, {Kuroda}, {Lacasse}, {Laing}, {Leitch},
  {Li}, {Lin}, {Liu}, {Liu}, {Lu}, {Marson}, {Martin-Cocher}, {Massingill},
  {Matulonis}, {McColl}, {McWhirter}, {Messias}, {Meyer-Zhao}, {Michalik},
  {Monta{\~n}a}, {Montgomerie}, {Mora-Klein}, {Muders}, {Nadolski}, {Navarro},
  {Neilsen}, {Nguyen}, {Nishioka}, {Norton}, {Nowak}, {Nystrom}, {Ogawa},
  {Oshiro}, {Oyama}, {Parsons}, {Paine}, {Pe{\~n}alver}, {Phillips}, {Poirier},
  {Pradel}, {Primiani}, {Raffin}, {Rahlin}, {Reiland}, {Risacher}, {Ruiz},
  {S{\'a}ez-Mada{\'\i}n}, {Sassella}, {Schellart}, {Shaw}, {Silva}, {Shiokawa},
  {Smith}, {Snow}, {Souccar}, {Sousa}, {Sridharan}, {Srinivasan}, {Stahm},
  {Stark}, {Story}, {Timmer}, {Vertatschitsch}, {Walther}, {Wei}, {Whitehorn},
  {Whitney}, {Woody}, {Wouterloot}, {Wright}, {Yamaguchi}, {Yu}, {Zeballos},
  {Zhang}, \& {Ziurys}}]{EHTC2019_1}
{Event Horizon Telescope Collaboration}, {Akiyama}, K., {Alberdi}, A., {et~al.}
  2019{\natexlab{a}}, \apjl, 875, L1

\bibitem[{{Event Horizon Telescope Collaboration}
  {et~al.}(2019{\natexlab{b}}){Event Horizon Telescope Collaboration},
  {Akiyama}, {Alberdi}, {Alef}, {Asada}, {Azulay}, {Baczko}, {Ball},
  {Balokovi{\'c}}, {Barrett}, {Bintley}, {Blackburn}, {Boland}, {Bouman},
  {Bower}, {Bremer}, {Brinkerink}, {Brissenden}, {Britzen}, {Broderick},
  {Broguiere}, {Bronzwaer}, {Byun}, {Carlstrom}, {Chael}, {Chan}, {Chatterjee},
  {Chatterjee}, {Chen}, {Chen}, {Cho}, {Christian}, {Conway}, {Cordes}, {Crew},
  {Cui}, {Davelaar}, {De Laurentis}, {Deane}, {Dempsey}, {Desvignes}, {Dexter},
  {Doeleman}, {Eatough}, {Falcke}, {Fish}, {Fomalont}, {Fraga-Encinas},
  {Friberg}, {Fromm}, {G{\'o}mez}, {Galison}, {Gammie}, {Garc{\'\i}a},
  {Gentaz}, {Georgiev}, {Goddi}, {Gold}, {Gu}, {Gurwell}, {Hada}, {Hecht},
  {Hesper}, {Ho}, {Ho}, {Honma}, {Huang}, {Huang}, {Hughes}, {Ikeda}, {Inoue},
  {Issaoun}, {James}, {Jannuzi}, {Janssen}, {Jeter}, {Jiang}, {Johnson},
  {Jorstad}, {Jung}, {Karami}, {Karuppusamy}, {Kawashima}, {Keating},
  {Kettenis}, {Kim}, {Kim}, {Kim}, {Kino}, {Koay}, {Koch}, {Koyama}, {Kramer},
  {Kramer}, {Krichbaum}, {Kuo}, {Lauer}, {Lee}, {Li}, {Li}, {Lindqvist}, {Liu},
  {Liuzzo}, {Lo}, {Lobanov}, {Loinard}, {Lonsdale}, {Lu}, {MacDonald}, {Mao},
  {Markoff}, {Marrone}, {Marscher}, {Mart{\'\i}-Vidal}, {Matsushita},
  {Matthews}, {Medeiros}, {Menten}, {Mizuno}, {Mizuno}, {Moran}, {Moriyama},
  {Moscibrodzka}, {M{\"u}ller}, {Nagai}, {Nagar}, {Nakamura}, {Narayan},
  {Narayanan}, {Natarajan}, {Neri}, {Ni}, {Noutsos}, {Okino}, {Olivares},
  {Ortiz-Le{\'o}n}, {Oyama}, {{\"O}zel}, {Palumbo}, {Patel}, {Pen}, {Pesce},
  {Pi{\'e}tu}, {Plambeck}, {PopStefanija}, {Porth}, {Prather},
  {Preciado-L{\'o}pez}, {Psaltis}, {Pu}, {Ramakrishnan}, {Rao}, {Rawlings},
  {Raymond}, {Rezzolla}, {Ripperda}, {Roelofs}, {Rogers}, {Ros}, {Rose},
  {Roshanineshat}, {Rottmann}, {Roy}, {Ruszczyk}, {Ryan}, {Rygl},
  {S{\'a}nchez}, {S{\'a}nchez-Arguelles}, {Sasada}, {Savolainen}, {Schloerb},
  {Schuster}, {Shao}, {Shen}, {Small}, {Sohn}, {SooHoo}, {Tazaki}, {Tiede},
  {Tilanus}, {Titus}, {Toma}, {Torne}, {Trent}, {Trippe}, {Tsuda}, {van
  Bemmel}, {van Langevelde}, {van Rossum}, {Wagner}, {Wardle}, {Weintroub},
  {Wex}, {Wharton}, {Wielgus}, {Wong}, {Wu}, {Young}, {Young}, {Younsi},
  {Yuan}, {Yuan}, {Zensus}, {Zhao}, {Zhao}, {Zhu}, {Algaba}, {Allardi},
  {Amestica}, {Bach}, {Beaudoin}, {Benson}, {Berthold}, {Blanchard},
  {Blundell}, {Bustamente}, {Cappallo}, {Castillo-Dom{\'\i}nguez}, {Chang},
  {Chang}, {Chang}, {Chen}, {Chilson}, {Chuter}, {C{\'o}rdova Rosado},
  {Coulson}, {Crawford}, {Crowley}, {David}, {Derome}, {Dexter}, {Dornbusch},
  {Dudevoir}, {Dzib}, {Eckert}, {Erickson}, {Everett}, {Faber}, {Farah},
  {Fath}, {Folkers}, {Forbes}, {Freund}, {G{\'o}mez-Ruiz}, {Gale}, {Gao},
  {Geertsema}, {Graham}, {Greer}, {Grosslein}, {Gueth}, {Halverson}, {Han},
  {Han}, {Hao}, {Hasegawa}, {Henning}, {Hern{\'a}ndez-G{\'o}mez},
  {Herrero-Illana}, {Heyminck}, {Hirota}, {Hoge}, {Huang}, {Impellizzeri},
  {Jiang}, {Kamble}, {Keisler}, {Kimura}, {Kono}, {Kubo}, {Kuroda}, {Lacasse},
  {Laing}, {Leitch}, {Li}, {Lin}, {Liu}, {Liu}, {Lu}, {Marson},
  {Martin-Cocher}, {Massingill}, {Matulonis}, {McColl}, {McWhirter}, {Messias},
  {Meyer-Zhao}, {Michalik}, {Monta{\~n}a}, {Montgomerie}, {Mora-Klein},
  {Muders}, {Nadolski}, {Navarro}, {Nguyen}, {Nishioka}, {Norton}, {Nystrom},
  {Ogawa}, {Oshiro}, {Oyama}, {Padin}, {Parsons}, {Paine}, {Pe{\~n}alver},
  {Phillips}, {Poirier}, {Pradel}, {Primiani}, {Raffin}, {Rahlin}, {Reiland},
  {Risacher}, {Ruiz}, {S{\'a}ez-Mada{\'\i}n}, {Sassella}, {Schellart}, {Shaw},
  {Silva}, {Shiokawa}, {Smith}, {Snow}, {Souccar}, {Sousa}, {Sridharan},
  {Srinivasan}, {Stahm}, {Stark}, {Story}, {Timmer}, {Vertatschitsch},
  {Walther}, {Wei}, {Whitehorn}, {Whitney}, {Woody}, {Wouterloot}, {Wright},
  {Yamaguchi}, {Yu}, {Zeballos}, \& {Ziurys}}]{EHTC2019_2}
{Event Horizon Telescope Collaboration}, {Akiyama}, K., {Alberdi}, A., {et~al.}
  2019{\natexlab{b}}, \apjl, 875, L2

\bibitem[{{Event Horizon Telescope Collaboration}
  {et~al.}(2019{\natexlab{c}}){Event Horizon Telescope Collaboration},
  {Akiyama}, {Alberdi}, {Alef}, {Asada}, {Azulay}, {Baczko}, {Ball},
  {Balokovi{\'c}}, {Barrett}, {Bintley}, {Blackburn}, {Boland}, {Bouman},
  {Bower}, {Bremer}, {Brinkerink}, {Brissenden}, {Britzen}, {Broderick},
  {Broguiere}, {Bronzwaer}, {Byun}, {Carlstrom}, {Chael}, {Chan}, {Chatterjee},
  {Chatterjee}, {Chen}, {Chen}, {Cho}, {Christian}, {Conway}, {Cordes}, {Crew},
  {Cui}, {Davelaar}, {De Laurentis}, {Deane}, {Dempsey}, {Desvignes}, {Dexter},
  {Doeleman}, {Eatough}, {Falcke}, {Fish}, {Fomalont}, {Fraga-Encinas},
  {Friberg}, {Fromm}, {G{\'o}mez}, {Galison}, {Gammie}, {Garc{\'\i}a},
  {Gentaz}, {Georgiev}, {Goddi}, {Gold}, {Gu}, {Gurwell}, {Hada}, {Hecht},
  {Hesper}, {Ho}, {Ho}, {Honma}, {Huang}, {Huang}, {Hughes}, {Ikeda}, {Inoue},
  {Issaoun}, {James}, {Jannuzi}, {Janssen}, {Jeter}, {Jiang}, {Johnson},
  {Jorstad}, {Jung}, {Karami}, {Karuppusamy}, {Kawashima}, {Keating},
  {Kettenis}, {Kim}, {Kim}, {Kim}, {Kino}, {Koay}, {Koch}, {Koyama}, {Kramer},
  {Kramer}, {Krichbaum}, {Kuo}, {Lauer}, {Lee}, {Li}, {Li}, {Lindqvist}, {Liu},
  {Liuzzo}, {Lo}, {Lobanov}, {Loinard}, {Lonsdale}, {Lu}, {MacDonald}, {Mao},
  {Markoff}, {Marrone}, {Marscher}, {Mart{\'\i}-Vidal}, {Matsushita},
  {Matthews}, {Medeiros}, {Menten}, {Mizuno}, {Mizuno}, {Moran}, {Moriyama},
  {Moscibrodzka}, {M{\"u}ller}, {Nagai}, {Nagar}, {Nakamura}, {Narayan},
  {Narayanan}, {Natarajan}, {Neri}, {Ni}, {Noutsos}, {Okino}, {Olivares},
  {Ortiz-Le{\'o}n}, {Oyama}, {{\"O}zel}, {Palumbo}, {Patel}, {Pen}, {Pesce},
  {Pi{\'e}tu}, {Plambeck}, {PopStefanija}, {Porth}, {Prather},
  {Preciado-L{\'o}pez}, {Psaltis}, {Pu}, {Ramakrishnan}, {Rao}, {Rawlings},
  {Raymond}, {Rezzolla}, {Ripperda}, {Roelofs}, {Rogers}, {Ros}, {Rose},
  {Roshanineshat}, {Rottmann}, {Roy}, {Ruszczyk}, {Ryan}, {Rygl},
  {S{\'a}nchez}, {S{\'a}nchez-Arguelles}, {Sasada}, {Savolainen}, {Schloerb},
  {Schuster}, {Shao}, {Shen}, {Small}, {Sohn}, {SooHoo}, {Tazaki}, {Tiede},
  {Tilanus}, {Titus}, {Toma}, {Torne}, {Trent}, {Trippe}, {Tsuda}, {van
  Bemmel}, {van Langevelde}, {van Rossum}, {Wagner}, {Wardle}, {Weintroub},
  {Wex}, {Wharton}, {Wielgus}, {Wong}, {Wu}, {Young}, {Young}, {Younsi},
  {Yuan}, {Yuan}, {Zensus}, {Zhao}, {Zhao}, {Zhu}, {Cappallo}, {Farah},
  {Folkers}, {Meyer-Zhao}, {Michalik}, {Nadolski}, {Nishioka}, {Pradel},
  {Primiani}, {Souccar}, {Vertatschitsch}, \& {Yamaguchi}}]{EHTC2019_3}
{Event Horizon Telescope Collaboration}, {Akiyama}, K., {Alberdi}, A., {et~al.}
  2019{\natexlab{c}}, \apjl, 875, L3

\bibitem[{{Event Horizon Telescope Collaboration}
  {et~al.}(2019{\natexlab{d}}){Event Horizon Telescope Collaboration},
  {Akiyama}, {Alberdi}, {Alef}, {Asada}, {Azulay}, {Baczko}, {Ball},
  {Balokovi{\'c}}, {Barrett}, {Bintley}, {Blackburn}, {Boland}, {Bouman},
  {Bower}, {Bremer}, {Brinkerink}, {Brissenden}, {Britzen}, {Broderick},
  {Broguiere}, {Bronzwaer}, {Byun}, {Carlstrom}, {Chael}, {Chan}, {Chatterjee},
  {Chatterjee}, {Chen}, {Chen}, {Cho}, {Christian}, {Conway}, {Cordes}, {Crew},
  {Cui}, {Davelaar}, {De Laurentis}, {Deane}, {Dempsey}, {Desvignes}, {Dexter},
  {Doeleman}, {Eatough}, {Falcke}, {Fish}, {Fomalont}, {Fraga-Encinas},
  {Freeman}, {Friberg}, {Fromm}, {G{\'o}mez}, {Galison}, {Gammie},
  {Garc{\'\i}a}, {Gentaz}, {Georgiev}, {Goddi}, {Gold}, {Gu}, {Gurwell},
  {Hada}, {Hecht}, {Hesper}, {Ho}, {Ho}, {Honma}, {Huang}, {Huang}, {Hughes},
  {Ikeda}, {Inoue}, {Issaoun}, {James}, {Jannuzi}, {Janssen}, {Jeter}, {Jiang},
  {Johnson}, {Jorstad}, {Jung}, {Karami}, {Karuppusamy}, {Kawashima},
  {Keating}, {Kettenis}, {Kim}, {Kim}, {Kim}, {Kino}, {Koay}, {Koch}, {Koyama},
  {Kramer}, {Kramer}, {Krichbaum}, {Kuo}, {Lauer}, {Lee}, {Li}, {Li},
  {Lindqvist}, {Liu}, {Liuzzo}, {Lo}, {Lobanov}, {Loinard}, {Lonsdale}, {Lu},
  {MacDonald}, {Mao}, {Markoff}, {Marrone}, {Marscher}, {Mart{\'\i}-Vidal},
  {Matsushita}, {Matthews}, {Medeiros}, {Menten}, {Mizuno}, {Mizuno}, {Moran},
  {Moriyama}, {Moscibrodzka}, {M{\"u}ller}, {Nagai}, {Nagar}, {Nakamura},
  {Narayan}, {Narayanan}, {Natarajan}, {Neri}, {Ni}, {Noutsos}, {Okino},
  {Olivares}, {Oyama}, {{\"O}zel}, {Palumbo}, {Patel}, {Pen}, {Pesce},
  {Pi{\'e}tu}, {Plambeck}, {PopStefanija}, {Porth}, {Prather},
  {Preciado-L{\'o}pez}, {Psaltis}, {Pu}, {Ramakrishnan}, {Rao}, {Rawlings},
  {Raymond}, {Rezzolla}, {Ripperda}, {Roelofs}, {Rogers}, {Ros}, {Rose},
  {Roshanineshat}, {Rottmann}, {Roy}, {Ruszczyk}, {Ryan}, {Rygl},
  {S{\'a}nchez}, {S{\'a}nchez-Arguelles}, {Sasada}, {Savolainen}, {Schloerb},
  {Schuster}, {Shao}, {Shen}, {Small}, {Sohn}, {SooHoo}, {Tazaki}, {Tiede},
  {Tilanus}, {Titus}, {Toma}, {Torne}, {Trent}, {Trippe}, {Tsuda}, {van
  Bemmel}, {van Langevelde}, {van Rossum}, {Wagner}, {Wardle}, {Weintroub},
  {Wex}, {Wharton}, {Wielgus}, {Wong}, {Wu}, {Young}, {Young}, {Younsi},
  {Yuan}, {Yuan}, {Zensus}, {Zhao}, {Zhao}, {Zhu}, {Farah}, {Meyer-Zhao},
  {Michalik}, {Nadolski}, {Nishioka}, {Pradel}, {Primiani}, {Souccar},
  {Vertatschitsch}, \& {Yamaguchi}}]{EHTC2019_4}
{Event Horizon Telescope Collaboration}, {Akiyama}, K., {Alberdi}, A., {et~al.}
  2019{\natexlab{d}}, \apjl, 875, L4

\bibitem[{{Event Horizon Telescope Collaboration}
  {et~al.}(2019{\natexlab{e}}){Event Horizon Telescope Collaboration},
  {Akiyama}, {Alberdi}, {Alef}, {Asada}, {Azulay}, {Baczko}, {Ball},
  {Balokovi{\'c}}, {Barrett}, {Bintley}, {Blackburn}, {Boland}, {Bouman},
  {Bower}, {Bremer}, {Brinkerink}, {Brissenden}, {Britzen}, {Broderick},
  {Broguiere}, {Bronzwaer}, {Byun}, {Carlstrom}, {Chael}, {Chan}, {Chatterjee},
  {Chatterjee}, {Chen}, {Chen}, {Cho}, {Christian}, {Conway}, {Cordes}, {Crew},
  {Cui}, {Davelaar}, {De Laurentis}, {Deane}, {Dempsey}, {Desvignes}, {Dexter},
  {Doeleman}, {Eatough}, {Falcke}, {Fish}, {Fomalont}, {Fraga-Encinas},
  {Friberg}, {Fromm}, {G{\'o}mez}, {Galison}, {Gammie}, {Garc{\'\i}a},
  {Gentaz}, {Georgiev}, {Goddi}, {Gold}, {Gu}, {Gurwell}, {Hada}, {Hecht},
  {Hesper}, {Ho}, {Ho}, {Honma}, {Huang}, {Huang}, {Hughes}, {Ikeda}, {Inoue},
  {Issaoun}, {James}, {Jannuzi}, {Janssen}, {Jeter}, {Jiang}, {Johnson},
  {Jorstad}, {Jung}, {Karami}, {Karuppusamy}, {Kawashima}, {Keating},
  {Kettenis}, {Kim}, {Kim}, {Kim}, {Kino}, {Koay}, {Koch}, {Koyama}, {Kramer},
  {Kramer}, {Krichbaum}, {Kuo}, {Lauer}, {Lee}, {Li}, {Li}, {Lindqvist}, {Liu},
  {Liuzzo}, {Lo}, {Lobanov}, {Loinard}, {Lonsdale}, {Lu}, {MacDonald}, {Mao},
  {Markoff}, {Marrone}, {Marscher}, {Mart{\'\i}-Vidal}, {Matsushita},
  {Matthews}, {Medeiros}, {Menten}, {Mizuno}, {Mizuno}, {Moran}, {Moriyama},
  {Moscibrodzka}, {M{\"u}ller}, {Nagai}, {Nagar}, {Nakamura}, {Narayan},
  {Narayanan}, {Natarajan}, {Neri}, {Ni}, {Noutsos}, {Okino}, {Olivares},
  {Oyama}, {{\"O}zel}, {Palumbo}, {Patel}, {Pen}, {Pesce}, {Pi{\'e}tu},
  {Plambeck}, {PopStefanija}, {Porth}, {Prather}, {Preciado-L{\'o}pez},
  {Psaltis}, {Pu}, {Ramakrishnan}, {Rao}, {Rawlings}, {Raymond}, {Rezzolla},
  {Ripperda}, {Roelofs}, {Rogers}, {Ros}, {Rose}, {Roshanineshat}, {Rottmann},
  {Roy}, {Ruszczyk}, {Ryan}, {Rygl}, {S{\'a}nchez}, {S{\'a}nchez-Arguelles},
  {Sasada}, {Savolainen}, {Schloerb}, {Schuster}, {Shao}, {Shen}, {Small},
  {Sohn}, {SooHoo}, {Tazaki}, {Tiede}, {Tilanus}, {Titus}, {Toma}, {Torne},
  {Trent}, {Trippe}, {Tsuda}, {van Bemmel}, {van Langevelde}, {van Rossum},
  {Wagner}, {Wardle}, {Weintroub}, {Wex}, {Wharton}, {Wielgus}, {Wong}, {Wu},
  {Young}, {Young}, {Younsi}, {Yuan}, {Yuan}, {Zensus}, {Zhao}, {Zhao}, {Zhu},
  {Farah}, {Meyer-Zhao}, {Michalik}, {Nadolski}, {Nishioka}, {Pradel},
  {Primiani}, {Souccar}, {Vertatschitsch}, \& {Yamaguchi}}]{EHTC2019_5}
{Event Horizon Telescope Collaboration}, {Akiyama}, K., {Alberdi}, A., {et~al.}
  2019{\natexlab{e}}, \apjl, 875, L6

\bibitem[{{Event Horizon Telescope Collaboration}
  {et~al.}(2019{\natexlab{f}}){Event Horizon Telescope Collaboration},
  {Akiyama}, {Alberdi}, {Alef}, {Asada}, {Azulay}, {Baczko}, {Ball},
  {Balokovi{\'c}}, {Barrett}, {Bintley}, {Blackburn}, {Boland}, {Bouman},
  {Bower}, {Bremer}, {Brinkerink}, {Brissenden}, {Britzen}, {Broderick},
  {Broguiere}, {Bronzwaer}, {Byun}, {Carlstrom}, {Chael}, {Chan}, {Chatterjee},
  {Chatterjee}, {Chen}, {Chen}, {Cho}, {Christian}, {Conway}, {Cordes}, {Crew},
  {Cui}, {Davelaar}, {De Laurentis}, {Deane}, {Dempsey}, {Desvignes}, {Dexter},
  {Doeleman}, {Eatough}, {Falcke}, {Fish}, {Fomalont}, {Fraga-Encinas},
  {Friberg}, {Fromm}, {G{\'o}mez}, {Galison}, {Gammie}, {Garc{\'\i}a},
  {Gentaz}, {Georgiev}, {Goddi}, {Gold}, {Gu}, {Gurwell}, {Hada}, {Hecht},
  {Hesper}, {Ho}, {Ho}, {Honma}, {Huang}, {Huang}, {Hughes}, {Ikeda}, {Inoue},
  {Issaoun}, {James}, {Jannuzi}, {Janssen}, {Jeter}, {Jiang}, {Johnson},
  {Jorstad}, {Jung}, {Karami}, {Karuppusamy}, {Kawashima}, {Keating},
  {Kettenis}, {Kim}, {Kim}, {Kim}, {Kino}, {Koay}, {Koch}, {Koyama}, {Kramer},
  {Kramer}, {Krichbaum}, {Kuo}, {Lauer}, {Lee}, {Li}, {Li}, {Lindqvist}, {Liu},
  {Liuzzo}, {Lo}, {Lobanov}, {Loinard}, {Lonsdale}, {Lu}, {MacDonald}, {Mao},
  {Markoff}, {Marrone}, {Marscher}, {Mart{\'\i}-Vidal}, {Matsushita},
  {Matthews}, {Medeiros}, {Menten}, {Mizuno}, {Mizuno}, {Moran}, {Moriyama},
  {Moscibrodzka}, {Mul{\ensuremath{\ddot{}}}ler}, {Nagai}, {Nagar}, {Nakamura},
  {Narayan}, {Narayanan}, {Natarajan}, {Neri}, {Ni}, {Noutsos}, {Okino},
  {Olivares}, {Oyama}, {{\"O}zel}, {Palumbo}, {Patel}, {Pen}, {Pesce},
  {Pi{\'e}tu}, {Plambeck}, {PopStefanija}, {Porth}, {Prather},
  {Preciado-L{\'o}pez}, {Psaltis}, {Pu}, {Ramakrishnan}, {Rao}, {Rawlings},
  {Raymond}, {Rezzolla}, {Ripperda}, {Roelofs}, {Rogers}, {Ros}, {Rose},
  {Roshanineshat}, {Rottmann}, {Roy}, {Ruszczyk}, {Ryan}, {Rygl},
  {S{\'a}nchez}, {S{\'a}nchez-Arguelles}, {Sasada}, {Savolainen}, {Schloerb},
  {Schuster}, {Shao}, {Shen}, {Small}, {Sohn}, {SooHoo}, {Tazaki}, {Tiede},
  {Tilanus}, {Titus}, {Toma}, {Torne}, {Trent}, {Trippe}, {Tsuda}, {van
  Bemmel}, {van Langevelde}, {van Rossum}, {Wagner}, {Wardle}, {Weintroub},
  {Wex}, {Wharton}, {Wielgus}, {Wong}, {Wu}, {Young}, {Young}, {Younsi},
  {Yuan}, {Yuan}, {Zensus}, {Zhao}, {Zhao}, {Zhu}, {Anczarski}, {Baganoff},
  {Eckart}, {Farah}, {Haggard}, {Meyer-Zhao}, {Michalik}, {Nadolski},
  {Neilsen}, {Nishioka}, {Nowak}, {Pradel}, {Primiani}, {Souccar},
  {Vertatschitsch}, {Yamaguchi}, \& {Zhang}}]{EHTC2019_6}
{Event Horizon Telescope Collaboration}, {Akiyama}, K., {Alberdi}, A., {et~al.}
  2019{\natexlab{f}}, \apjl, 875, L5

\bibitem[{{Event Horizon Telescope Collaboration}
  {et~al.}(2022{\natexlab{f}}){Event Horizon Telescope Collaboration},
  {Akiyama}, {Alberdi}, {Alef}, {Carlos Algaba}, {Anantua}, {Asada}, {Azulay},
  {Bach}, {Baczko}, {Ball}, {Balokovi{\'c}}, {Barrett}, {Baub{\"o}ck},
  {Benson}, {Bintley}, {Blackburn}, {Blundell}, {Bouman}, {Bower}, {Boyce},
  {Bremer}, {Brinkerink}, {Brissenden}, {Britzen}, {Broderick}, {Broguiere},
  {Bronzwaer}, {Bustamante}, {Byun}, {Carlstrom}, {Ceccobello}, {Chael},
  {Chan}, {Chatterjee}, {Chatterjee}, {Chen}, {Chen}, {Cheng}, {Cho},
  {Christian}, {Conroy}, {Conway}, {Cordes}, {Crawford}, {Crew}, {Cruz-Osorio},
  {Cui}, {Davelaar}, {De Laurentis}, {Deane}, {Dempsey}, {Desvignes}, {Dexter},
  {Dhruv}, {Doeleman}, {Dougal}, {Dzib}, {Eatough}, {Emami}, {Falcke}, {Farah},
  {Fish}, {Fomalont}, {Ford}, {Fraga-Encinas}, {Freeman}, {Friberg}, {Fromm},
  {Fuentes}, {Galison}, {Gammie}, {Garc{\'\i}a}, {Gentaz}, {Georgiev}, {Goddi},
  {Gold}, {G{\'o}mez-Ruiz}, {G{\'o}mez}, {Gu}, {Gurwell}, {Hada}, {Haggard},
  {Haworth}, {Hecht}, {Hesper}, {Heumann}, {Ho}, {Ho}, {Honma}, {Huang},
  {Huang}, {Hughes}, {Ikeda}, {Violette Impellizzeri}, {Inoue}, {Issaoun},
  {James}, {Jannuzi}, {Janssen}, {Jeter}, {Jiang}, {Jim{\'e}nez-Rosales},
  {Johnson}, {Jorstad}, {Joshi}, {Jung}, {Karami}, {Karuppusamy}, {Kawashima},
  {Keating}, {Kettenis}, {Kim}, {Kim}, {Kim}, {Kim}, {Kino}, {Koay},
  {Kocherlakota}, {Kofuji}, {Koch}, {Koyama}, {Kramer}, {Kramer}, {Krichbaum},
  {Kuo}, {Bella}, {Lauer}, {Lee}, {Lee}, {Leung}, {Levis}, {Li}, {Lico},
  {Lindahl}, {Lindqvist}, {Lisakov}, {Liu}, {Liu}, {Liuzzo}, {Lo}, {Lobanov},
  {Loinard}, {Lonsdale}, {Lu}, {Mao}, {Marchili}, {Markoff}, {Marrone},
  {Marscher}, {Mart{\'\i}-Vidal}, {Matsushita}, {Matthews}, {Medeiros},
  {Menten}, {Michalik}, {Mizuno}, {Mizuno}, {Moran}, {Moriyama},
  {Moscibrodzka}, {M{\"u}ller}, {Mus}, {Musoke}, {Myserlis}, {Nadolski},
  {Nagai}, {Nagar}, {Nakamura}, {Narayan}, {Narayanan}, {Natarajan},
  {Nathanail}, {Navarro Fuentes}, {Neilsen}, {Neri}, {Ni}, {Noutsos}, {Nowak},
  {Oh}, {Okino}, {Olivares}, {Ortiz-Le{\'o}n}, {Oyama}, {{\"O}zel}, {Palumbo},
  {Filippos Paraschos}, {Park}, {Parsons}, {Patel}, {Pen}, {Pesce},
  {Pi{\'e}tu}, {Plambeck}, {PopStefanija}, {Porth}, {P{\"o}tzl}, {Prather},
  {Preciado-L{\'o}pez}, {Psaltis}, {Pu}, {Ramakrishnan}, {Rao}, {Rawlings},
  {Raymond}, {Rezzolla}, {Ricarte}, {Ripperda}, {Roelofs}, {Rogers}, {Ros},
  {Romero-Ca{\~n}izales}, {Roshanineshat}, {Rottmann}, {Roy}, {Ruiz},
  {Ruszczyk}, {Rygl}, {S{\'a}nchez}, {S{\'a}nchez-Arg{\"u}elles},
  {S{\'a}nchez-Portal}, {Sasada}, {Satapathy}, {Savolainen}, {Schloerb},
  {Schonfeld}, {Schuster}, {Shao}, {Shen}, {Small}, {Sohn}, {SooHoo},
  {Souccar}, {Sun}, {Tazaki}, {Tetarenko}, {Tiede}, {Tilanus}, {Titus},
  {Torne}, {Traianou}, {Trent}, {Trippe}, {Turk}, {van Bemmel}, {van
  Langevelde}, {van Rossum}, {Vos}, {Wagner}, {Ward-Thompson}, {Wardle},
  {Weintroub}, {Wex}, {Wharton}, {Wielgus}, {Wiik}, {Witzel}, {Wondrak},
  {Wong}, {Wu}, {Yamaguchi}, {Yoon}, {Young}, {Young}, {Younsi}, {Yuan},
  {Yuan}, {Zensus}, {Zhang}, {Zhao}, {Zhao}, {Chan}, {Qiu}, {Ressler}, \&
  {White}}]{EHTC2022_5}
{Event Horizon Telescope Collaboration}, {Akiyama}, K., {Alberdi}, A., {et~al.}
  2022{\natexlab{f}}, \apjl, 930, L16

\bibitem[{{Event Horizon Telescope Collaboration}
  {et~al.}(2021{\natexlab{a}}){Event Horizon Telescope Collaboration},
  {Akiyama}, {Algaba}, {Alberdi}, {Alef}, {Anantua}, {Asada}, {Azulay},
  {Baczko}, {Ball}, {Balokovi{\'c}}, {Barrett}, {Benson}, {Bintley},
  {Blackburn}, {Blundell}, {Boland}, {Bouman}, {Bower}, {Boyce}, {Bremer},
  {Brinkerink}, {Brissenden}, {Britzen}, {Broderick}, {Broguiere}, {Bronzwaer},
  {Byun}, {Carlstrom}, {Chael}, {Chan}, {Chatterjee}, {Chatterjee}, {Chen},
  {Chen}, {Chesler}, {Cho}, {Christian}, {Conway}, {Cordes}, {Crawford},
  {Crew}, {Cruz-Osorio}, {Cui}, {Davelaar}, {De Laurentis}, {Deane}, {Dempsey},
  {Desvignes}, {Dexter}, {Doeleman}, {Eatough}, {Falcke}, {Farah}, {Fish},
  {Fomalont}, {Ford}, {Fraga-Encinas}, {Freeman}, {Friberg}, {Fromm},
  {Fuentes}, {Galison}, {Gammie}, {Garc{\'\i}a}, {Gentaz}, {Georgiev}, {Goddi},
  {Gold}, {G{\'o}mez}, {G{\'o}mez-Ruiz}, {Gu}, {Gurwell}, {Hada}, {Haggard},
  {Hecht}, {Hesper}, {Ho}, {Ho}, {Honma}, {Huang}, {Huang}, {Hughes}, {Ikeda},
  {Inoue}, {Issaoun}, {James}, {Jannuzi}, {Janssen}, {Jeter}, {Jiang},
  {Jimenez-Rosales}, {Johnson}, {Jorstad}, {Jung}, {Karami}, {Karuppusamy},
  {Kawashima}, {Keating}, {Kettenis}, {Kim}, {Kim}, {Kim}, {Kim}, {Kino},
  {Koay}, {Kofuji}, {Koch}, {Koyama}, {Kramer}, {Kramer}, {Krichbaum}, {Kuo},
  {Lauer}, {Lee}, {Levis}, {Li}, {Li}, {Lindqvist}, {Lico}, {Lindahl}, {Liu},
  {Liu}, {Liuzzo}, {Lo}, {Lobanov}, {Loinard}, {Lonsdale}, {Lu}, {MacDonald},
  {Mao}, {Marchili}, {Markoff}, {Marrone}, {Marscher}, {Mart{\'\i}-Vidal},
  {Matsushita}, {Matthews}, {Medeiros}, {Menten}, {Mizuno}, {Mizuno}, {Moran},
  {Moriyama}, {Moscibrodzka}, {M{\"u}ller}, {Musoke}, {Mej{\'\i}as},
  {Michalik}, {Nadolski}, {Nagai}, {Nagar}, {Nakamura}, {Narayan}, {Narayanan},
  {Natarajan}, {Nathanail}, {Neilsen}, {Neri}, {Ni}, {Noutsos}, {Nowak},
  {Okino}, {Olivares}, {Ortiz-Le{\'o}n}, {Oyama}, {{\"O}zel}, {Palumbo},
  {Park}, {Patel}, {Pen}, {Pesce}, {Pi{\'e}tu}, {Plambeck}, {PopStefanija},
  {Porth}, {P{\"o}tzl}, {Prather}, {Preciado-L{\'o}pez}, {Psaltis}, {Pu},
  {Ramakrishnan}, {Rao}, {Rawlings}, {Raymond}, {Rezzolla}, {Ricarte},
  {Ripperda}, {Roelofs}, {Rogers}, {Ros}, {Rose}, {Roshanineshat}, {Rottmann},
  {Roy}, {Ruszczyk}, {Rygl}, {S{\'a}nchez}, {S{\'a}nchez-Arguelles}, {Sasada},
  {Savolainen}, {Schloerb}, {Schuster}, {Shao}, {Shen}, {Small}, {Sohn},
  {SooHoo}, {Sun}, {Tazaki}, {Tetarenko}, {Tiede}, {Tilanus}, {Titus}, {Toma},
  {Torne}, {Trent}, {Traianou}, {Trippe}, {van Bemmel}, {van Langevelde}, {van
  Rossum}, {Wagner}, {Ward-Thompson}, {Wardle}, {Weintroub}, {Wex}, {Wharton},
  {Wielgus}, {Wong}, {Wu}, {Yoon}, {Young}, {Young}, {Younsi}, {Yuan}, {Yuan},
  {Zensus}, {Zhao}, \& {Zhao}}]{EHT2021_1}
{Event Horizon Telescope Collaboration}, {Akiyama}, K., {Algaba}, J.~C.,
  {et~al.} 2021{\natexlab{a}}, \apjl, 910, L12

\bibitem[{{Event Horizon Telescope Collaboration}
  {et~al.}(2021{\natexlab{b}}){Event Horizon Telescope Collaboration},
  {Akiyama}, {Algaba}, {Alberdi}, {Alef}, {Anantua}, {Asada}, {Azulay},
  {Baczko}, {Ball}, {Balokovi{\'c}}, {Barrett}, {Benson}, {Bintley},
  {Blackburn}, {Blundell}, {Boland}, {Bouman}, {Bower}, {Boyce}, {Bremer},
  {Brinkerink}, {Brissenden}, {Britzen}, {Broderick}, {Broguiere}, {Bronzwaer},
  {Byun}, {Carlstrom}, {Chael}, {Chan}, {Chatterjee}, {Chatterjee}, {Chen},
  {Chen}, {Chesler}, {Cho}, {Christian}, {Conway}, {Cordes}, {Crawford},
  {Crew}, {Cruz-Osorio}, {Cui}, {Davelaar}, {De Laurentis}, {Deane}, {Dempsey},
  {Desvignes}, {Dexter}, {Doeleman}, {Eatough}, {Falcke}, {Farah}, {Fish},
  {Fomalont}, {Ford}, {Fraga-Encinas}, {Friberg}, {Fromm}, {Fuentes},
  {Galison}, {Gammie}, {Garc{\'\i}a}, {Gelles}, {Gentaz}, {Georgiev}, {Goddi},
  {Gold}, {G{\'o}mez}, {G{\'o}mez-Ruiz}, {Gu}, {Gurwell}, {Hada}, {Haggard},
  {Hecht}, {Hesper}, {Himwich}, {Ho}, {Ho}, {Honma}, {Huang}, {Huang},
  {Hughes}, {Ikeda}, {Inoue}, {Issaoun}, {James}, {Jannuzi}, {Janssen},
  {Jeter}, {Jiang}, {Jimenez-Rosales}, {Johnson}, {Jorstad}, {Jung}, {Karami},
  {Karuppusamy}, {Kawashima}, {Keating}, {Kettenis}, {Kim}, {Kim}, {Kim},
  {Kim}, {Kino}, {Koay}, {Kofuji}, {Koch}, {Koyama}, {Kramer}, {Kramer},
  {Krichbaum}, {Kuo}, {Lauer}, {Lee}, {Levis}, {Li}, {Li}, {Lindqvist}, {Lico},
  {Lindahl}, {Liu}, {Liu}, {Liuzzo}, {Lo}, {Lobanov}, {Loinard}, {Lonsdale},
  {Lu}, {MacDonald}, {Mao}, {Marchili}, {Markoff}, {Marrone}, {Marscher},
  {Mart{\'\i}-Vidal}, {Matsushita}, {Matthews}, {Medeiros}, {Menten}, {Mizuno},
  {Mizuno}, {Moran}, {Moriyama}, {Moscibrodzka}, {M{\"u}ller}, {Musoke}, {Mus
  Mej{\'\i}as}, {Michalik}, {Nadolski}, {Nagai}, {Nagar}, {Nakamura},
  {Narayan}, {Narayanan}, {Natarajan}, {Nathanail}, {Neilsen}, {Neri}, {Ni},
  {Noutsos}, {Nowak}, {Okino}, {Olivares}, {Ortiz-Le{\'o}n}, {Oyama},
  {{\"O}zel}, {Palumbo}, {Park}, {Patel}, {Pen}, {Pesce}, {Pi{\'e}tu},
  {Plambeck}, {PopStefanija}, {Porth}, {P{\"o}tzl}, {Prather},
  {Preciado-L{\'o}pez}, {Psaltis}, {Pu}, {Ramakrishnan}, {Rao}, {Rawlings},
  {Raymond}, {Rezzolla}, {Ricarte}, {Ripperda}, {Roelofs}, {Rogers}, {Ros},
  {Rose}, {Roshanineshat}, {Rottmann}, {Roy}, {Ruszczyk}, {Rygl},
  {S{\'a}nchez}, {S{\'a}nchez-Arguelles}, {Sasada}, {Savolainen}, {Schloerb},
  {Schuster}, {Shao}, {Shen}, {Small}, {Sohn}, {SooHoo}, {Sun}, {Tazaki},
  {Tetarenko}, {Tiede}, {Tilanus}, {Titus}, {Toma}, {Torne}, {Trent},
  {Traianou}, {Trippe}, {van Bemmel}, {van Langevelde}, {van Rossum}, {Wagner},
  {Ward-Thompson}, {Wardle}, {Weintroub}, {Wex}, {Wharton}, {Wielgus}, {Wong},
  {Wu}, {Yoon}, {Young}, {Young}, {Younsi}, {Yuan}, {Yuan}, {Zensus}, {Zhao},
  \& {Zhao}}]{EHT2021_2}
{Event Horizon Telescope Collaboration}, {Akiyama}, K., {Algaba}, J.~C.,
  {et~al.} 2021{\natexlab{b}}, \apjl, 910, L13

\bibitem[{{Gabuzda} {et~al.}(2017){Gabuzda}, {Roche}, {Kirwan}, {Knuettel},
  {Nagle}, \& {Houston}}]{Gabuzda2017}
{Gabuzda}, D.~C., {Roche}, N., {Kirwan}, A., {et~al.} 2017, \mnras, 472, 1792

\bibitem[{{Giroletti} {et~al.}(2012){Giroletti}, {Hada}, {Giovannini},
  {Casadio}, {Beilicke}, {Cesarini}, {Cheung}, {Doi}, {Krawczynski}, {Kino},
  {Lee}, \& {Nagai}}]{Giroletti2012}
{Giroletti}, M., {Hada}, K., {Giovannini}, G., {et~al.} 2012, \aap, 538, L10

\bibitem[{{Goddi} {et~al.}(2019{\natexlab{a}}){Goddi}, {Crew}, {Impellizzeri},
  {Mart{\'\i}-Vidal}, {Matthews}, {Messias}, {Rottmann}, {Alef}, {Blackburn},
  {Bronzwaer}, {Chan}, {Davelaar}, {Deane}, {Dexter}, {Doeleman}, {Falcke},
  {Fish}, {Fraga-Encinas}, {Fromm}, {Herrero-Illana}, {Issaoun}, {James},
  {Janssen}, {Kramer}, {Krichbaum}, {De Laurentis}, {Liuzzo}, {Mizuno},
  {Moscibrodzka}, {Natarajan}, {Porth}, {Rezzolla}, {Rygl}, {Roelofs}, {Ros},
  {Roy}, {Shao}, {van Langevelde}, {van Bemmel}, {Tilanus}, {Torne}, {Wielgus},
  {Younsi}, {Zensus}, \& {Event Horizon Telescope Collaboration}}]{Goddi2019}
{Goddi}, C., {Crew}, G., {Impellizzeri}, V., {et~al.} 2019{\natexlab{a}}, The
  Messenger, 177, 25

\bibitem[{{Goddi} {et~al.}(2021){Goddi}, {Mart{\'\i}-Vidal}, {Messias},
  {Bower}, {Broderick}, {Dexter}, {Marrone}, {Moscibrodzka}, {Nagai}, {Algaba},
  {Asada}, {Crew}, {G{\'o}mez}, {Impellizzeri}, {Janssen}, {Kadler},
  {Krichbaum}, {Lico}, {Matthews}, {Nathanail}, {Ricarte}, {Ros}, {Younsi},
  {Akiyama}, {Alberdi}, {Alef}, {Anantua}, {Azulay}, {Baczko}, {Ball},
  {Balokovi{\'c}}, {Barrett}, {Benson}, {Bintley}, {Blackburn}, {Blundell},
  {Boland}, {Bouman}, {Boyce}, {Bremer}, {Brinkerink}, {Brissenden}, {Britzen},
  {Broguiere}, {Bronzwaer}, {Byun}, {Carlstrom}, {Chael}, {Chan}, {Chatterjee},
  {Chatterjee}, {Chen}, {Chen}, {Chesler}, {Cho}, {Christian}, {Conway},
  {Cordes}, {Crawford}, {Cruz-Osorio}, {Cui}, {Davelaar}, {De Laurentis},
  {Deane}, {Dempsey}, {Desvignes}, {Doeleman}, {Eatough}, {Falcke}, {Farah},
  {Fish}, {Fomalont}, {Ford}, {Fraga-Encinas}, {Freeman}, {Friberg}, {Fromm},
  {Fuentes}, {Galison}, {Gammie}, {Garc{\'\i}a}, {Gentaz}, {Georgiev}, {Gold},
  {G{\'o}mez-Ruiz}, {Gu}, {Gurwell}, {Hada}, {Haggard}, {Hecht}, {Hesper},
  {Ho}, {Ho}, {Honma}, {Huang}, {Huang}, {Hughes}, {Inoue}, {Issaoun}, {James},
  {Jannuzi}, {Jeter}, {Jiang}, {Jimenez-Rosales}, {Johnson}, {Jorstad}, {Jung},
  {Karami}, {Karuppusamy}, {Kawashima}, {Keating}, {Kettenis}, {Kim}, {Kim},
  {Kim}, {Kim}, {Kino}, {Koay}, {Kofuji}, {Koch}, {Koyama}, {Kramer}, {Kramer},
  {Kuo}, {Lauer}, {Lee}, {Levis}, {Li}, {Li}, {Lindqvist}, {Lindahl}, {Liu},
  {Liu}, {Liuzzo}, {Lo}, {Lobanov}, {Loinard}, {Lonsdale}, {Lu}, {MacDonald},
  {Mao}, {Marchili}, {Markoff}, {Marscher}, {Matsushita}, {Medeiros}, {Menten},
  {Mizuno}, {Mizuno}, {Moran}, {Moriyama}, {M{\"u}ller}, {Musoke},
  {Mej{\'\i}as}, {Nagar}, {Nakamura}, {Narayan}, {Narayanan}, {Natarajan},
  {Neilsen}, {Neri}, {Ni}, {Noutsos}, {Nowak}, {Okino}, {Olivares},
  {Ortiz-Le{\'o}n}, {Oyama}, {{\"O}zel}, {Palumbo}, {Park}, {Patel}, {Pen},
  {Pesce}, {Pi{\'e}tu}, {Plambeck}, {PopStefanija}, {Porth}, {P{\"o}tzl},
  {Prather}, {Preciado-L{\'o}pez}, {Psaltis}, {Pu}, {Ramakrishnan}, {Rao},
  {Rawlings}, {Raymond}, {Rezzolla}, {Ripperda}, {Roelofs}, {Rogers}, {Rose},
  {Roshanineshat}, {Rottmann}, {Roy}, {Ruszczyk}, {Rygl}, {S{\'a}nchez},
  {S{\'a}nchez-Arguelles}, {Sasada}, {Savolainen}, {Schloerb}, {Schuster},
  {Shao}, {Shen}, {Small}, {Sohn}, {SooHoo}, {Sun}, {Tazaki}, {Tetarenko},
  {Tiede}, {Tilanus}, {Titus}, {Toma}, {Torne}, {Trent}, {Traianou}, {Trippe},
  {van Bemmel}, {van Langevelde}, {van Rossum}, {Wagner}, {Ward-Thompson},
  {Wardle}, {Weintroub}, {Wex}, {Wharton}, {Wielgus}, {Wong}, {Wu}, {Yoon},
  {Young}, {Young}, {Yuan}, {Yuan}, {Zensus}, {Zhao}, {Zhao}, {Bruni},
  {Gopakumar}, {Hern{\'a}ndez-G{\'o}mez}, {Herrero-Illana}, {Ingram},
  {Komossa}, {Kovalev}, {Muders}, {Perucho}, {R{\"o}sch}, \&
  {Valtonen}}]{Goddi2021}
{Goddi}, C., {Mart{\'\i}-Vidal}, I., {Messias}, H., {et~al.} 2021, \apjl, 910,
  L14

\bibitem[{{Goddi} {et~al.}(2019{\natexlab{b}}){Goddi}, {Mart{\'\i}-Vidal},
  {Messias}, {Crew}, {Herrero-Illana}, {Impellizzeri}, {Rottmann}, {Wagner},
  {Fomalont}, \& {Matthews}}]{QA2Paper}
{Goddi}, C., {Mart{\'\i}-Vidal}, I., {Messias}, H., {et~al.}
  2019{\natexlab{b}}, \pasp, 131, 075003

\bibitem[{{Hovatta} {et~al.}(2019){Hovatta}, {O'Sullivan},
  {Mart{\'{\i}}-Vidal}, {Savolainen}, \& {Tchekhovskoy}}]{Hovatta2019}
{Hovatta}, T., {O'Sullivan}, S., {Mart{\'{\i}}-Vidal}, I., {Savolainen}, T., \&
  {Tchekhovskoy}, A. 2019, \aap, 623, A111

\bibitem[{{Kravchenko} {et~al.}(2020){Kravchenko}, {Giroletti}, {Hada},
  {Meier}, {Nakamura}, {Park}, \& {Walker}}]{Kravchenko2020}
{Kravchenko}, E., {Giroletti}, M., {Hada}, K., {et~al.} 2020, \aap, 637, L6

\bibitem[{{Laing}(1980)}]{Laing1980}
{Laing}, R.~A. 1980, \mnras, 193, 439

\bibitem[{{Lee} {et~al.}(2015){Lee}, {Kang}, {Byun}, {Chapman}, {Novak},
  {Trippe}, {Algaba}, \& {Kino}}]{Lee2015}
{Lee}, S.-S., {Kang}, S., {Byun}, D.-Y., {et~al.} 2015, \apjl, 808, L26

\bibitem[{{Mart{\'\i}-Vidal} {et~al.}(2015){Mart{\'\i}-Vidal}, {Muller},
  {Vlemmings}, {Horellou}, \& {Aalto}}]{IMV2015}
{Mart{\'\i}-Vidal}, I., {Muller}, S., {Vlemmings}, W., {Horellou}, C., \&
  {Aalto}, S. 2015, Science, 348, 311

\bibitem[{{Mart{\'{\i}}-Vidal} {et~al.}(2016){Mart{\'{\i}}-Vidal}, {Roy},
  {Conway}, \& {Zensus}}]{PCPaper}
{Mart{\'{\i}}-Vidal}, I., {Roy}, A., {Conway}, J., \& {Zensus}, A.~J. 2016,
  \aap, 587, A143

\bibitem[{{Mart{\'\i}-Vidal} {et~al.}(2014){Mart{\'\i}-Vidal}, {Vlemmings},
  {Muller}, \& {Casey}}]{UVMULTIFIT}
{Mart{\'\i}-Vidal}, I., {Vlemmings}, W.~H.~T., {Muller}, S., \& {Casey}, S.
  2014, \aap, 563, A136

\bibitem[{{Matthews} \& {Crew}(2024)}]{app2_report_2024}
{Matthews}, L.~D. \& {Crew}, G.~B. 2024, {Enabling New Science with the ALMA
  Phasing System -- Phase 2 (APP2)},
  \href{https://science.nrao.edu/facilities/alma/science_sustainability/APP2_Final_Report_v4.1.pdf}{Version
  4.1}

\bibitem[{{Matthews} {et~al.}(2018){Matthews}, {Crew}, {Doeleman}, {Lacasse},
  {Saez}, {Alef}, {Akiyama}, {Amestica}, {Anderson}, {Barkats}, {Baudry},
  {Brogui{\`e}re}, {Escoffier}, {Fish}, {Greenberg}, {Hecht}, {Hiriart},
  {Hirota}, {Honma}, {Ho}, {Impellizzeri}, {Inoue}, {Kohno}, {Lopez},
  {Mart{\'{\i}}-Vidal}, {Messias}, {Meyer-Zhao}, {Mora-Klein}, {Nagar},
  {Nishioka}, {Oyama}, {Pankratius}, {Perez}, {Phillips}, {Pradel}, {Rottmann},
  {Roy}, {Ruszczyk}, {Shillue}, {Suzuki}, \& {Treacy}}]{APPPaper}
{Matthews}, L.~D., {Crew}, G.~B., {Doeleman}, S.~S., {et~al.} 2018, \pasp, 130,
  015002

\bibitem[{{Mizuno} {et~al.}(2015){Mizuno}, {G{\'o}mez}, {Nishikawa}, {Meli},
  {Hardee}, \& {Rezzolla}}]{Mizuno2015}
{Mizuno}, Y., {G{\'o}mez}, J.~L., {Nishikawa}, K.-I., {et~al.} 2015, \apj, 809,
  38

\bibitem[{{Myserlis} \& {Contopoulos}(2021)}]{Myserlis2021}
{Myserlis}, I. \& {Contopoulos}, I. 2021, \aap, 649, A94

\bibitem[{{Nagai} {et~al.}(2016){Nagai}, {Nakanishi}, {Paladino}, {Hull},
  {Cortes}, {Moellenbrock}, {Fomalont}, {Asada}, \& {Hada}}]{Nagai2016}
{Nagai}, H., {Nakanishi}, K., {Paladino}, R., {et~al.} 2016, \apj, 824, 132

\bibitem[{{Pasetto} {et~al.}(2021){Pasetto}, {Carrasco-Gonz{\'a}lez},
  {G{\'o}mez}, {Mart{\'\i}}, {Perucho}, {O'Sullivan}, {Anderson},
  {D{\'\i}az-Gonz{\'a}lez}, {Fuentes}, \& {Wardle}}]{Pasetto2021}
{Pasetto}, A., {Carrasco-Gonz{\'a}lez}, C., {G{\'o}mez}, J.~L., {et~al.} 2021,
  \apjl, 923, L5

\bibitem[{{Peng} {et~al.}(2024){Peng}, {Lu}, {Goddi}, {Krichbaum}, {Li}, {Liu},
  {Kim}, {Nakamura}, {Yuan}, {Chen}, {Mart{\'\i}-Vidal}, \& {Shen}}]{Peng2024}
{Peng}, S., {Lu}, R.-S., {Goddi}, C., {et~al.} 2024, \apj, 975, 103

\bibitem[{{Plambeck} {et~al.}(2014){Plambeck}, {Bower}, {Rao}, {Marrone},
  {Jorstad}, {Marscher}, {Doeleman}, {Fish}, \& {Johnson}}]{Plambeck2014}
{Plambeck}, R.~L., {Bower}, G.~C., {Rao}, R., {et~al.} 2014, \apj, 797, 66

\bibitem[{{Raymond} {et~al.}(2024){Raymond}, {Doeleman}, {Asada}, {Blackburn},
  {Bower}, {Bremer}, {Broguiere}, {Chen}, {Crew}, {Dornbusch}, {Fish},
  {Garc{\'\i}a}, {Gentaz}, {Goddi}, {Han}, {Hecht}, {Huang}, {Janssen},
  {Keating}, {Koay}, {Krichbaum}, {Lo}, {Matsushita}, {Matthews}, {Moran},
  {Norton}, {Patel}, {Pesce}, {Ramakrishnan}, {Rottmann}, {Roy}, {S{\'a}nchez},
  {Tilanus}, {Titus}, {Torne}, {Wagner}, {Weintroub}, {Wielgus}, {Young},
  {Akiyama}, {Albentosa-Ru{\'\i}z}, {Alberdi}, {Alef}, {Algaba}, {Anantua},
  {Azulay}, {Bach}, {Baczko}, {Ball}, {Balokovic}, {Bandyopadhyay}, {Barrett},
  {Baub{\"o}ck}, {Benson}, {Bintley}, {Blundell}, {Bouman}, {Boyce},
  {Brissenden}, {Britzen}, {Broderick}, {Bronzwaer}, {Bustamante}, {Carlstrom},
  {Chael}, {Chan}, {Chang}, {Chatterjee}, {Chatterjee}, {Chen}, {Cheng}, {Cho},
  {Christian}, {Conroy}, {Conway}, {Crawford}, {Cruz-Osorio}, {Cui}, {Dahale},
  {Davelaar}, {De Laurentis}, {Deane}, {Dempsey}, {Desvignes}, {Dexter},
  {Dhruv}, {Dihingia}, {Dzib}, {Eatough}, {Emami}, {Falcke}, {Farah},
  {Fomalont}, {Fontana}, {Ford}, {Foschi}, {Fraga-Encinas}, {Freeman},
  {Friberg}, {Fromm}, {Fuentes}, {Galison}, {Gammie}, {Georgiev}, {Gold},
  {G{\'o}mez-Ruiz}, {G{\'o}mez}, {Gu}, {Gurwell}, {Hada}, {Haggard}, {Hesper},
  {Heumann}, {Ho}, {Ho}, {Honma}, {Huang}, {Huang}, {Hughes}, {Ikeda},
  {Impellizzeri}, {Inoue}, {Issaoun}, {James}, {Jannuzi}, {Jeter}, {Jiang},
  {Jim{\'e}nez-Rosales}, {Johnson}, {Jorstad}, {Jones}, {Joshi}, {Jung},
  {Karuppusamy}, {Kawashima}, {Kettenis}, {Kim}, {Kim}, {Kim}, {Kim}, {Kino},
  {Kocherlakota}, {Kofuji}, {Koch}, {Koyama}, {Kramer}, {Kramer}, {Kramer},
  {Kubo}, {Kuo}, {La Bella}, {Lee}, {Levis}, {Li}, {Lico}, {Lindahl},
  {Lindqvist}, {Lisakov}, {Liu}, {Liu}, {Liuzzo}, {Lobanov}, {Loinard},
  {Lonsdale}, {Lowitz}, {Lu}, {MacDonald}, {Mahieu}, {Maier}, {Mao},
  {Marchili}, {Markoff}, {Marrone}, {Marscher}, {Mart{\'\i}-Vidal}, {Medeiros},
  {Menten}, {Mizuno}, {Mizuno}, {Montgomery}, {Moriyama}, {Moscibrodzka},
  {Mulaudzi}, {M{\"u}ller}, {M{\"u}ller}, {Mus}, {Musoke}, {Myserlis}, {Nagai},
  {Nagar}, {Nakamura}, {Narayanan}, {Natarajan}, {Nathanail}, {Navarro
  Fuentes}, {Neilsen}, {Ni}, {Nowak}, {Oh}, {Okino}, {Olivares S{\'a}nchez},
  {Oyama}, {{\"O}zel}, {Palumbo}, {Paraschos}, {Park}, {Parsons}, {Pen},
  {Pi{\'e}tu}, {PopStefanija}, {Porth}, {Prather}, {Principe}, {Psaltis}, {Pu},
  {Raffin}, {Rao}, {Rawlings}, {Ricarte}, {Ripperda}, {Roelofs},
  {Romero-Ca{\~n}izales}, {Ros}, {Roshanineshat}, {Ruiz}, {Ruszczyk}, {Rygl},
  {S{\'a}nchez-Arg{\"u}elles}, {S{\'a}nchez-Portal}, {Sasada}, {Satapathy},
  {Savolainen}, {Schloerb}, {Schonfeld}, {Schuster}, {Shao}, {Shen}, {Small},
  {Sohn}, {SooHoo}, {Sosapanta Salas}, {Souccar}, {Stanway}, {Sun}, {Tazaki},
  {Tetarenko}, {Tiede}, {Toma}, {Toscano}, {Traianou}, {Trent}, {Trippe},
  {Turk}, {van Bemmel}, {van Langevelde}, {van Rossum}, {Vos}, {Ward-Thompson},
  {Wardle}, {Washington}, {Wharton}, {Wiik}, {Witzel}, {Wondrak}, {Wong}, {Wu},
  {Yadlapalli}, {Yamaguchi}, {Yfantis}, {Yoon}, {Younsi}, {Yu}, {Yuan}, {Yuan},
  {Zensus}, {Zhang}, {Zhao}, \& {Zhao}}]{Raymond2024}
{Raymond}, A.~W., {Doeleman}, S.~S., {Asada}, K., {et~al.} 2024, \aj, 168, 130

\bibitem[{{Remijan} {et~al.}(2019){Remijan}, {Biggs}, {Cortes}, {Dent}, {Di
  Francesco}, {Fomalont}, \& {Hales}}]{Remijan2019}
{Remijan}, A., {Biggs}, A., {Cortes}, P., {et~al.} 2019, {ALMA Technical
  Handbook}

\bibitem[{{Taylor} {et~al.}(2009){Taylor}, {Stil}, \& {Sunstrum}}]{Taylor2009}
{Taylor}, A.~R., {Stil}, J.~M., \& {Sunstrum}, C. 2009, \apj, 702, 1230

\bibitem[{{Tchekhovskoy} {et~al.}(2011){Tchekhovskoy}, {Narayan}, \&
  {McKinney}}]{Tchekhovskoy2011}
{Tchekhovskoy}, A., {Narayan}, R., \& {McKinney}, J.~C. 2011, \mnras, 418, L79

\end{thebibliography}
%=======================================================================
%
\begin{appendix}
\onecolumn

\section{Stokes parameters per ALMA frequency band (SPW)}
\label{app:stokes}
Table~\ref{tab:EHT_uvmf_RM} in the main text reports the polarization quantities averaged across the four SPWs.  Here we report  the polarimetric quantities (Stokes\,$IQUV$, LP, and EVPA) for each SPW in Table~\ref{tab:EHT_uvmf_spw}. 
The quoted uncertainties include the thermal error, the $1\sigma$ systematic error associated with Stokes~$I$ leakage into Stokes~$Q$, $U$ (0.03\% of Stokes~$I$), and Stokes~$V$ (0.6\% of Stokes~$I$), as recommended by the ALMA observatory \citep[see ALMA Technical Handbook;][]{Remijan2019}. The total uncertainties are computed by combining these contributions in quadrature. 
The uncertainty in LP is primarily dominated by systematic errors, except for the weakest sources, where thermal noise becomes significant. Unlike \citet{Goddi2021}, we did not apply an LP bias correction because our sample does not include low-polarization  sources with LP$<<1$\%. 
We note that Stokes  $V$ is assumed to be zero during polarization calibration \citep[see][]{QA2Paper}, and no additional calibration step was applied to reliably constrain circular polarization (CP) in this experiment. Consequently, we do not claim CP detections for the observed sources. For an assessment of the reliability of Stokes $V$ detections in interferometric observations using linearly polarized feeds, we refer the reader to Appendix G in \citet{Goddi2021}.

\begin{table*}
\caption{Polarization parameters of targeted AGN sources per frequency band (SPW).} \label{tab:EHT_uvmf_spw}
\begin{tabular}{ccccccc} 
\hline\hline
Frequency & I & Q & U & V & LP & EVPA\\
(GHz) & (Jy) & (mJy) & (mJy) & (mJy) & (\%) & (deg)\\
\hline\hline
\multicolumn{7}{c}{VLBI scans}\\
\hline
\multicolumn{7}{c}{3C279}\\
335.60 & 8.8086 $\pm$ 0.0003 & -63.2 $\pm$ 2.6 & -923.0 $\pm$ 2.6 & -52 $\pm$ 53 & 10.50 $\pm$ 0.03 & -46.960 $\pm$ 0.082 \\
337.54 & 8.7961 $\pm$ 0.0003 & -59.9 $\pm$ 2.6 & -927.2 $\pm$ 2.6 & -49 $\pm$ 53 & 10.56 $\pm$ 0.03 & -46.849 $\pm$ 0.081 \\
347.60 & 8.6754 $\pm$ 0.0003 & -50.1 $\pm$ 2.6 & -929.0 $\pm$ 2.6 & -43 $\pm$ 52 & 10.72 $\pm$ 0.03 & -46.542 $\pm$ 0.080 \\
349.60 & 8.6327 $\pm$ 0.0003 & -48.2 $\pm$ 2.6 & -923.8 $\pm$ 2.6 & -53 $\pm$ 52 & 10.72 $\pm$ 0.03 & -46.492 $\pm$ 0.080 \\
\hline
\multicolumn{7}{c}{3C273}\\
335.60 & 4.2364 $\pm$ 0.0001 & -45.4 $\pm$ 1.3 & -53.1 $\pm$ 1.3 & -44 $\pm$ 25 & 1.65 $\pm$ 0.03 & -65.28 $\pm$ 0.52 \\
337.54 & 4.2165 $\pm$ 0.0001 & -45.1 $\pm$ 1.3 & -54.2 $\pm$ 1.3 & -41 $\pm$ 25 & 1.67 $\pm$ 0.03 & -64.88 $\pm$ 0.52 \\
347.60 & 4.1226 $\pm$ 0.0001 & -41.1 $\pm$ 1.2 & -55.0 $\pm$ 1.2 & -36 $\pm$ 25 & 1.67 $\pm$ 0.03 & -63.39 $\pm$ 0.52 \\
349.60 & 4.1019 $\pm$ 0.0002 & -40.9 $\pm$ 1.2 & -55.4 $\pm$ 1.2 & -43 $\pm$ 25 & 1.68 $\pm$ 0.03 & -63.22 $\pm$ 0.52 \\
\hline
\multicolumn{7}{c}{M87}\\
335.60 & 0.9944 $\pm$ 0.0002 & 25.22 $\pm$ 0.34 & 12.04 $\pm$ 0.34 & -14.9 $\pm$ 6.0 & 2.81 $\pm$ 0.03 & 12.76 $\pm$ 0.35 \\
337.54 & 0.9881 $\pm$ 0.0002 & 25.33 $\pm$ 0.33 & 12.03 $\pm$ 0.33 & -14.3 $\pm$ 5.9 & 2.84 $\pm$ 0.03 & 12.70 $\pm$ 0.34 \\
347.60 & 0.9526 $\pm$ 0.0002 & 24.00 $\pm$ 0.32 & 11.37 $\pm$ 0.33 & -12.2 $\pm$ 5.7 & 2.79 $\pm$ 0.03 & 12.67 $\pm$ 0.35 \\
349.60 & 0.9451 $\pm$ 0.0002 & 23.66 $\pm$ 0.33 & 10.79 $\pm$ 0.33 & -13.6 $\pm$ 5.7 & 2.75 $\pm$ 0.04 & 12.26 $\pm$ 0.37 \\
\hline
\multicolumn{7}{c}{4C01.28}\\
335.60 & 1.5159 $\pm$ 0.0002 & 32.89 $\pm$ 0.50 & -126.68 $\pm$ 0.50 & 1.0 $\pm$ 9.1 & 8.63 $\pm$ 0.03 & -37.72 $\pm$ 0.11 \\
337.54 & 1.5102 $\pm$ 0.0002 & 32.80 $\pm$ 0.49 & -126.78 $\pm$ 0.49 & 1.0 $\pm$ 9.1 & 8.67 $\pm$ 0.03 & -37.75 $\pm$ 0.11 \\
347.60 & 1.4915 $\pm$ 0.0002 & 32.61 $\pm$ 0.49 & -125.31 $\pm$ 0.49 & 1.1 $\pm$ 8.9 & 8.68 $\pm$ 0.03 & -37.71 $\pm$ 0.11 \\
349.60 & 1.4906 $\pm$ 0.0002 & 32.67 $\pm$ 0.50 & -125.56 $\pm$ 0.50 & 1.3 $\pm$ 8.9 & 8.70 $\pm$ 0.03 & -37.71 $\pm$ 0.11 \\
\hline
\multicolumn{7}{c}{J1146+3958}\\
335.60 & 0.4007 $\pm$ 0.0003 & -10.78 $\pm$ 0.29 & 1.24 $\pm$ 0.29 & -7.0 $\pm$ 2.4 & 2.71 $\pm$ 0.07 & 86.73 $\pm$ 0.76 \\
337.54 & 0.3975 $\pm$ 0.0002 & -10.65 $\pm$ 0.27 & 1.34 $\pm$ 0.27 & -6.2 $\pm$ 2.4 & 2.70 $\pm$ 0.07 & 86.42 $\pm$ 0.71 \\
347.60 & 0.3921 $\pm$ 0.0003 & -10.49 $\pm$ 0.28 & 0.79 $\pm$ 0.27 & -5.8 $\pm$ 2.4 & 2.68 $\pm$ 0.07 & 87.85 $\pm$ 0.74 \\
349.60 & 0.3953 $\pm$ 0.0003 & -10.03 $\pm$ 0.30 & 0.67 $\pm$ 0.30 & -7.0 $\pm$ 2.4 & 2.54 $\pm$ 0.08 & 88.08 $\pm$ 0.86 \\
\hline
\multicolumn{7}{c}{J1512-0905}\\
335.60 & 1.7563 $\pm$ 0.0004 & -12.30 $\pm$ 0.63 & -14.67 $\pm$ 0.62 & -18 $\pm$ 10 & 1.09 $\pm$ 0.04 & -64.99 $\pm$ 0.94 \\
337.54 & 1.7541 $\pm$ 0.0003 & -11.27 $\pm$ 0.61 & -14.16 $\pm$ 0.61 & -16 $\pm$ 10 & 1.03 $\pm$ 0.03 & -64.25 $\pm$ 0.97 \\
347.60 & 1.7340 $\pm$ 0.0004 & -12.91 $\pm$ 0.61 & -14.21 $\pm$ 0.61 & -14 $\pm$ 10 & 1.11 $\pm$ 0.04 & -66.14 $\pm$ 0.92 \\
349.60 & 1.7254 $\pm$ 0.0004 & -12.90 $\pm$ 0.63 & -12.91 $\pm$ 0.63 & -18 $\pm$ 10 & 1.06 $\pm$ 0.04 & -67.48 $\pm$ 0.99 \\
\hline
\multicolumn{7}{c}{J1337-1257}\\
335.60 & 3.1590 $\pm$ 0.0004 & 505.4 $\pm$ 1.0 & -132.5 $\pm$ 1.0 & -105 $\pm$ 19 & 16.54 $\pm$ 0.03 & -7.344 $\pm$ 0.055 \\
337.54 & 3.1469 $\pm$ 0.0004 & 504.46 $\pm$ 0.99 & -129.79 $\pm$ 0.99 & -99 $\pm$ 19 & 16.55 $\pm$ 0.03 & -7.214 $\pm$ 0.055 \\
347.60 & 3.1199 $\pm$ 0.0004 & 508.98 $\pm$ 0.99 & -132.01 $\pm$ 0.99 & -88 $\pm$ 19 & 16.85 $\pm$ 0.03 & -7.270 $\pm$ 0.054 \\
349.60 & 3.1042 $\pm$ 0.0004 & 503.87 $\pm$ 1.00 & -133.33 $\pm$ 1.00 & -104 $\pm$ 19 & 16.79 $\pm$ 0.03 & -7.411 $\pm$ 0.055 \\
\hline\hline
\multicolumn{7}{c}{non-VLBI scans$^{a}$}\\
\hline\hline
\multicolumn{7}{c}{J1246-0730}\\
335.60 & 0.2826 $\pm$ 0.0003 & 1.50 $\pm$ 0.36 & -0.49 $\pm$ 0.36 & 0.7 $\pm$ 1.7 & 0.6 $\pm$ 0.1 & -9.1 $\pm$ 6.5 \\
337.54 & 0.2822 $\pm$ 0.0003 & 1.35 $\pm$ 0.35 & -1.03 $\pm$ 0.35 & 0.5 $\pm$ 1.7 & 0.6 $\pm$ 0.1 & -18.6 $\pm$ 6.0 \\
347.60 & 0.2776 $\pm$ 0.0004 & 1.16 $\pm$ 0.36 & -1.25 $\pm$ 0.36 & 0.0 $\pm$ 1.7 & 0.6 $\pm$ 0.1 & -23.5 $\pm$ 6.1 \\
349.60 & 0.2761 $\pm$ 0.0004 & 1.17 $\pm$ 0.38 & -0.76 $\pm$ 0.38 & 0.4 $\pm$ 1.7 & 0.5 $\pm$ 0.1 & -16.6 $\pm$ 7.9 \\
\hline
\multicolumn{7}{c}{4C01.28}\\
335.60 & 1.5126 $\pm$ 0.0002 & 33.13 $\pm$ 0.52 & -126.08 $\pm$ 0.52 & -0.0 $\pm$ 9.1 & 8.62 $\pm$ 0.03 & -37.64 $\pm$ 0.11 \\
337.54 & 1.5079 $\pm$ 0.0002 & 32.94 $\pm$ 0.52 & -126.40 $\pm$ 0.52 & -0.0 $\pm$ 9.1 & 8.66 $\pm$ 0.03 & -37.70 $\pm$ 0.11 \\
347.60 & 1.4920 $\pm$ 0.0003 & 32.82 $\pm$ 0.51 & -125.19 $\pm$ 0.51 & 0.1 $\pm$ 9.0 & 8.67 $\pm$ 0.03 & -37.66 $\pm$ 0.11 \\
349.60 & 1.4906 $\pm$ 0.0003 & 32.66 $\pm$ 0.52 & -125.53 $\pm$ 0.52 & 0.1 $\pm$ 8.9 & 8.70 $\pm$ 0.03 & -37.71 $\pm$ 0.12 \\
\hline\hline
\end{tabular}

\tablefoottext{a}{Non-VLBI scans on 3C273 were not included in this polarization analysis.} 
\end{table*}
%------------------------------------------------------------------------------------------------------------------------------------------------
\section{Comparison of Stokes parameters with the AMAPOLA polarimetric Grid Survey} 
\label{app:amapola_comp}
%------------------------------------------------------------------------------------------------------------------------------------------------

\begin{figure*}[ht!]
\centering
%\hspace{-1.8cm}
\includegraphics[width=0.3\linewidth]{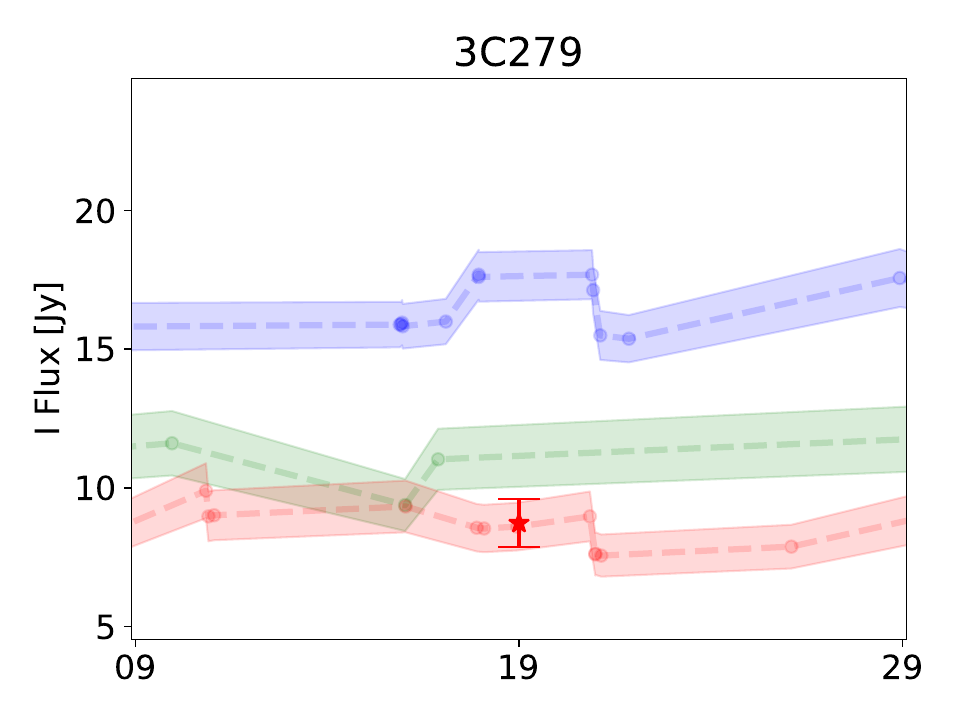} \hspace{-0.3cm}
\includegraphics[width=0.3\linewidth]{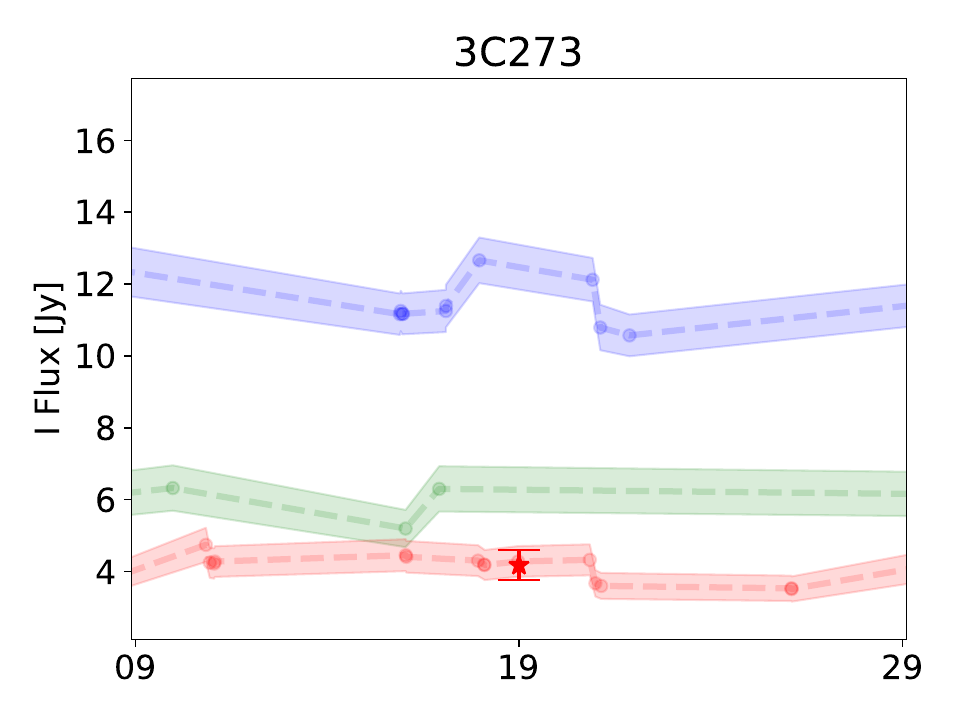}  \hspace{-0.3cm}
\includegraphics[width=0.3\linewidth]{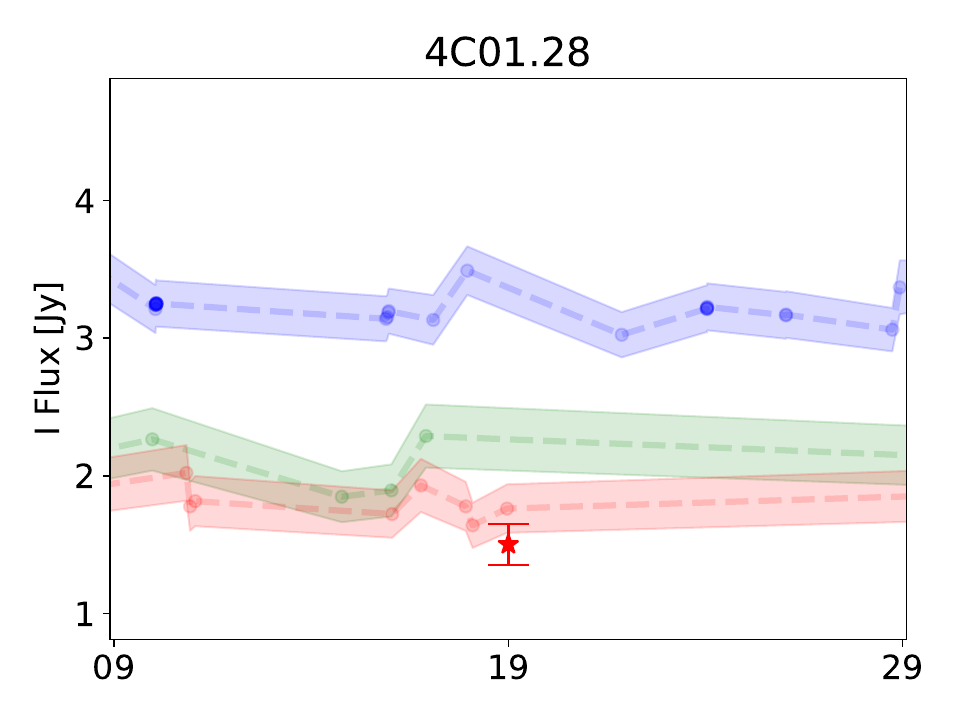}  \hspace{-0.3cm}
\includegraphics[width=0.3\linewidth]{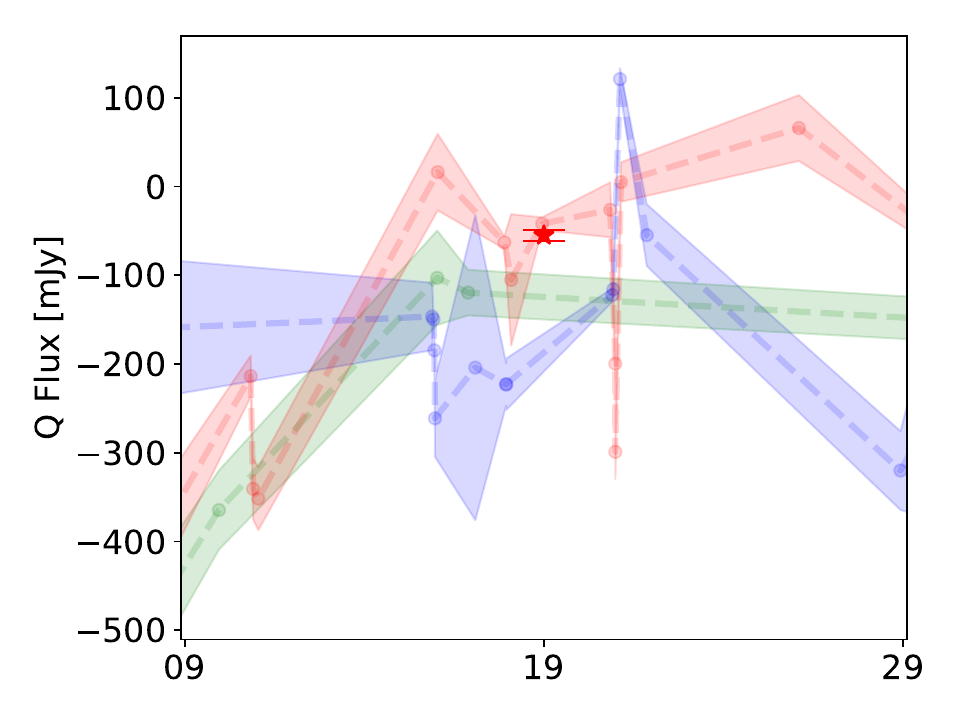} \hspace{-0.3cm}
\includegraphics[width=0.3\linewidth]{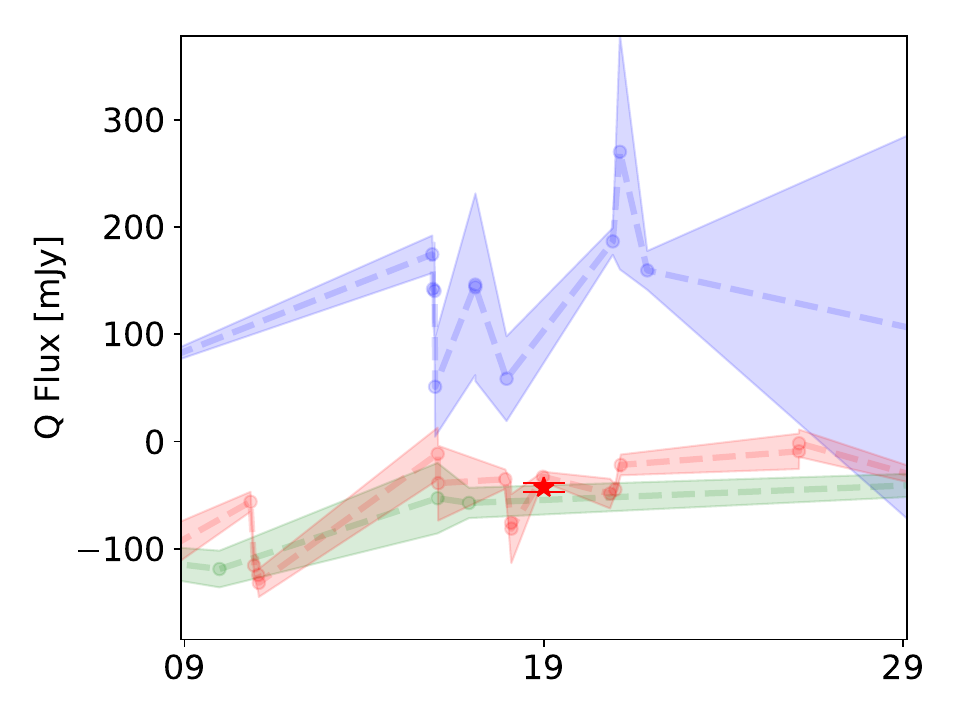}  \hspace{-0.3cm}
\includegraphics[width=0.3\linewidth]{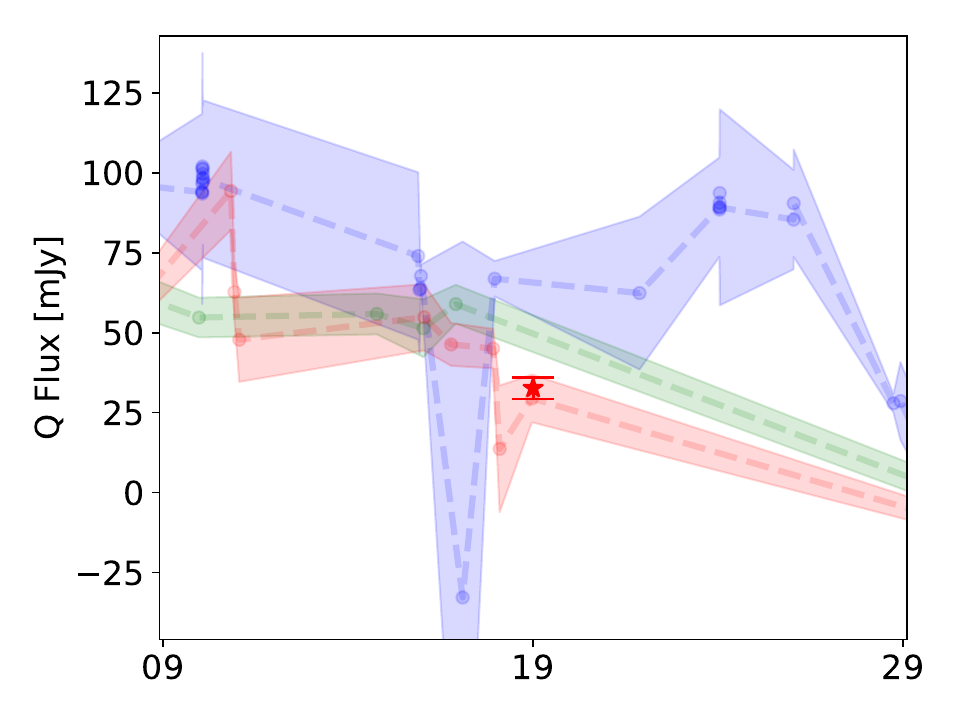}  \hspace{-0.3cm}
\includegraphics[width=0.3\linewidth]{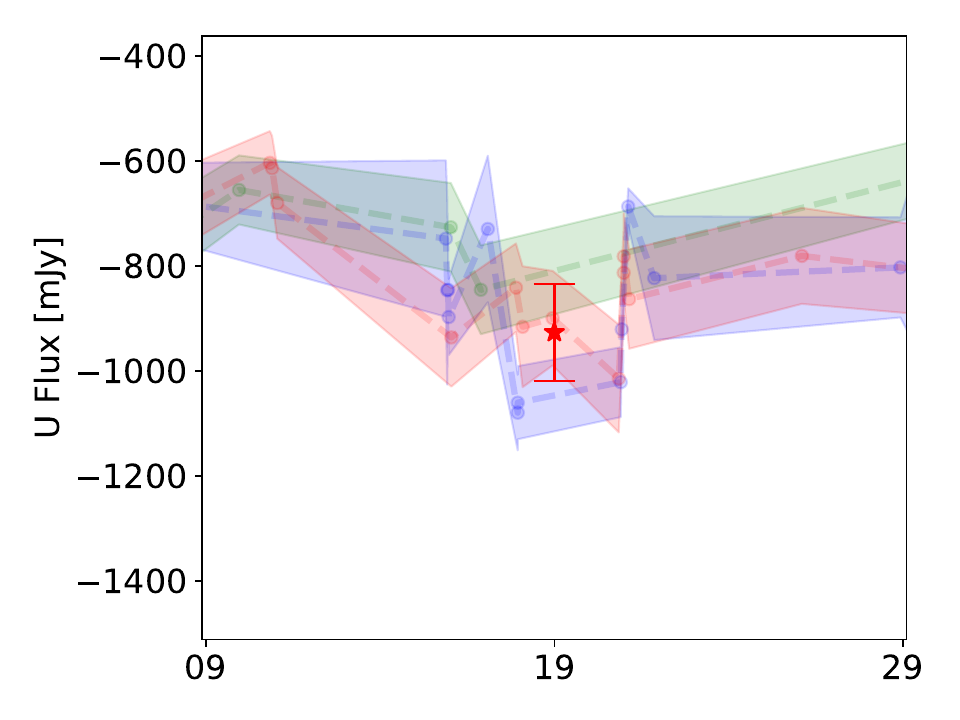} \hspace{-0.3cm}
\includegraphics[width=0.3\linewidth]{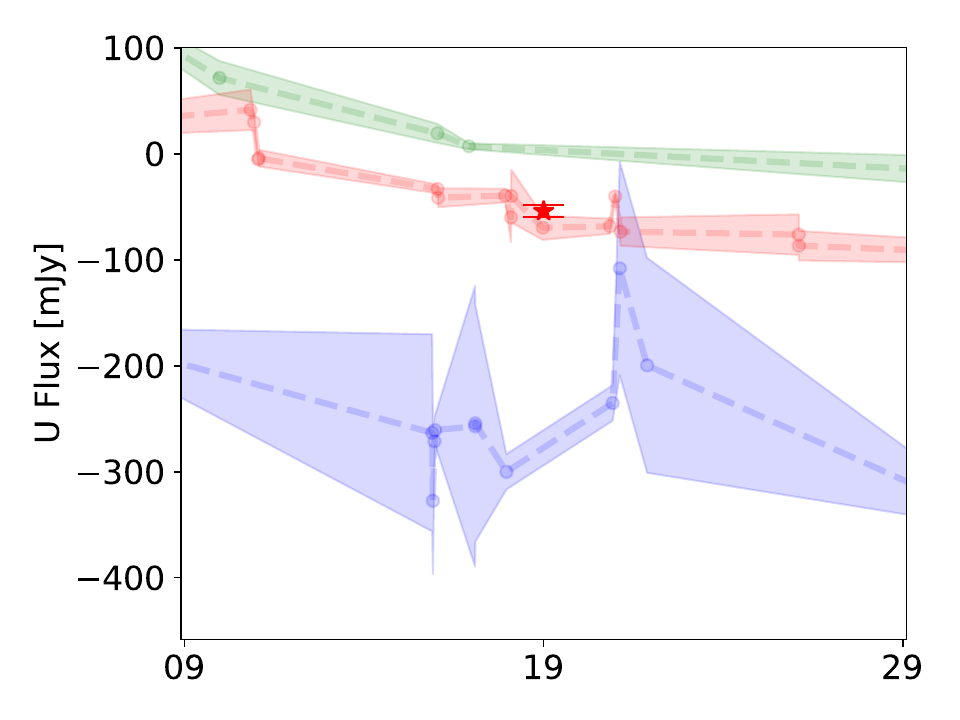}  \hspace{-0.3cm}
\includegraphics[width=0.3\linewidth]{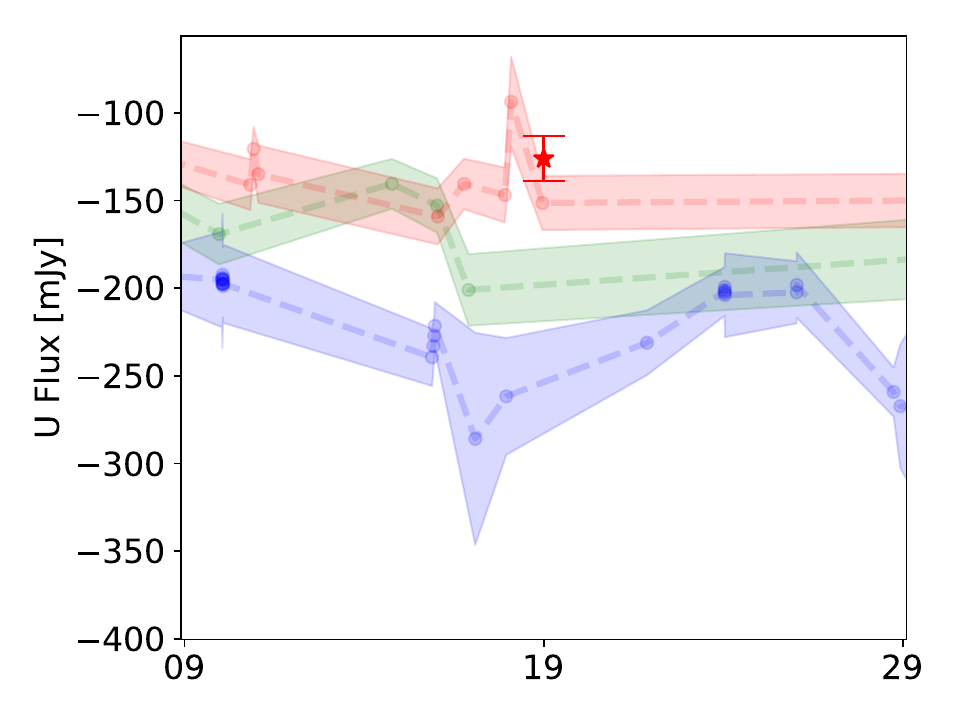}  \hspace{-0.3cm}
\includegraphics[width=0.3\linewidth]{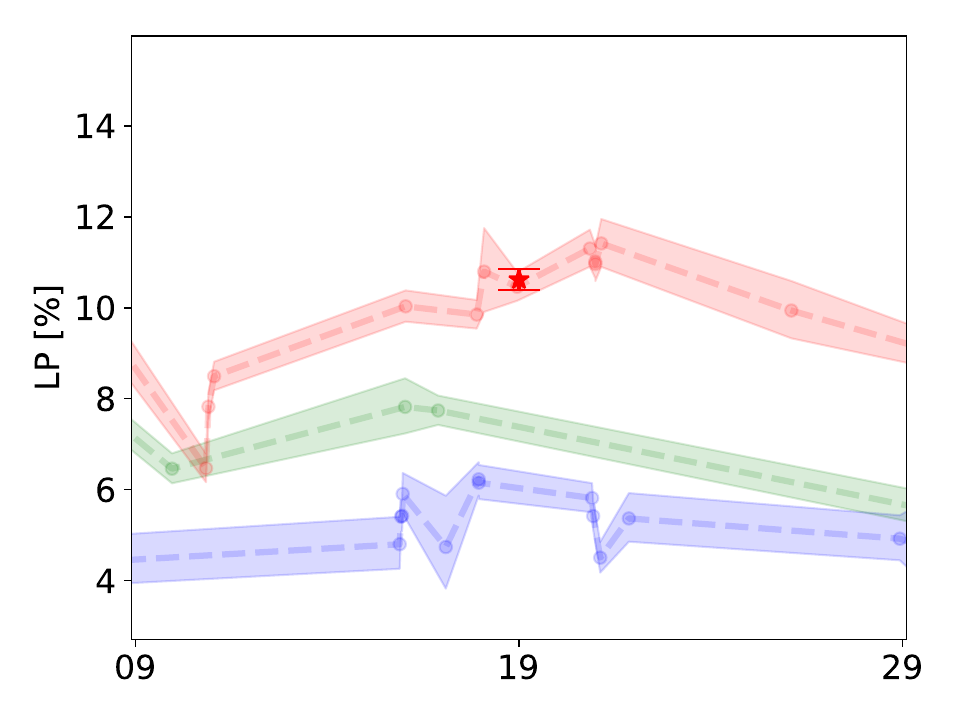} \hspace{-0.3cm}
\includegraphics[width=0.3\linewidth]{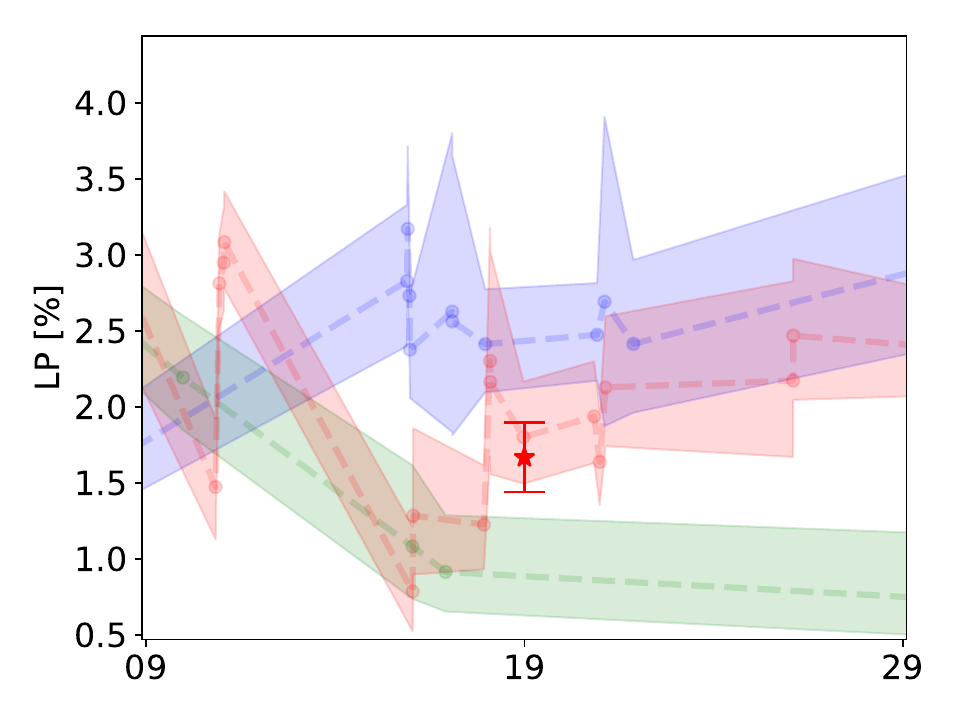}  \hspace{-0.3cm}
\includegraphics[width=0.3\linewidth]{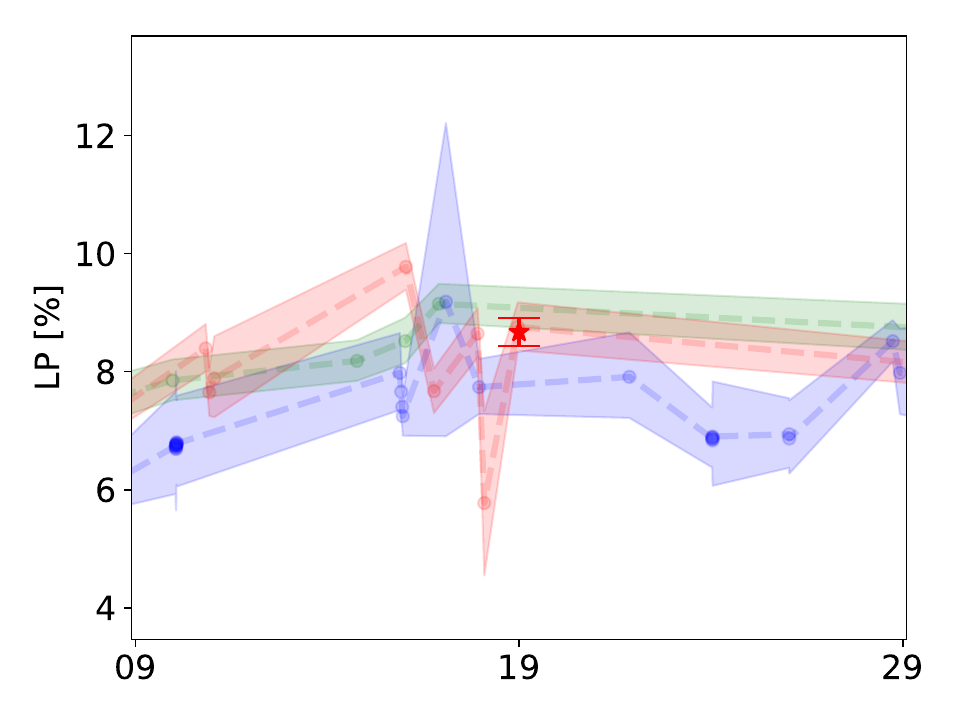}  \hspace{-0.3cm}
\includegraphics[width=0.3\linewidth]{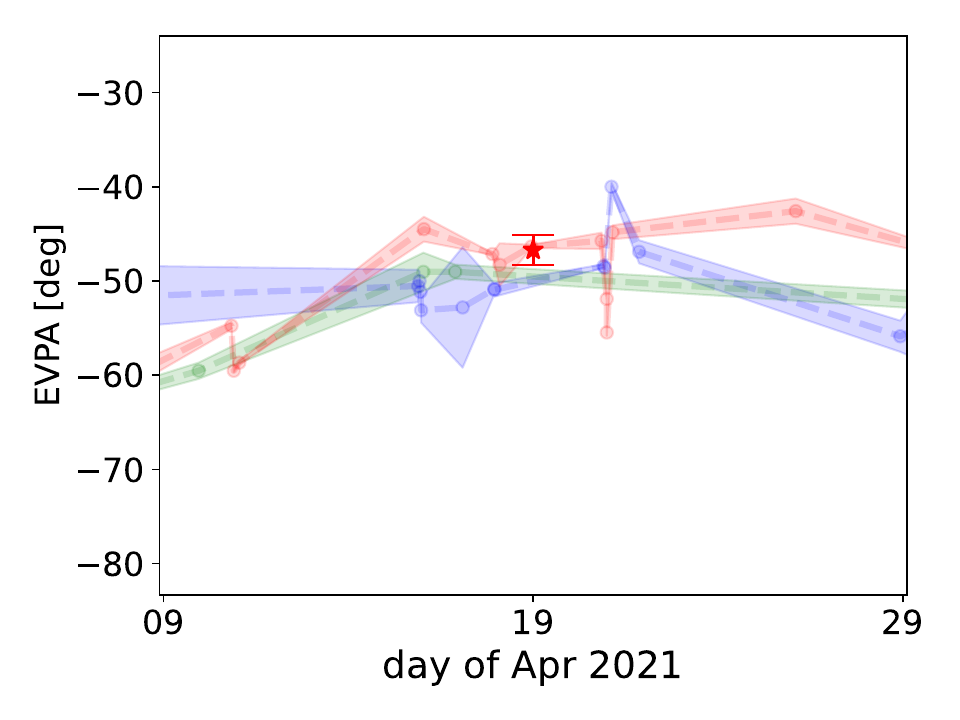} \hspace{-0.3cm}
\includegraphics[width=0.3\linewidth]{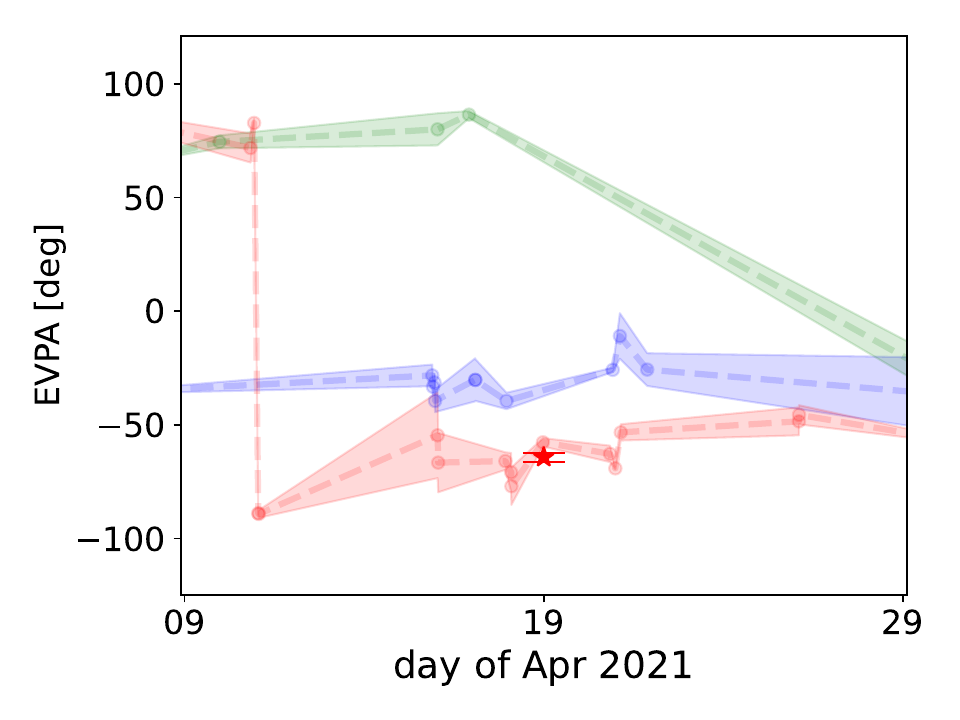}  \hspace{-0.3cm}
\includegraphics[width=0.3\linewidth]{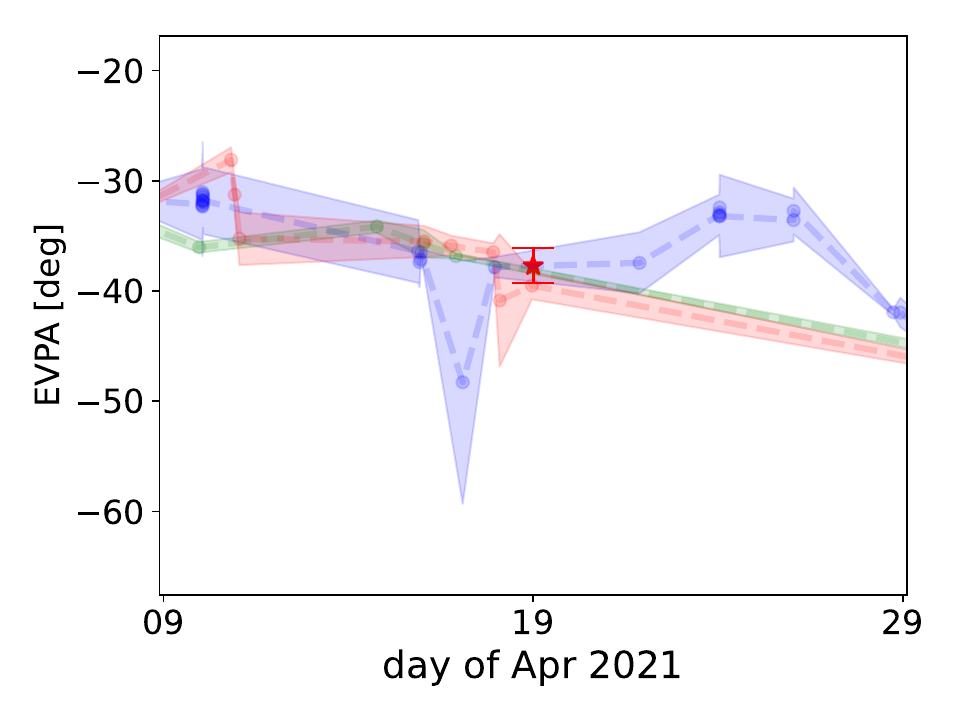}  \hspace{-0.3cm}
\caption{
Comparison of the polarimetric results obtained for all sources observed in VLBI mode on April 19, 2021, with those retrieved from the AMAPOLA polarimetric analysis of GS data. Each row shows a specific parameter (from top to bottom: Stokes~$I$, $Q$, $U$, LP, and EVPA), while each column corresponds to a different source (labels in the top panel; see also Fig.~\ref{fig:stokescomp_gs_2} for more sources).
The measurements from the ALMA-VLBI observations are indicated as red stars with associated error bars. The shaded regions correspond to AMAPOLA's $\pm1\sigma$ uncertainties for Band\,3 (97.5\,GHz; blue shading), Band\,6 (221.1\,GHz; green shading), and Band\,7 (343.4\,GHz; red shading). These uncertainties are derived from the ACA GS data, while their temporal evolution (lines) is interpolated between individual GS measurements.
This figure and Fig.~\ref{fig:stokescomp_gs_2} highlight that, for most cases, the ALMA-VLBI measurements agree well with the AMAPOLA trends, helping confirm the reliability of the polarization calibration across bands.}
\label{fig:stokescomp_gs_1}
\end{figure*}

\begin{figure*}[ht!]
\centering
%\hspace{-1.8cm}
\includegraphics[width=0.25\linewidth]{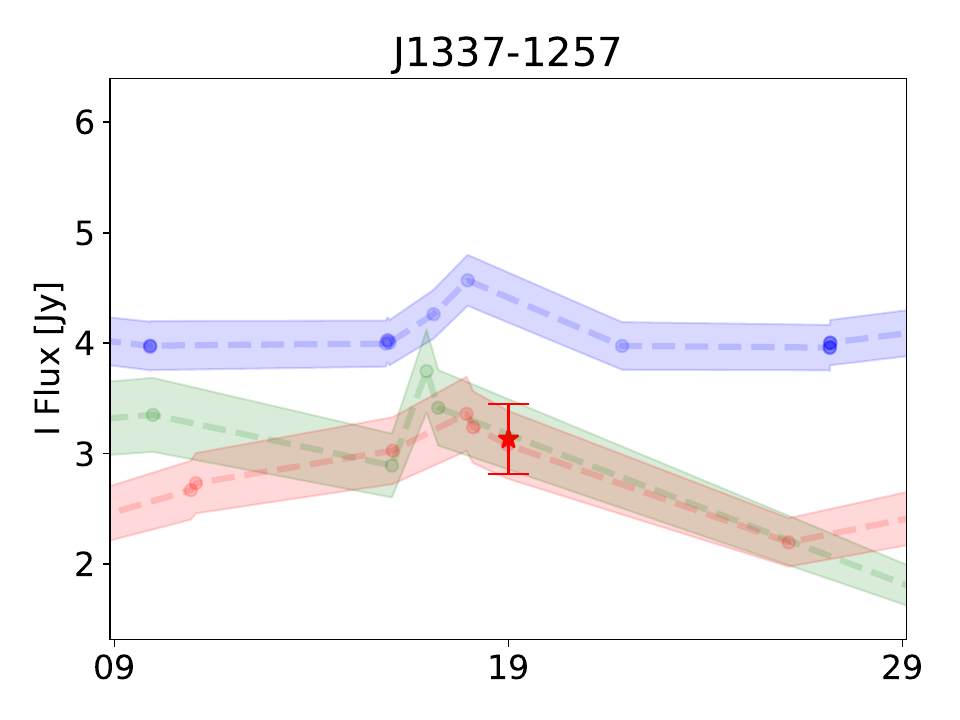} \hspace{-0.3cm}
\includegraphics[width=0.25\linewidth]{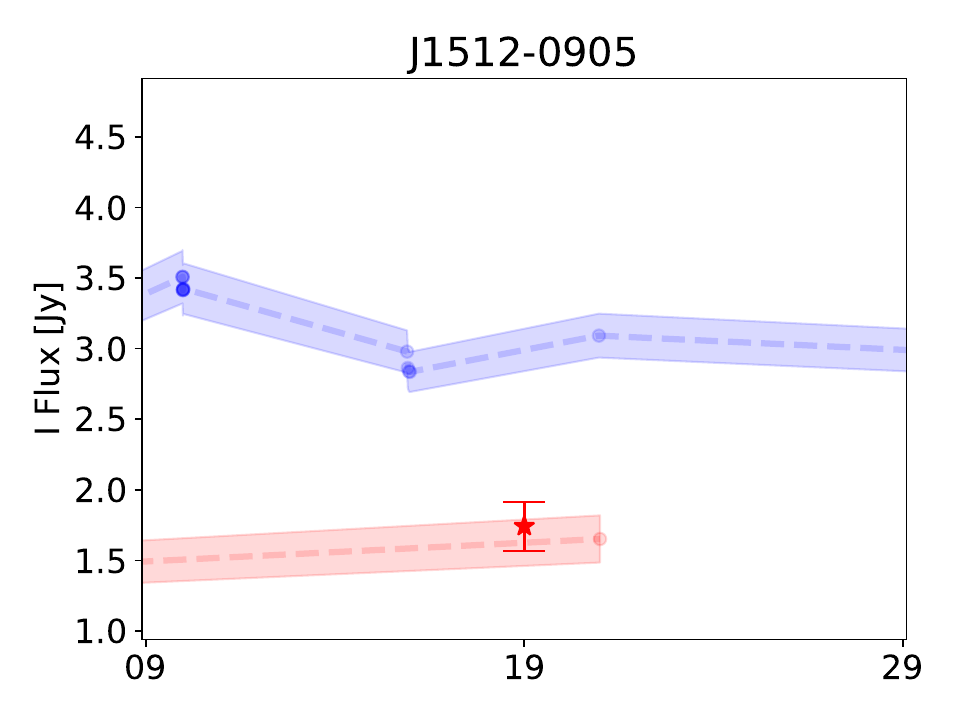}  \hspace{-0.3cm}
\includegraphics[width=0.25\linewidth]{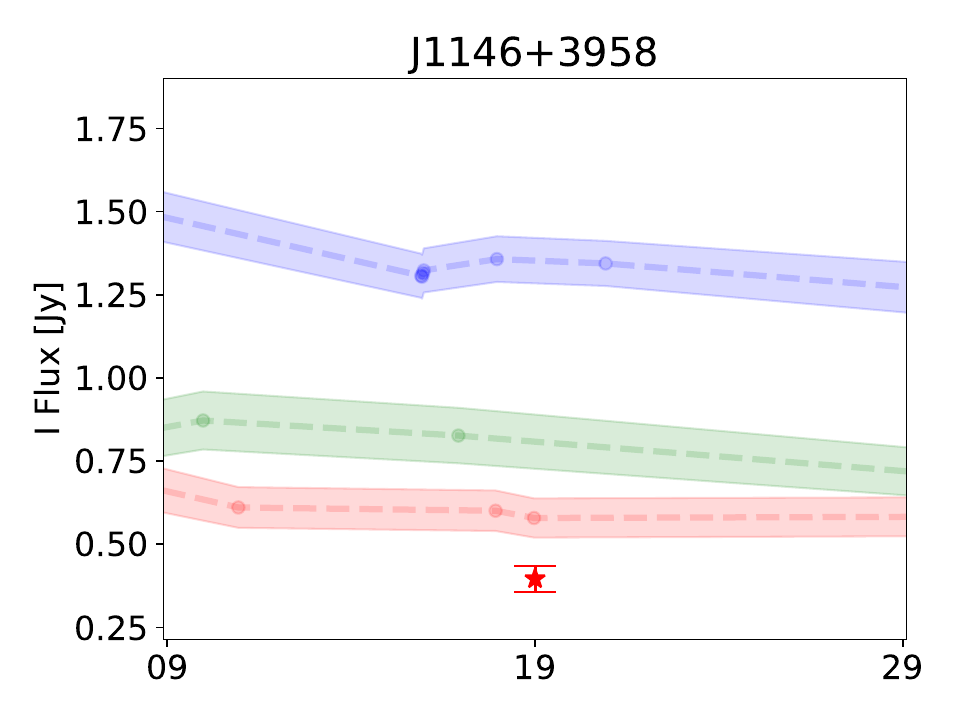}  \hspace{-0.3cm}
\includegraphics[width=0.25\linewidth]{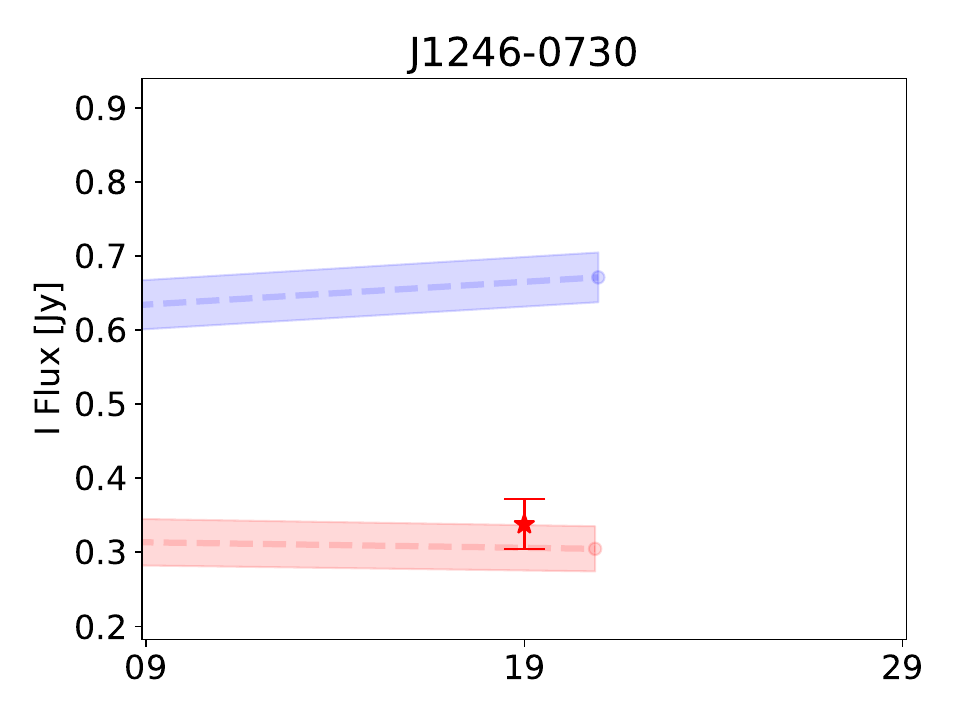}  
\includegraphics[width=0.25\linewidth]{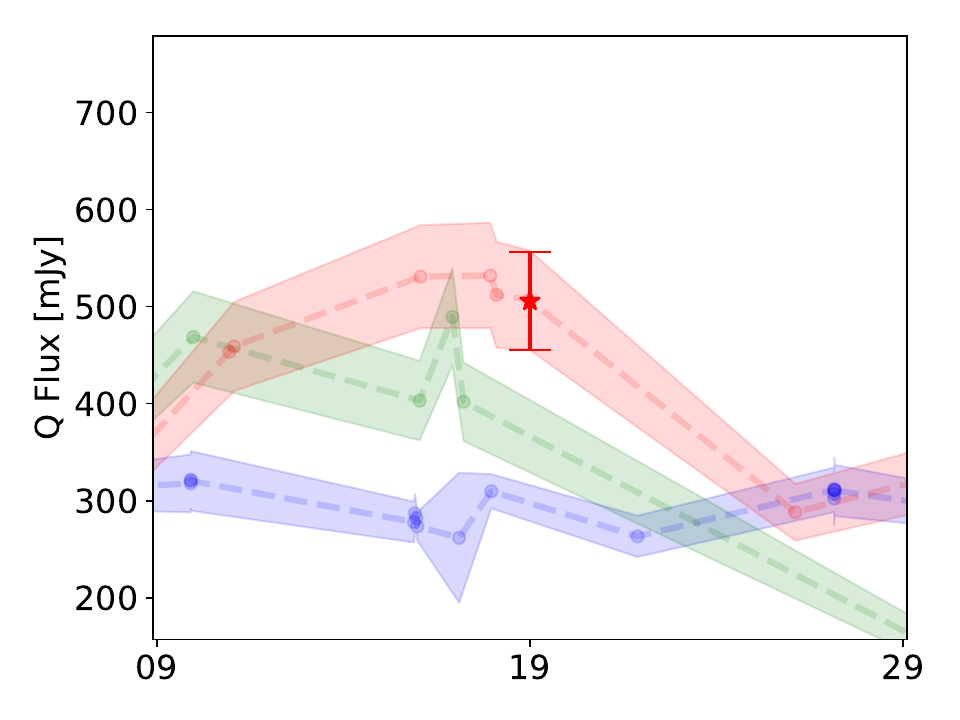} \hspace{-0.3cm}
\includegraphics[width=0.25\linewidth]{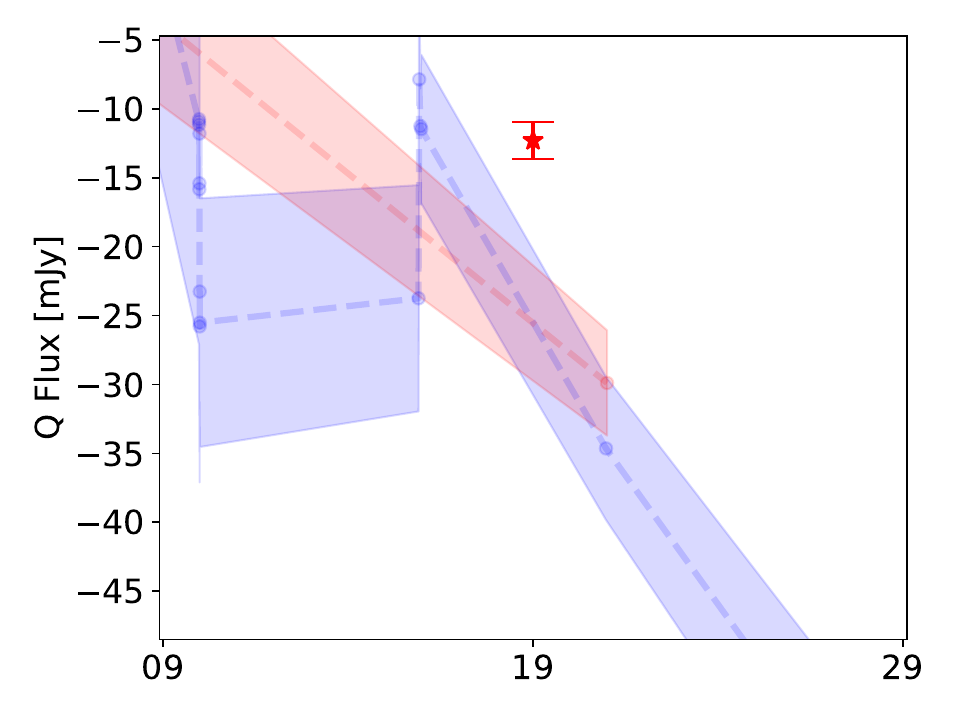}  \hspace{-0.3cm}
\includegraphics[width=0.25\linewidth]{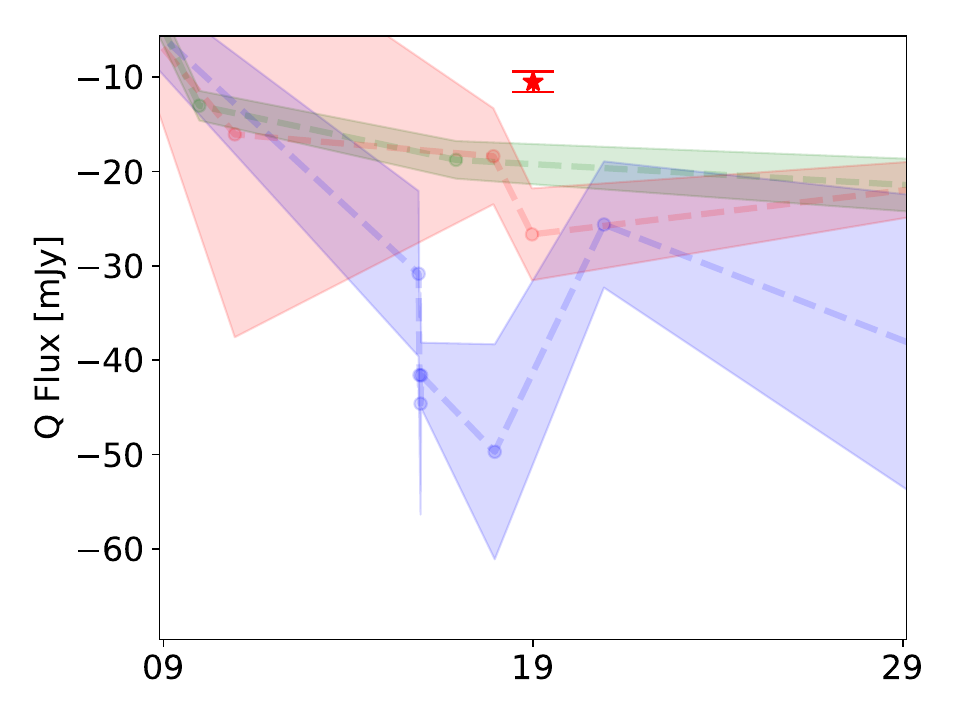}  \hspace{-0.3cm}
\includegraphics[width=0.25\linewidth]{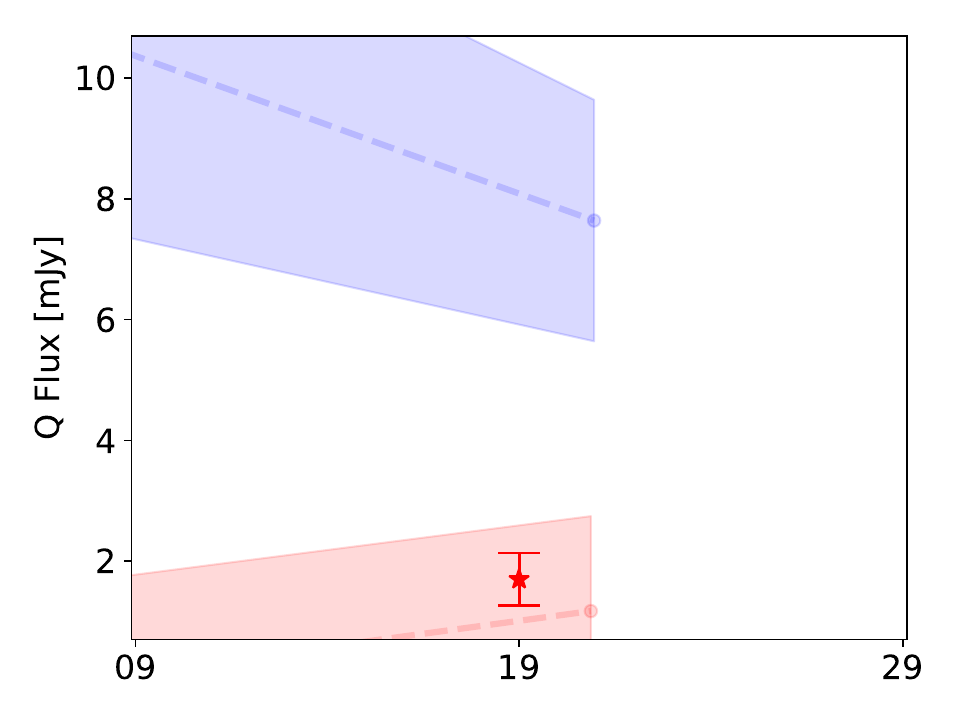}  
\includegraphics[width=0.25\linewidth]{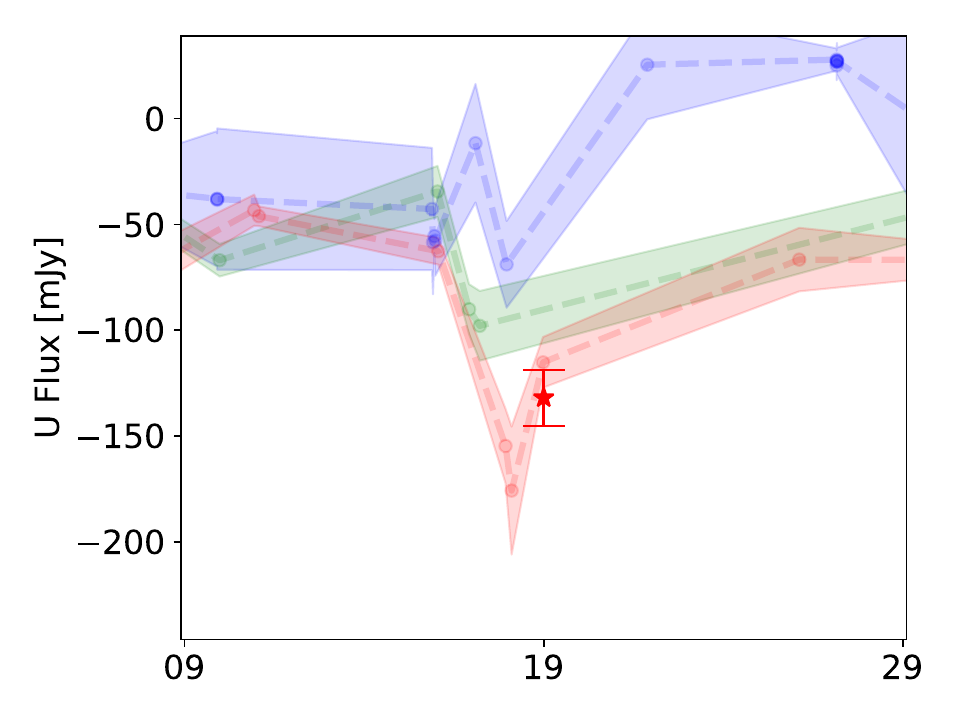} \hspace{-0.3cm}
\includegraphics[width=0.25\linewidth]{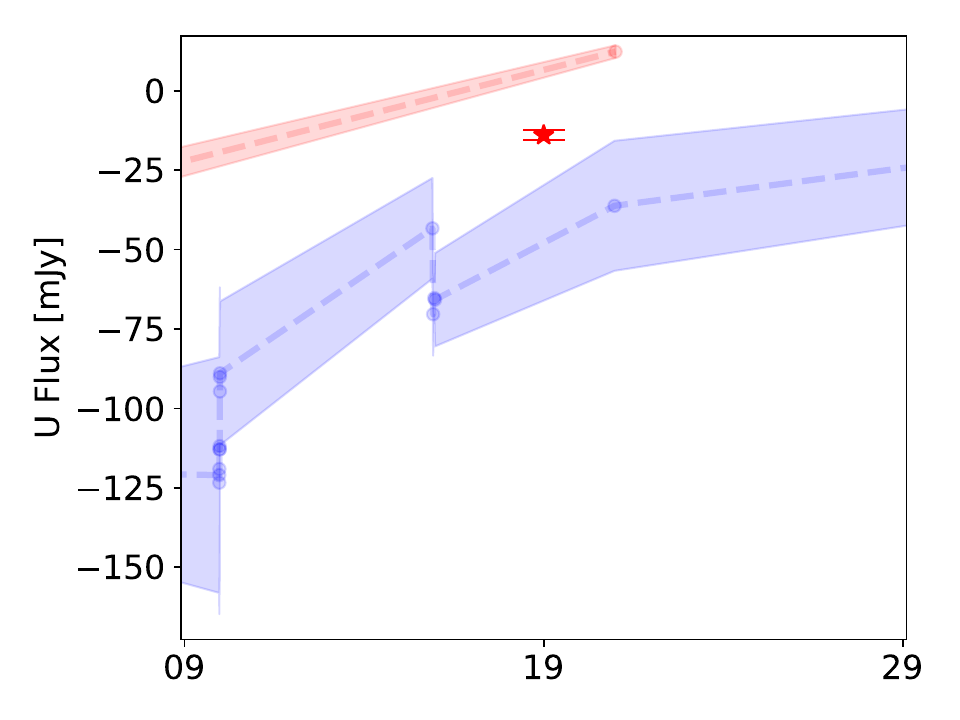}  \hspace{-0.3cm}
\includegraphics[width=0.25\linewidth]{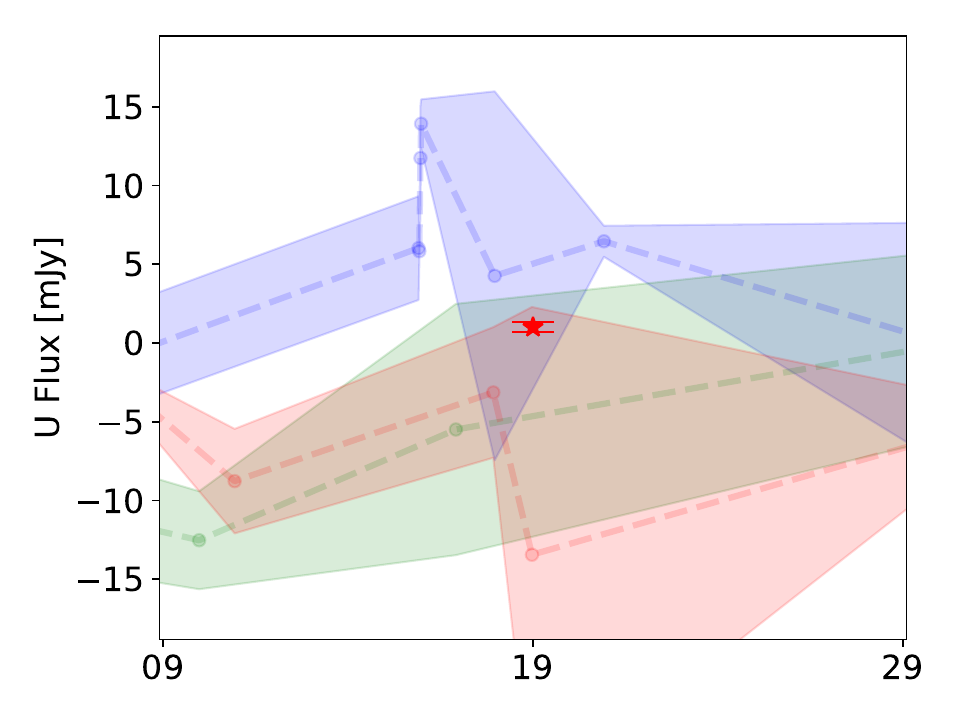}  \hspace{-0.3cm}
\includegraphics[width=0.25\linewidth]{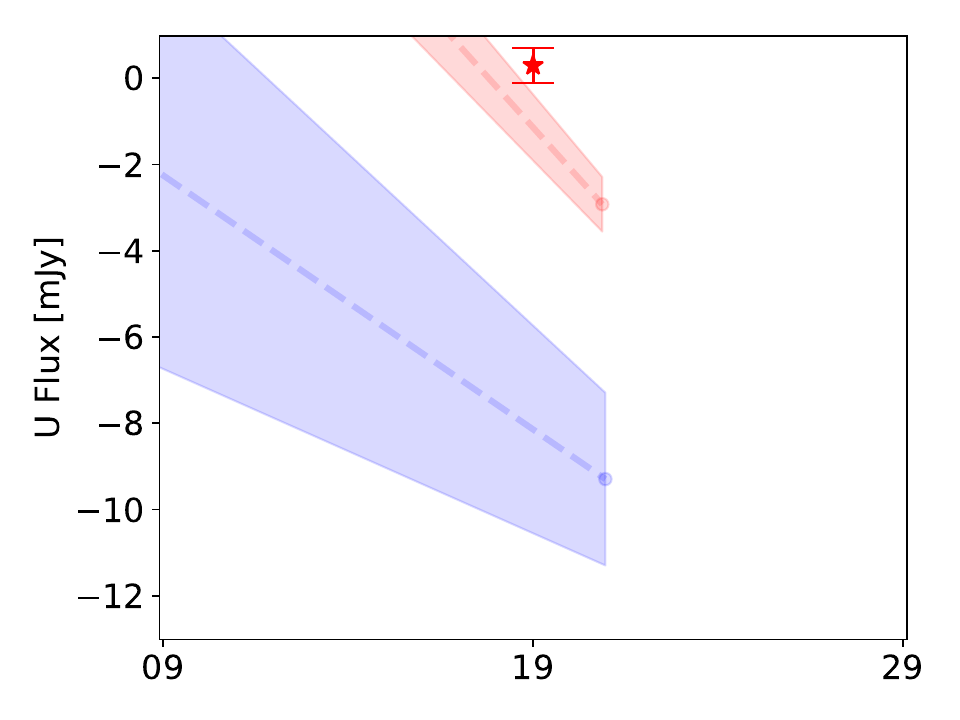}  
\includegraphics[width=0.25\linewidth]{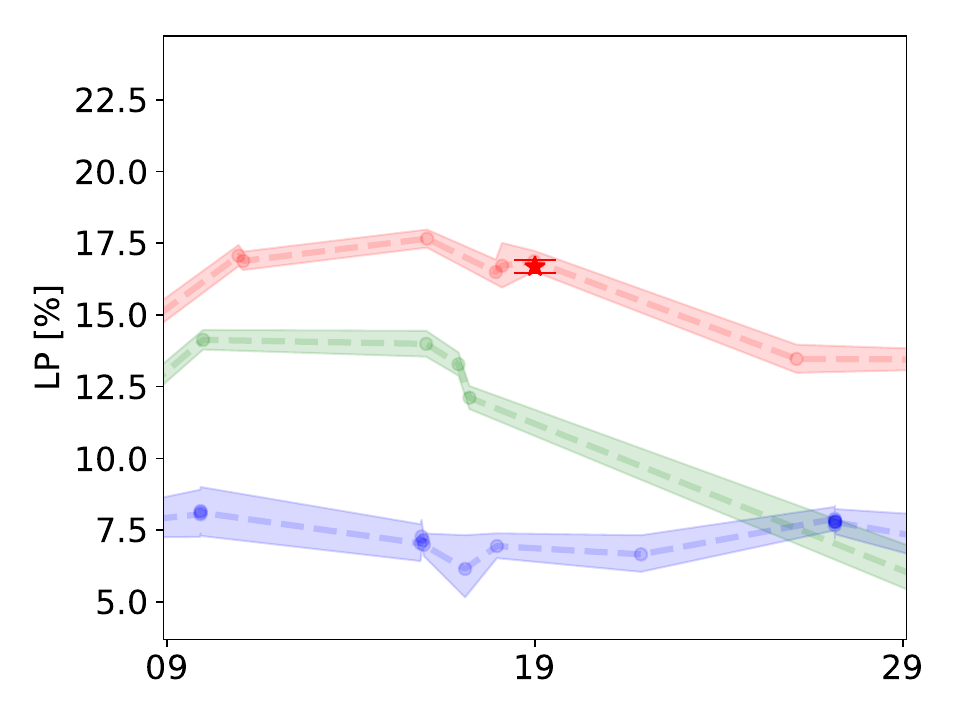} \hspace{-0.3cm}
\includegraphics[width=0.25\linewidth]{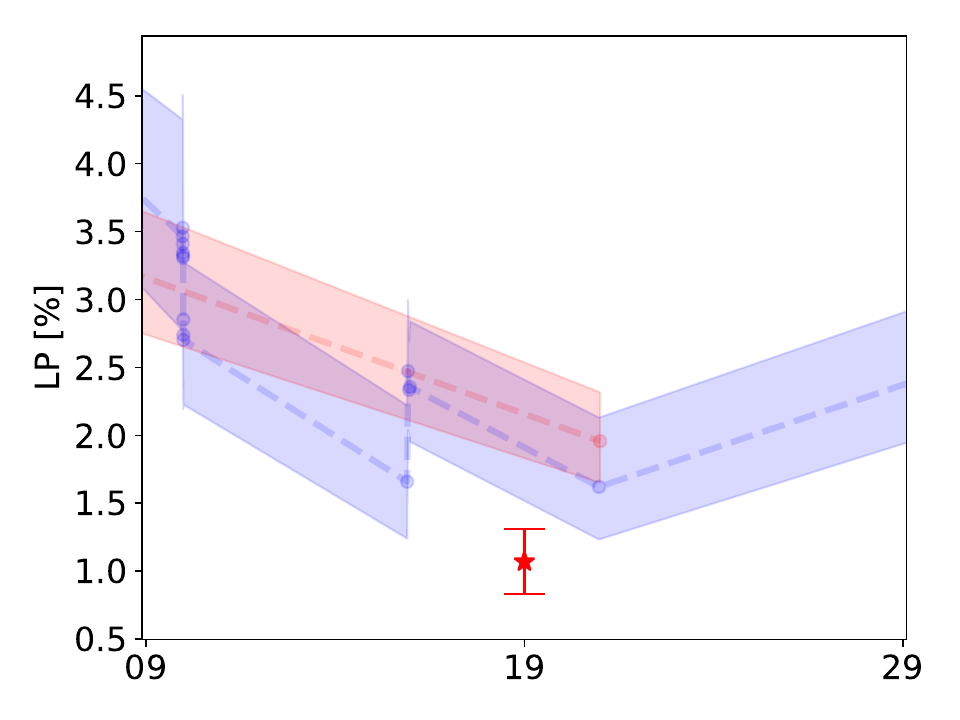}  \hspace{-0.3cm}
\includegraphics[width=0.25\linewidth]{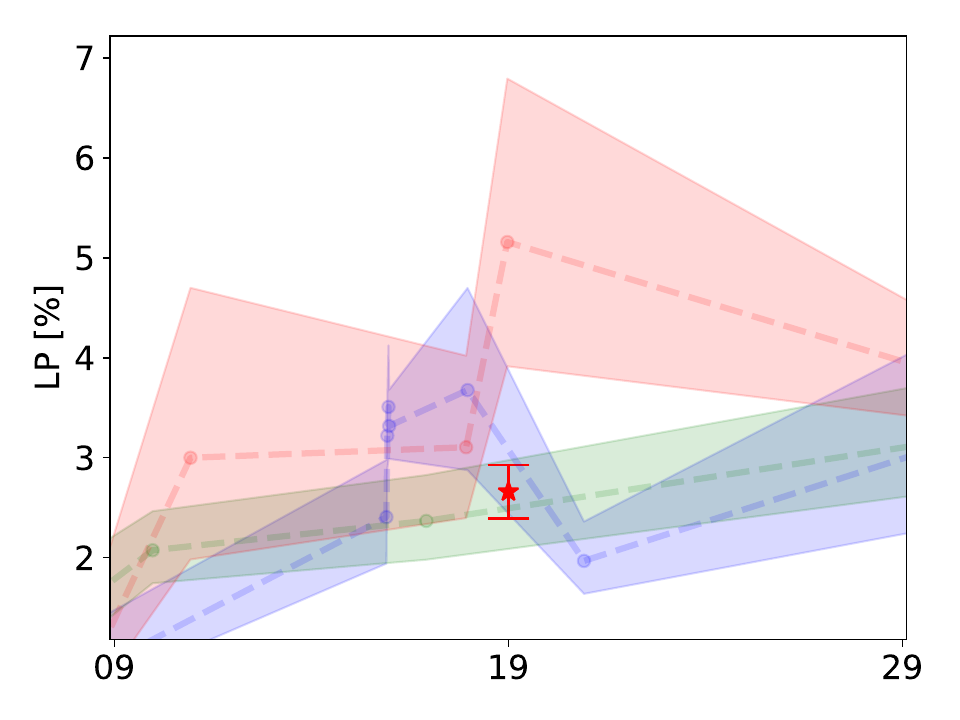}  \hspace{-0.3cm}
\includegraphics[width=0.25\linewidth]{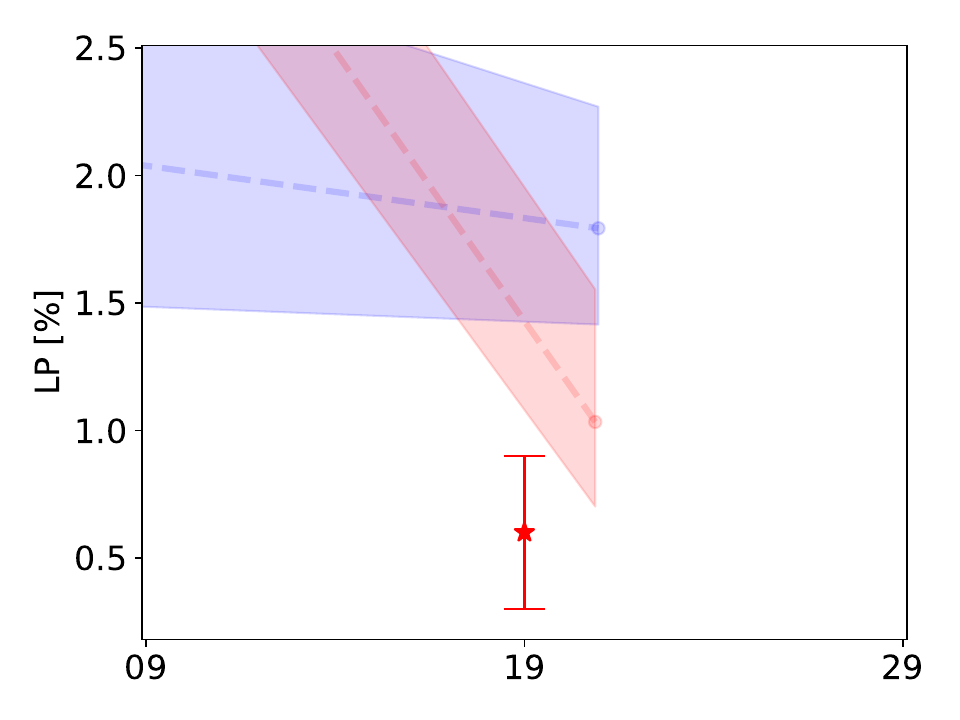}  

\includegraphics[width=0.25\linewidth]{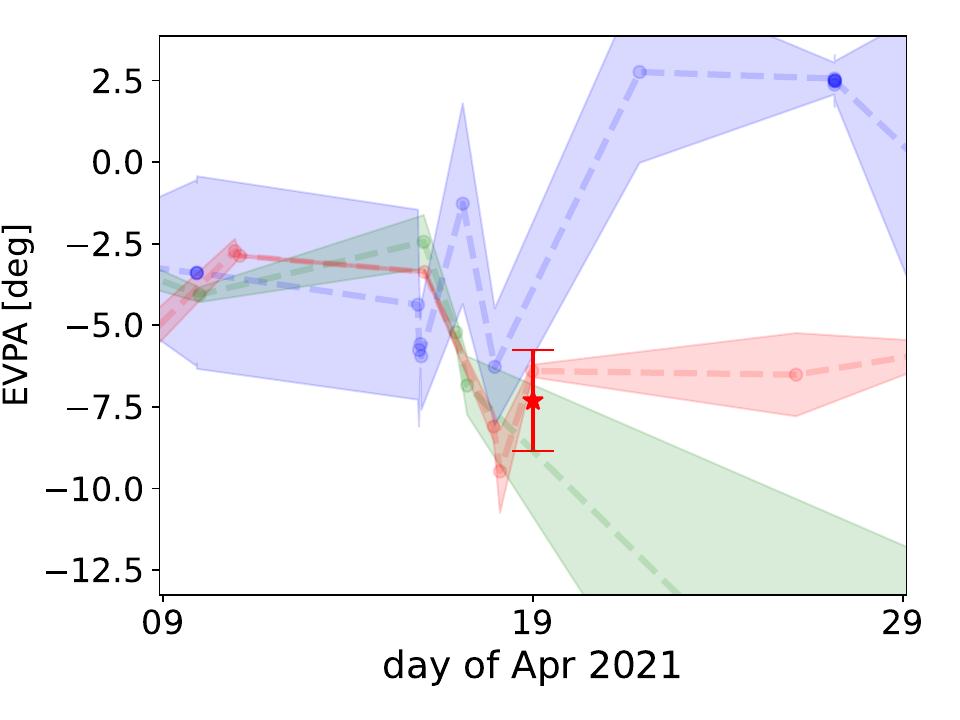} \hspace{-0.3cm}
\includegraphics[width=0.25\linewidth]{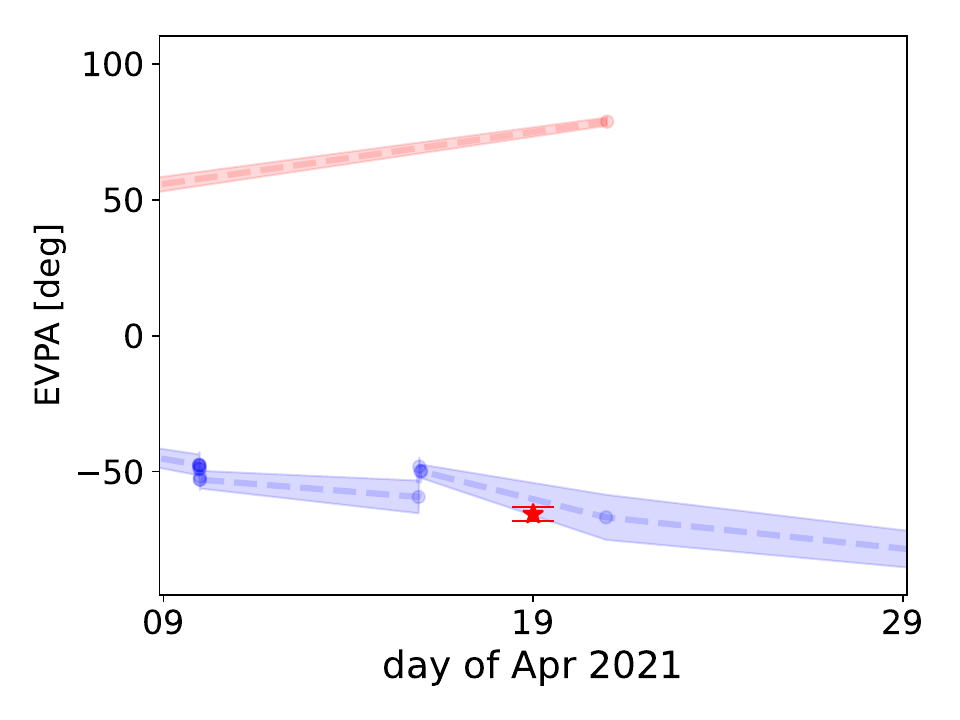}  \hspace{-0.3cm}
\includegraphics[width=0.25\linewidth]{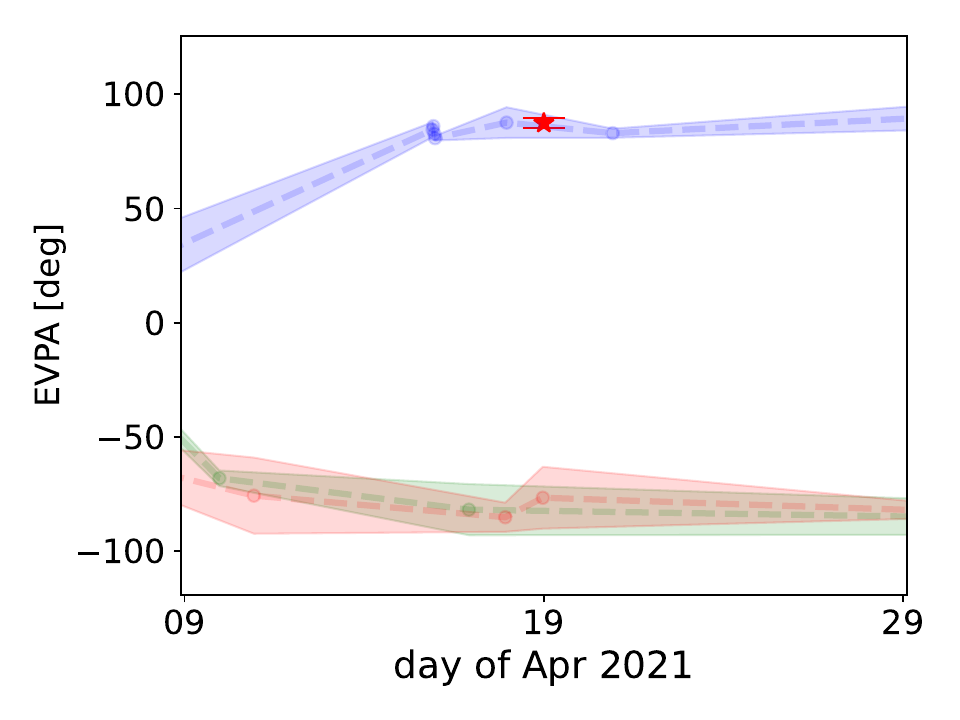}  \hspace{-0.3cm}
\includegraphics[width=0.25\linewidth]{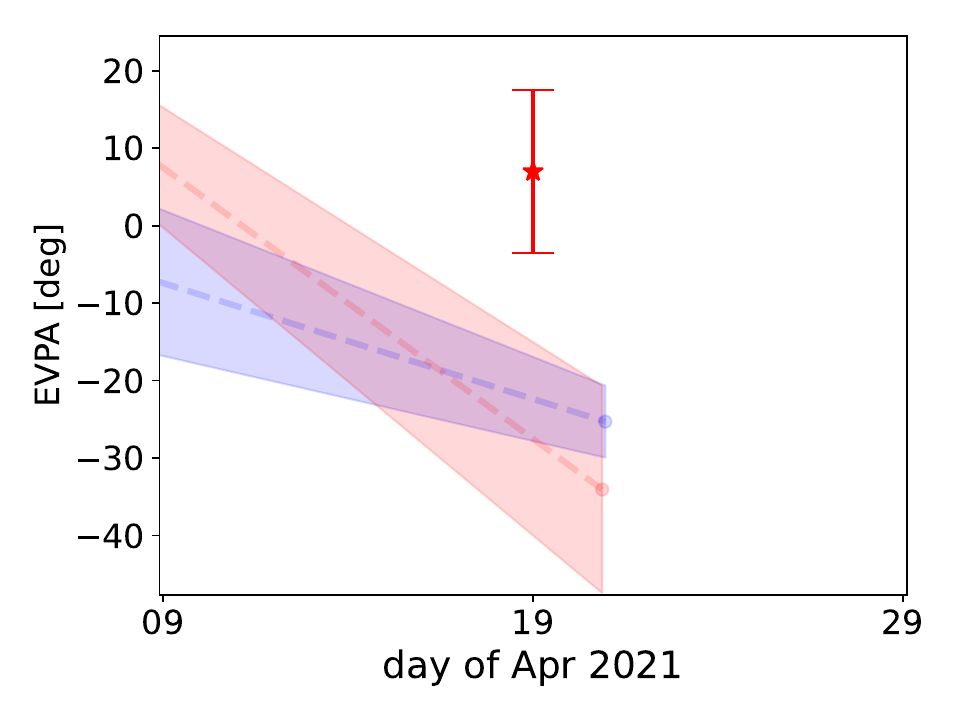}  

\caption{
Same as Fig.~\ref{fig:stokescomp_gs_1} but for different sources. The apparent discrepancy between the Band 7 GS prediction and our measurement of the EVPA for J1512-0905 and J1146+3958 is attributed to the $\pm \pi$ ambiguity in estimating the  
polarization direction \citep[e.g.,][]{Taylor2009}.
}
\label{fig:stokescomp_gs_2}
\end{figure*}

Figures~\ref{fig:stokescomp_gs_1} and \ref{fig:stokescomp_gs_2} illustrate the polarimetric parameters reported in Table~\ref{tab:EHT_uvmf_spw} (green data points and error bars), specifically Stokes $I$, $Q$, $U$, LP, and EVPA. The shaded $\pm1\sigma$ regions represent the time-variance of the same parameters as measured by AMAPOLA over a 20-day period surrounding the VLBI observations (April 9 to 29, 2021). Inflections in the shaded trend lines indicate the times of GS observations. Blue corresponds to Band~3 measurements, while green and red corresponds to Bands 6 and 7, respectively.
The figures show that most Band~7 measurements derived from our current study fall within the red-shaded regions, demonstrating general consistency with the AMAPOLA trends. The observed discrepancies may align with inter-GS cadence variability or differential time-variability between frequency bands, as also seen in some AMAPOLA monitoring cases.

We conclude that, despite differences in array configurations and data reduction methods, the ALMA-VLBI results are consistent with AMAPOLA. This consistency validates the accuracy of flux density and polarization calibration in VLBI mode for Band~7, extending the reliability previously demonstrated in Bands~3 and 6 \citep{QA2Paper,Goddi2021}.

%------
%------
%------

\end{appendix}

\end{document}